\documentclass[10pt,letterpaper]{article}
\usepackage[top=0.85in,footskip=0.75in]{geometry}
\usepackage{comment}
\usepackage{adjustbox}

\usepackage{amsmath,amssymb}

\usepackage{changepage}

\usepackage[utf8x]{inputenc}

\usepackage{textcomp,marvosym}

\usepackage{cite}

\usepackage{nameref,hyperref}

\usepackage{microtype}
\DisableLigatures[f]{encoding = *, family = * }

\usepackage[table]{xcolor}

\usepackage{array}
\usepackage{float}
\usepackage{ulem}

\newcolumntype{+}{!{\vrule width 2pt}}

\newlength\savedwidth

\newcommand{\isEquivTo}[1]{\underset{#1}{\sim}}

\raggedright
\usepackage[aboveskip=1pt,labelfont=bf,labelsep=period,justification=raggedright,singlelinecheck=off]{caption}

\makeatletter
\renewcommand{\@biblabel}[1]{\quad#1.}
\makeatother

\usepackage{lastpage,fancyhdr,graphicx}

\usepackage{multirow} 

\usepackage{epstopdf}
\fancyheadoffset[L]{2.25in}
\fancyfootoffset[L]{2.25in}
\definecolor{mygreen}{rgb}{0.0, 0.5, 0.0}
\newcommand{\lu}[1]{\textcolor{mygreen}{#1}}

\begin{document}
\vspace*{0.2in}

\begin{flushleft}
{\Large
\textbf\newline{Applicability of Hubbert model to global mining industry: Interpretations and insights}}




\medbreak
\medbreak
Lucas Riondet\textsuperscript{1,2,3*},
Daniel Suchet\textsuperscript{4},
Olivier Vidal\textsuperscript{5}
José Halloy\textsuperscript{3,**}
\\
\bigskip
\textbf{1} Univ. Grenoble Alpes, CNRS, Grenoble INP, G-SCOP, 38000 Grenoble, France 
\\
\textbf{2} I2M Bordeaux, UMR 5295, Institut de Chambéry, 73370 Le Bourget du Lac, France 
\\
\textbf{3} Université Paris Cité, CNRS, LIED UMR 8236, F-75006 Paris, France.
\\
\textbf{4} Institut du Photovoltaïque d'Ile de France IPVF UMR 9006, CNRS, Ecole Polytechnique, 91120 Palaiseau, France.
\\
\textbf{5} Univ. Grenoble Alpes, CNRS, Institut des sciences de la Terre, Grenoble, France

\bigskip

%
%





* lucas.riondet@grenoble-inp.fr
** jose.halloy@u-paris.fr

\bigskip
This work has been submitted to PLOS Sustainability and Transformation.

\end{flushleft}

\section*{Abstract}

The Hubert's model has been introduced in 1956 as a phenomenological description of the time evolution of US oil fields production. It has since then acquired a vast notoriety as a conceptual approach to resource depletion. It is often invoked nowadays in the context of the energy transition to question the limitations induced by the finitude of mineral stocks. Yet, its validity is often controversial despite its popularity. This paper offers a pedagogical introduction to the model, assesses its ability to describe the current evolution of 20 mining elements, and discusses the nature and robustness of conclusions drawn from Hubbert's model considered either as a for cast or as a foresight tool. We propose a novel way to represent graphically these conclusions as a "Hubbert's map" which offers direct visualization of their main features.

\pagebreak
\tableofcontents
\section{Introduction}

Access and use of resources represent a key issue which has been exacerbated along the rising and the growing complexity of modern societies. 
A popular way of looking at this issue may take roots in the writings of the economist \textit{Thomas Malthus (1766-1834)} \cite{Malthus1803} and of the mathematician philosopher \textit{Nicolas de Condorcet (1743-1794)} \cite{Condorcet1797}, both conceptualizing that resources needs could not be covered in the future.
In their works, population is considered as the main factor causing the food resources decline and agricultural outputs were linearly increasing. Establishing that population tended to rise faster than yields, Malthus predicted massive food shortage for the following century, while Condorcet submitted the concept of "progress" which should make production means improving at the same time as population growth. 
Their conclusions differ broadly, due to their specific conception of societies and of technological impacts. In other words, their models are not only scientific objects but also support a societal vision.	
During the 20th century technical progress has brought about rising of production and then has turned the limitation of producing rate problematic to concern about limitation of resources themselves.     
Moreover, technical progress also has made fossil energy production a centerpiece of society, supplanting then conditioning food supply.
With the increasing share of fossil fuel, draining from a finite stock of resources, the issue raised by Malthus and Condorcet naturally turned towards the energy sector. This concern is first epitomized in 1865 by \emph{the Coal question} of W. Jevons, which foresees many of our contemporary challenges \cite{Jevons1865}. But the most celebrated approach is undoubtedly the 1956 work of Marion King Hubbert \cite{Hubbert1956}. While crude oil extractions in the US were booming, with an output doubling every 9 years, Hubbert addressed the question of the continuation of the observed trends. Considering a model building on Verhulst equation also called \emph{logistic curve} \cite{verhulst1845} \cite{Vogels1975Verhulst}, Hubbert envisioned a peak of the US conventional oil production 10 years in the future, followed by a rapid decrease. The simplicity of the model, the audaciousness of the conclusions and the remarkable accuracy of this trajectory up to the mid-2000' ensured the reputation of this work \cite{Hook2011}. 

Since this seminal study, a large amount of work has been dedicated to investigate the ramification of a depletion dynamics with various types of models. Based on system dynamics, the \emph{The limit to Growth} \cite{Meadows1972} defended a holistic approach on a global scale and promoted to widen the scope of exhaustion models, including to mineral resources. The model was based on curves similar to the Verhulst one, the Gompertz function, and has popularized depletion curves. In the continuity, the cumulative property of depletion curves implying that the sum of several production peaks gives a general peak shape is still being used to refine peak resources predictions on regional and global scale \cite{Northey2014} \cite{Bardi2009}.

Moreover, since 2000s and in the context of the energy transition mineral resources and/or demand have been increasingly studied with forecasting models using logistic curves or at least sigmoid curves \cite{Vidal2017} \cite{Sverdrup2014} \cite{Capellan2020}. Thus, even if a maximum of interest for oil peak happened during 2010s \cite{Bardi2019}, depletion curves are still prevalent and used to support decisions, notably in energy transition management.

As a consequence, and with recycling processes being minor so far in terms of mass flow, the mining sector evolution provides a good description of the current management of non-energy resources and its supposed depletion.

\subsection{Objectives}
Due to its history and the availability of data on global scale, the topic of mining production is a good study case for interrogating the use of future oriented model and investigate the possible interpretation of the results.


In that respect, and given the popularity of the concept of peak production, the Hubbert model has been chosen to illustrate two possible interpretative approaches: the forecasting and foresight perspectives. 

In a forecasting interpretation, the future already exists and is inevitable. As such it can be predicted with regard to uncertainties associated to the model and simplifying assumptions. 

In a foresight interpretation, the model depicts potential futures (i.e scenarios) and assesses their plausibility as well as their desirability. It is a decision support thinking. The group of scenarios we consider here is to continue mining in a business-as-usual mode. The only constraint considered that is limiting is the geological availability of the chemical elements considered, all things being equal. We therefore consider that the demand for mining products will continue to grow, that energy, technological or geopolitical factors are not limiting in these scenarios. Of course, these are simple and basic scenarios that could be enriched later in other studies. They highlight the geological constraint in a world where the demand for mineral resources does not weaken, which corresponds to the extrapolations of the current outlook.

Therefore, this work proposes a pedagogical case study of the application of Hubbert model on a global scale and on the mining industry. First, we define the basic assumptions and mathematical properties of the model. Then, we explore its applicability to 20 mining elements including copper, with a methodology similar to the one used historically. This application uses mining resources estimation and is associated to a predictive perspective. In addition, Hubbert models with varying input parameters are applied to highlight an equivalence zone where models fit historical data with the same accuracy. This work is closer to a foresight implementation of the model. A new representation has been built to facilitate a prospective reading of the model. Finally, the outcomes and the difference between the two approaches are exposed in the discussion part.




%

\subsection{Defining the Hubbert model}

The Hubbert model offers a phenomenological description for the time evolution of the cumulative production $Q(t)$ (or equivalently, the annual extraction  $P(t)=dQ/dt$) from a finite stock. One strength of the models resides in its utmost simplicity. The model is based on the logistic growth equation proposed by P.F. Verhulst \cite{verhulst1845}.

\noindent The three mathematical hypothesis of the model are:
\begin{enumerate}
\item There is an amount $K$, defined as the maximum quantity of resources extracted throughout the course of the study such as: 
\begin{equation}
K = \int_{-\infty}^{\infty} P(t) \cdot dt = Q(t=+\infty) - Q(t= -\infty) 
\end{equation}
where $Q(t= -\infty) = 0$. Note that the stock considered here represents the quantity that \textit{will} be effectively extracted at the end of time, and which is assumed to be fixed. This quantity is also called the \emph{Ultimately Recoverable Resources} (URR).

\item The annual production $P(t)$ is proportional to the quantity already extracted $Q(t)$ and to the quantity still to be produced, which are the remaining resources $R(t)=K-Q(t)$.  

\item The proportionality coefficient 1/(K$\tau)$ is constant, with $\tau$ a characteristic time parameter related to the dynamic of the production. 

\end{enumerate}

\noindent With these three assumptions, we define that the cumulative production $Q(t)$ satisfies the Verhulst's differential equation, as follows:

\begin{equation}
 \underbrace{\frac{d Q(t)}{dt}}_{Annual \ production} = \cfrac{1}{K} \cdot \cfrac{1}{\tau} \cdot 	\underbrace{Q(t)}_{Cumulative \ production} \cdot \underbrace{(K - Q(t))}_{Remaining \ resources} 
\label{Verhulst_eq}
\end{equation}

The solution to Verhulst's equation is the so-called logistic function and the resulting expressions of the production variables $Q(t)$, $P(t)$ and $R(t)$ constitute what we refer to the \emph{Hubbert model}, presented on Eq \eqref{Hubbert_model}.    

\begin{equation}
\forall t \in [t_i, \infty[,  \hspace{1cm} \left\{ 
\begin{array}{ll} 
Q(t) = \cfrac{K}{1+ \left( \cfrac{K}{Q_{0}}-1 \right) \cdot \exp{\left(-\cfrac{(t-t_0)}{\tau}\right)}} 
\\
 P(t) = \cfrac{dQ(t)}{dt}
 \\
 R(t) = K - Q(t)
\end{array}
\right.
\label{Hubbert_model}
\end{equation}

The model is then fully characterized with the two parameters $\tau$ and $K$, and one initial condition at time $t_0$ for the cumulative production such as $Q(t_0) = Q_0$.

\begin{figure}[!h]
    \centering
    \includegraphics[height = 6cm]{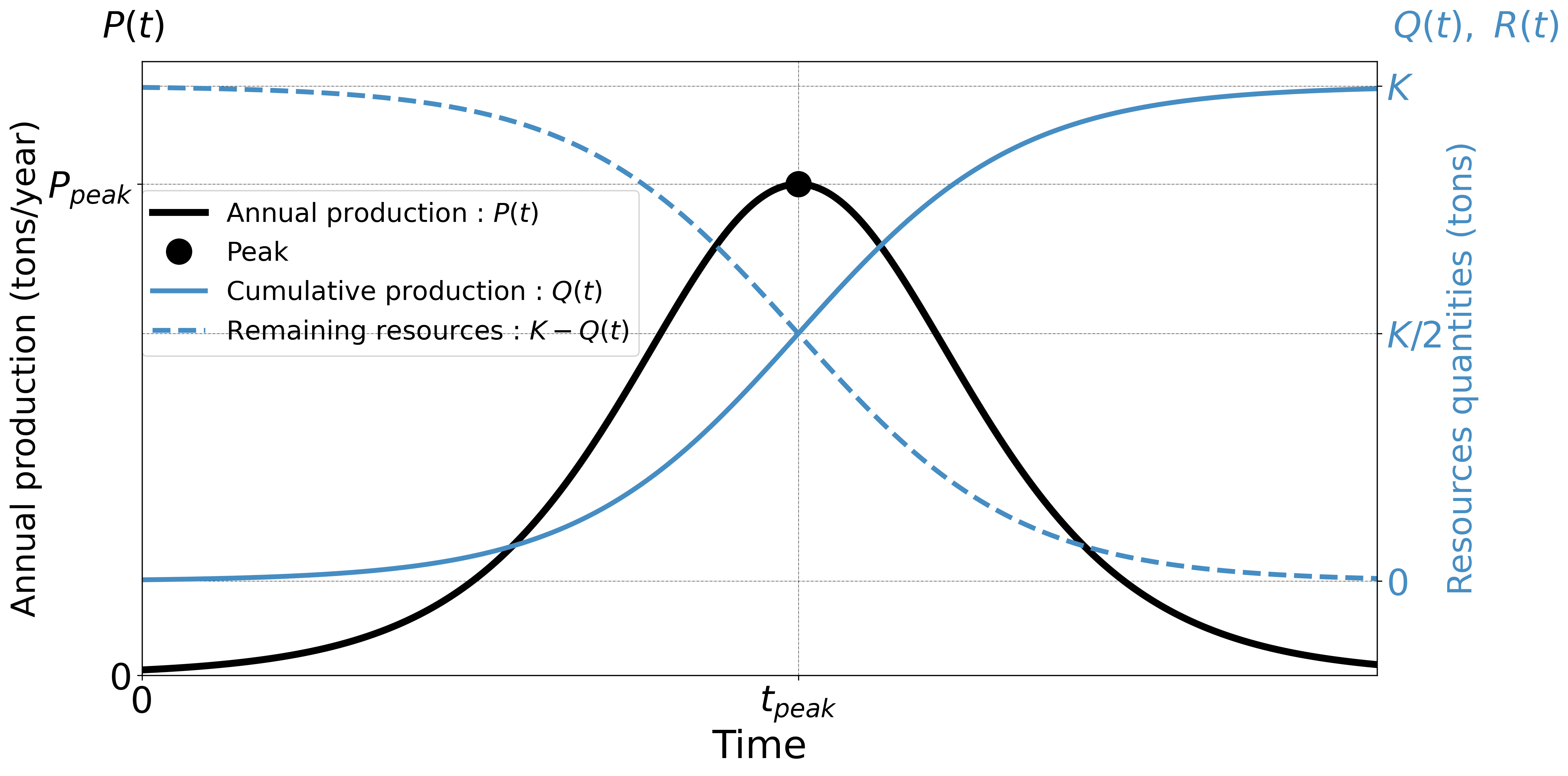}
    
    \caption{{\bf Typical time evolution of the Hubbert model solution}. The figure shows normalized values. The cumulative production, $Q(t)$, shows a sigmoidal shape (blue curve, right scale). The scale is normalized by $K$. Its asymptotic value is $K$. The remaining resources, $R(t)$, is a symmetrical sigmoid with an asymptotic value of $0$ (dashed blue curve). The annual production (black line, left scale), $P(t)$, presents a symmetrical bell shape with a maximum that will be referred too as a "peak" (black dot). The scale is normalized by the peak value.}
    \label{fig:classic_Hubbert}
\end{figure}

The typical time evolution of these functions are drawn in Fig \ref{fig:classic_Hubbert} with the initial condition $Q(0)= 0$.

\medbreak

Several stages can be identified in this dynamics. In the first stage of extraction (at times such that $K>>Q(t)$), production increases exponentially with a growth rate given by {\large $\mathbf{\tau}$}. When cumulative extraction becomes non negligible compared to $K$, production slows down, and reaches a maximum (the so-called "peak"). Note that the peak is reached when the cumulative production $Q(t_{\rm peak})$ is equal to half of $K$ - or equivalently, when a quantity $K/2$ remains to be extracted. The peak does not imply the depletion of the resource under study, but indicates the beginning of the reduction of the annual production. Finally, symmetrically to the first phase, production decreases exponentially at the same rate $\tau$ and tends towards zero when the cumulative production approaches the maximum extractable quantity.

The "peak" is the most striking feature of the model. Its properties can be easily deduced from the basic equation Eq \eqref{Hubbert_model}. The date $t_{x_{(\%)}}$ at which the cumulative production represents a fraction $0< x_{(\%)}<1 $ of the stock $K$ is given by
\begin{equation}
Q(t_{x_{(\%)}}) = x_{(\%)} \cdot K = \cfrac{K}{1+\left( \cfrac{K}{Q_{0}}-1 \right) \cdot \exp{ \left( - \cfrac{(t_{x_{(\%)}}-t_0)}{\tau} \right) }} 
\end{equation}
leading to:
\begin{equation}
t_{x_{(\%)}} = t_0 + \tau \cdot ln \left( \left( \frac{x_{(\%)}}{(1-x_{(\%)})} \right) \cdot \left( \frac{K}{Q_{0} } -1 \right) \right)
\end{equation}

Therefore, the date of the Hubbert's peak, corresponding to $x = 50\ \% $, is equal to: 
\begin{equation}
t_{peak} =  t_{(x = 50 \ \%) } =  t_0 + \tau \cdot ln \left(  \frac{K}{Q_{0} } -1 \right)
\label{eq:peak_date_eq}
\end{equation}

\begin{figure}[h!]
    \centering
  \includegraphics[height = 6cm]{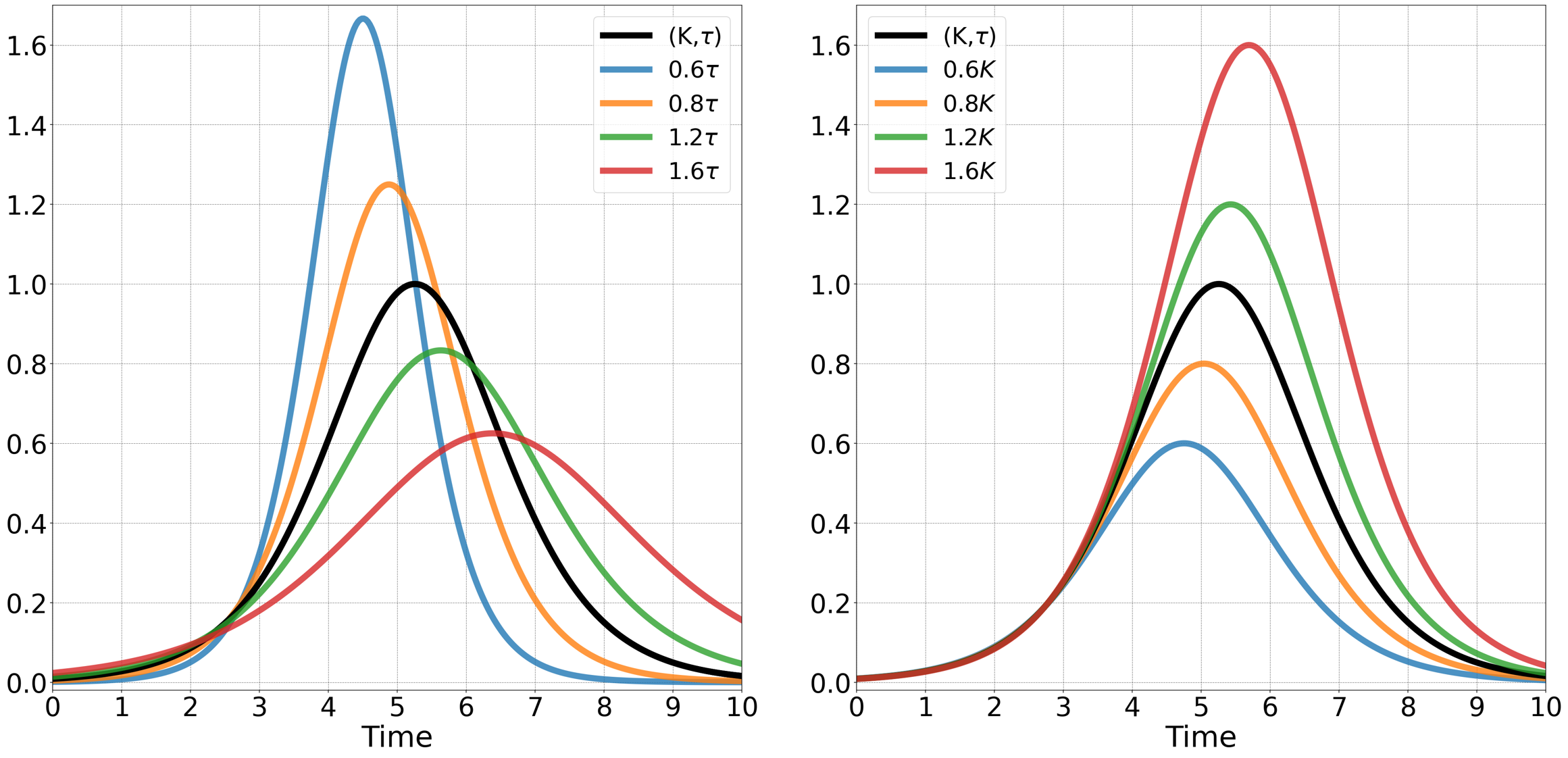}
  
    \caption{{\bf The parameters influence the value of the peak and its date.} The y-axes are values relative to the maximum value defined by the black curve. A model with higher values of $\tau$ has a later peak date and a lower maximum value, and conversely (left). Larger values of $K$ imply higher peak values and later peak dates, and conversely (right).}
   
    \label{fig:variations_Hubbert}
\end{figure}

An important consequence of this result is that the date of the peak depends linearly on {\large $ \tau $} and logarithmically on $\cfrac{K}{Q_{0}}$ as shown in Fig \ref{fig:variations_Hubbert}. 

Moreover, the value of the peak, which is the value of annual production when half of total resources $K$ has been extracted, can be expressed as follows:
\begin{equation}
P_{peak} =  P(t_{(x = 50 \ \%) }) =  \frac{K}{4 \cdot \tau}
\label{eq:peak_eq}
\end{equation}
The peak value depends linearly on the total amount of resources $K$ and is inversely proportional to $\tau$, the parameter reflecting the extraction rate.

\section{How to identify variables and parameters in the mining context?}
Hubert's model provides a phenomenological description of the temporal evolution of extraction rates from a finite stock K. This section focuses on the application of the Hubbert model to the global mining industry to examine the production of specific commodities. To do this, we have defined the measured quantities that will be associated with the model variable.

\subsection{Evaluation of $\mathbf{P(t)}$}
The variable $P(t)$ refers to the annual production of a chemical element, in ore grade and in metric tons (1000 kg), over long time series until 2019. These values are provided by the United States Geological Survey (USGS). 

\subsection{Evaluation of $\mathbf{Q(t) \ and \ Q_0}$}
A long time series of annual production data gives the cumulative production $Q(t)$ in metric tons, (sometimes since 1900). The value of $Q_0$ does not have to be exact for the model to correlate with the data. By varying its value, the production curve slides. It can be considered as a fitting parameter of the model.

\subsection{Evaluation of $\mathbf{\tau}$}
Parameter $\mathbf{\tau}$ is measurable during the first stage while parameter $K$ has a negligible influence (see equation \ref{DLQ}).

\begin{equation}
Q(t) \isEquivTo{t_0} Q_0 \cdot \exp{\left(\cfrac{(t-t_0)}{\tau}\right)}
 \label{DLQ}
\end{equation}

It is a past trend parameter and a low value of $\tau$ corresponds to intensive mining extraction or high mining rate. It is expressed in years.

\subsection{Evaluation of $\mathbf{K}$}
The "ultimately recoverable resource" $K$ is by definition a value that will only be known exactly in the future, at the end of mining. It embodies the foresight component of the model. We present the link between the $K$ parameter and the "Resources" and "Reserves" which are quantities defined for the mining industry. They are commonly defined as follows by the USGS (Mineral Commodity Summary, Appendix C \cite{USGS2021}):

\emph{Resources}: "A concentration of naturally occurring solid, liquid, or gaseous material in or on the Earth’s crust in such form and amount that economic extraction of a commodity from the concentration is currently or potentially feasible."

\emph{Reserves}: "[The part of resources] which could be economically extracted or produced at the time of determination. The term reserves need not signify that extraction facilities are in place and operative. Reserves include only recoverable materials; thus, terms such as “extractable reserves” and “recoverable reserves” are redundant [...]."

\begin{figure}[h!]
    \centering
    \includegraphics[height = 6cm]{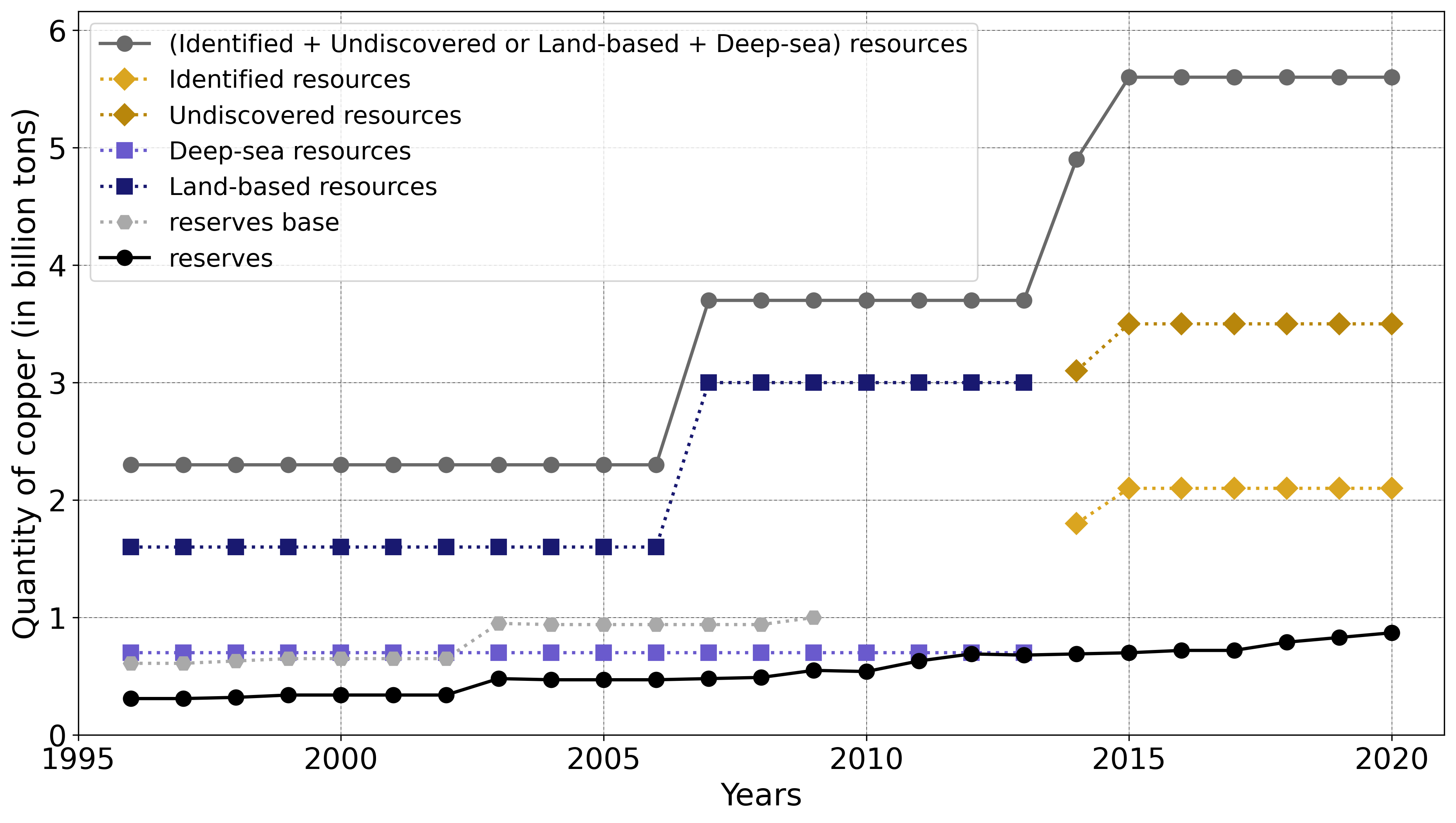}
    \caption{{\bf Time evolution of quantities related to global mining of copper from 1996 to 2020}. Extracted from USGS Annual reports \cite{USGS2021}.}
    \label{fig:mining_quantities}
\end{figure}

It should be noted that \emph{resources} and \emph{reserves} change over time, often with an upward trend and in stages, as it is illustrated on Fig \ref{fig:mining_quantities} with copper data from 1996 to 2020. Moreover, intermediate categories between these two bounds such as \emph{deep-sea resources} and \emph{Land-based resources} were discontinued in favor of \emph{undiscovered resources} and \emph{identified resources} in 2014. The \emph{reserves base} stopped being documented in 2010 for all the mining elements. These quantities depend on mining exploration and technological improvement enabling to convert \emph{resources} into \emph{reserves}.

In the following the terms \emph{resources} and \emph{reserves} will refer to the last available value.

Therefore, depending on whether we assume $K$ to be resource-based or reserve-based, the resulting peak will be affected. However, the date of the peak depends logarithmically on $K$, which mitigates the influence of the uncertainty of K on this date (see Eq. \ref{eq:peak_date_eq}). We investigate the interest of this specificity of the model from a foresight perspective in the discussion part. 


\section{The Hubbert map representation}

As shown in Figures \ref{fig:classic_Hubbert} and \ref{fig:variations_Hubbert}, historical applications of the Hubbert model use temporal representations. The emphasis is on the production trend and the intensity of the peak, while the parameters are rarely represented. Thus, the result is analysed as a prediction, which is facilitated by the possibility to directly compare the theoretical curve and the historical data with correlation analysis methods.

\begin{figure}[h]
    \centering
   \includegraphics[height = 6cm]{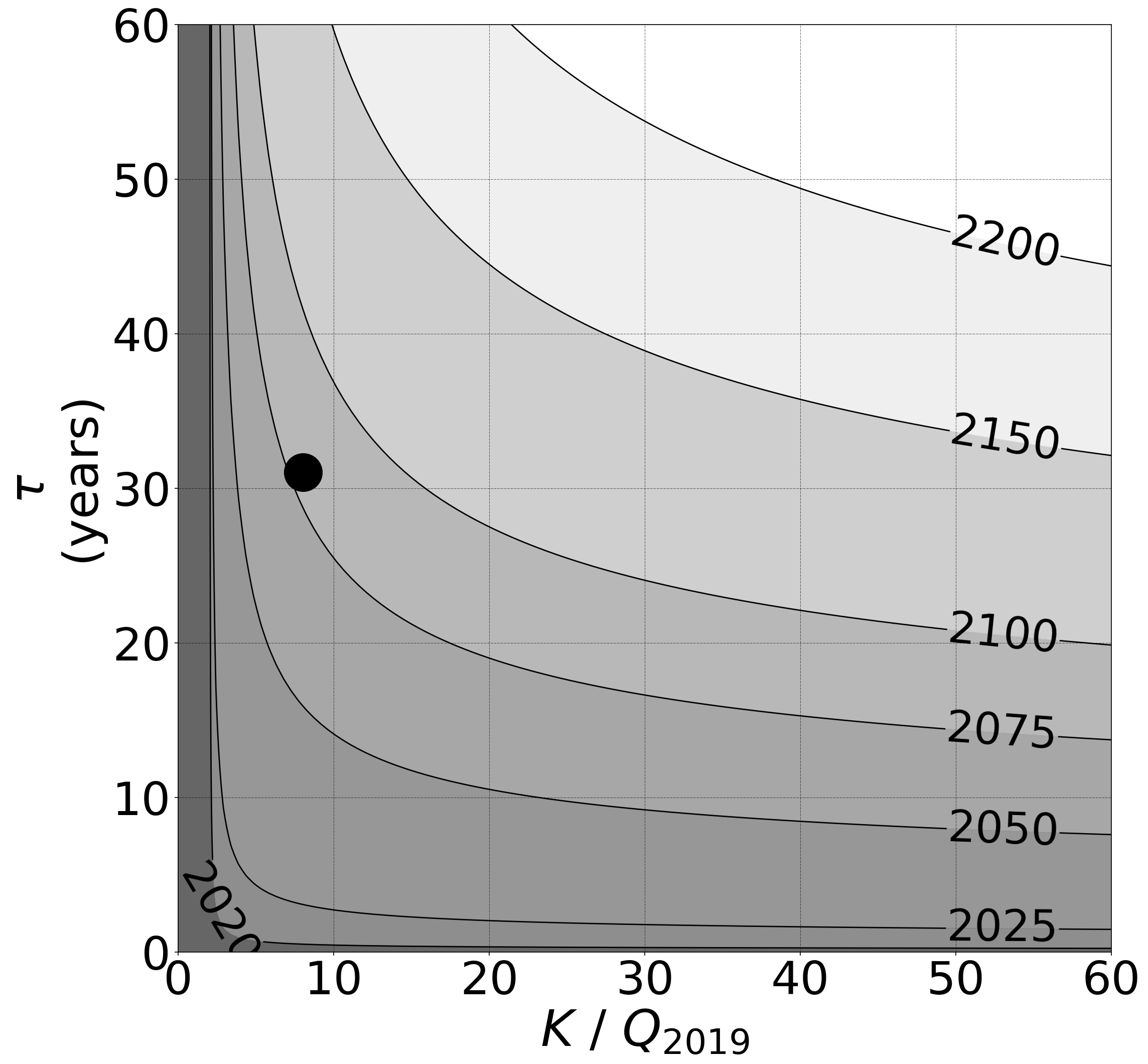}
    \caption{{\bf Defining the Hubbert model map}. The x-axis represents the relative value of $K/Q_{2019}$. The y-axis represents the characteristic time,$\tau$, that depends on the extraction rate. The lines represents the iso-dates of the peaks. The gray scale indicate the distance of the peak to the present, the darker the closer the peak is to 2020. The black dot corresponds to the peak defined in Fig \ref{fig:classic_Hubbert}. Normalizing $K$ by $Q_{2019}$ allows to place all the elements on the same map.}
    \label{fig:Hubbert_plane}
\end{figure}

We propose a new representation of the Hubbert model result in the form of a plane. This plane, in Fig \ref{fig:Hubbert_plane}, exposes the structure of the model defined by Eq \eqref{eq:peak_date_eq}. Consequently, the parameter $\tau$ and the parameter $K$ (normalized by initial condition) are respectively the x and y axes of the plane. Thus, each application of a model is then depicted as a point on this plane with its parameters as coordinates. 

In addition, this bi-parametric space shows different "iso-date" curves which correspond to the location of points that have the same peak date. The date of the peak of a model application is therefore deduced with respect to the nearest iso-date(s). The range of parameters and consequently the time horizon of the simulation (up to 2200) has been determined after applying the model to many chemical elements and with a long-term perspective. 

As mentioned above, the value of the parameter $K$ is normalized by the cumulated production of the chemical element estimated in 2019. This normalization makes possible to place several models applied to mining elements with different scale of production on the same chart.

It also allows to better illustrate effect of parameters variation on the peak date and thus to generate scenarios, as developed in the discussion section.

Finally, this normalized "Hubbert model map" can be taken as focused on the peak date estimation while making explicit the identification of the underlying parameters, rather than on the production trend and peak intensity. Nevertheless a similar plane can be created to represent the peak intensity, instead of the date, using equation \ref{eq:peak_eq} with  $K$ and $\tau$ as respectively x and y axes.

\section{Materials and methods}
\subsection{USGS data}
Copper production data have been gathered from the tabular presented by (a) between 1900 and 1993, and from annual reports delivered by (b) between 1994 and 2019. Resources and reserves were compiled from the latest data available in annual reports from (b).

Same process has been implemented on other elements with respective pages on USGS website (see appendix table  \ref{appendix:rawdata}). 

\medbreak
The time reference $t_0$ has been set equal to $2019$ because it corresponds to the latest available data of production from USGS statistics.

\medbreak
\noindent (a) U.S. Geological Survey, 2021, National Minerals Information Center, Copper Statistics and Information, accessed Mai 19, 2022 at URL \url{https://www.usgs.gov/centers/nmic/copper-statistics-and-information}

\noindent (b) U.S. Geological Survey, 2021, Historical Statistics for Mineral and Material Commodities in the United States, accessed Mai 19, 2022 at URL \url{https://www.usgs.gov/centers/nmic/historical-statistics-mineral-and-material-commodities-united-states}

\subsection{Computation and curve fitting}

We used Python version 3.8.2 and we implemented two programs based on the package scipy.optimize (curve\_fit), one optimizing the fit with three free parameters ($Q_{2019},\tau,K$) the other with two parameters ($Q_{2019},\tau$) and the third set as follow:   

\begin{gather} \label{eq:K}
   K(Resources) = Resources(t_{measure}) + \sum \limits_{t = 1900}^{t_{measure}} P(t)\\
    K(Reserves) = Reserves(t_{measure}) + \sum \limits_{t = 1900}^{t_{measure}} P(t)
\end{gather}

With $t_{measure}$ the date corresponding to the latest data available.

\medbreak

Remark: When $t_{measure} = 2021$, $P(t_{measure})$ and $P(t_{measure} -1)$ do not exist yet, because USGS report delivers annual production with a delay of two years. It results on an underestimation of K as presented in Fig \ref{appendix:Koverestimation} of the supporting information. Appendix \ref{appendix:Koverestimation} presents an illustration of this mathematical artifact. 

\medbreak

Thus, a correlation coefficient $0<R^2<1$, using the method of least squares, is computed after each method implementation and for defining equivalent zones. It is calculated using the following equation:

\begin{equation}
R^2 = \frac{1}{N} \cdot \sum_{t=1}^{N} (X_i - Y_i)^2
 \label{R2}
\end{equation}

With $X$ and $Y$ being respectively the production data and simulated production vectors and $N$ equals to the length of the production data vector. A $R^2$ equal to 1 means the model is perfectly correlated to data. Figures have been created with the package matplotlib.pyplot (plot,contour and coutourf).

The results of these curve fitting in presented in the supporting information, in temporal representations (see Appendices \ref{appendix:Hubbertian} \ref{appendix:Exponential} \ref{appendix:Multiple} and \ref{appendix:others}).

\pagebreak

\section{Results}

\subsection{A detailed application to Copper}
Copper production is a case in point for studying a Hubbert model. From a practical point of view, production data is available over a long period of time. From a methodological point of view, the situation resembles that of American oil production when Marion King Hubbert published his model in 1956. Copper production has been increasing steadily since 1900 and has been maintained until today with an exponential trend. Moreover, copper is an important raw material for infrastructure and for many consumer products, so its availability is crucial and even more so in the context of the energy transition.   


\begin{figure}[H]
    \centering

   \includegraphics[height = 6cm]{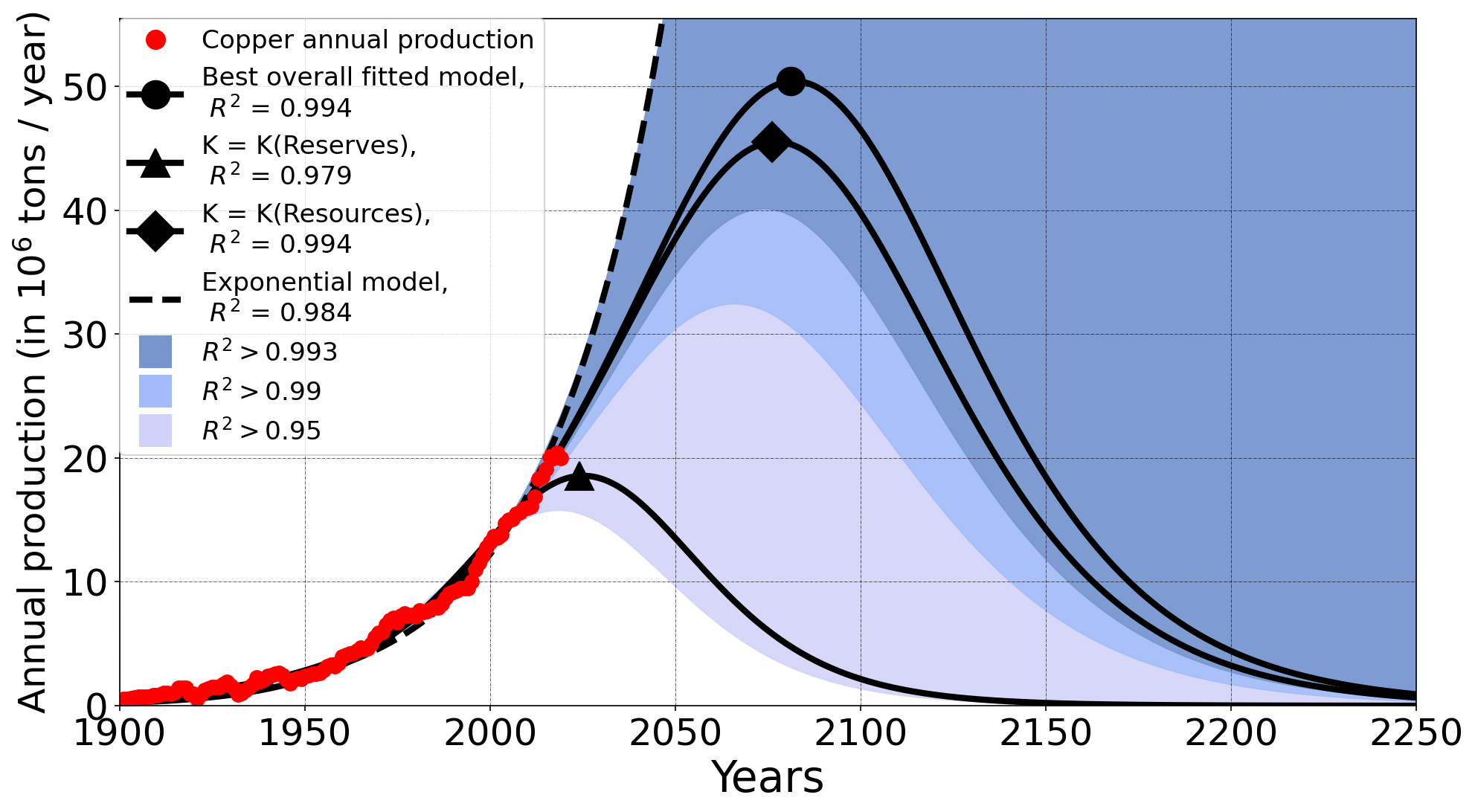}
    \caption{{\bf Estimating copper peak}. The red dots represent the copper USGS data amount extracted per year in $10^6$ tons. The dashed curve corresponds to an exponential fit of the data. The $R^2$ value is indicated on the graph. The black curves show three different Hubbert model fits of the data.  The curve with the peak indicated by a triangle is obtained by fixing $K$ as the reserves estimated by the USGS. The curve with the peak indicated by a diamond is obtained by fixing $K$ as the resources given by the USGS. The curve with the peak as a black dot represents the best fitted function (with the largest value of $R^2$) considering the parameters, $\tau, K$, and the initial condition, $Q_0$, as free. The blue zones show region that are equivalent zone of correlation \textit{i.e.} having the same $R^2$ values for the fit.}
    \label{fig:Copper_time}
\end{figure}

Fig \ref{fig:Copper_time} shows the comparison of several model fits to historical global annual copper production data, including an exponential curve. We set the value, $Q_{2019}$, which corresponds to the total amount of copper extracted until 2019. The value of $K$ is set either as the value reported by the USGS for reserves (triangle) or resources (diamond), or left as a free adjustment parameter (circle). The value of $\tau$ is adjusted to minimize the value of $R^2$ when fitting the model.
The noteworthy point of this picture is that any model characterized by the duplet $(\tau,Q_{2019}) \simeq (31,7\cdot 10^{8})$ is equivalently correlated to data (threshold: $R^2 = 0.95$) for mostly any $K$ value greater than the Reserves value (light blue area on Fig \ref{fig:Copper_time}). Increasing the minimum threshold of the correlative coefficient leads to reduce the equivalent zone size. If the correlation threshold is set at $0.99$ then the confidence zone is considerably reduced  (dark blue area on Fig \ref{fig:Copper_time}).

\medbreak

\begin{figure}[H]
    \centering
  
    \includegraphics[height = 6.5cm]{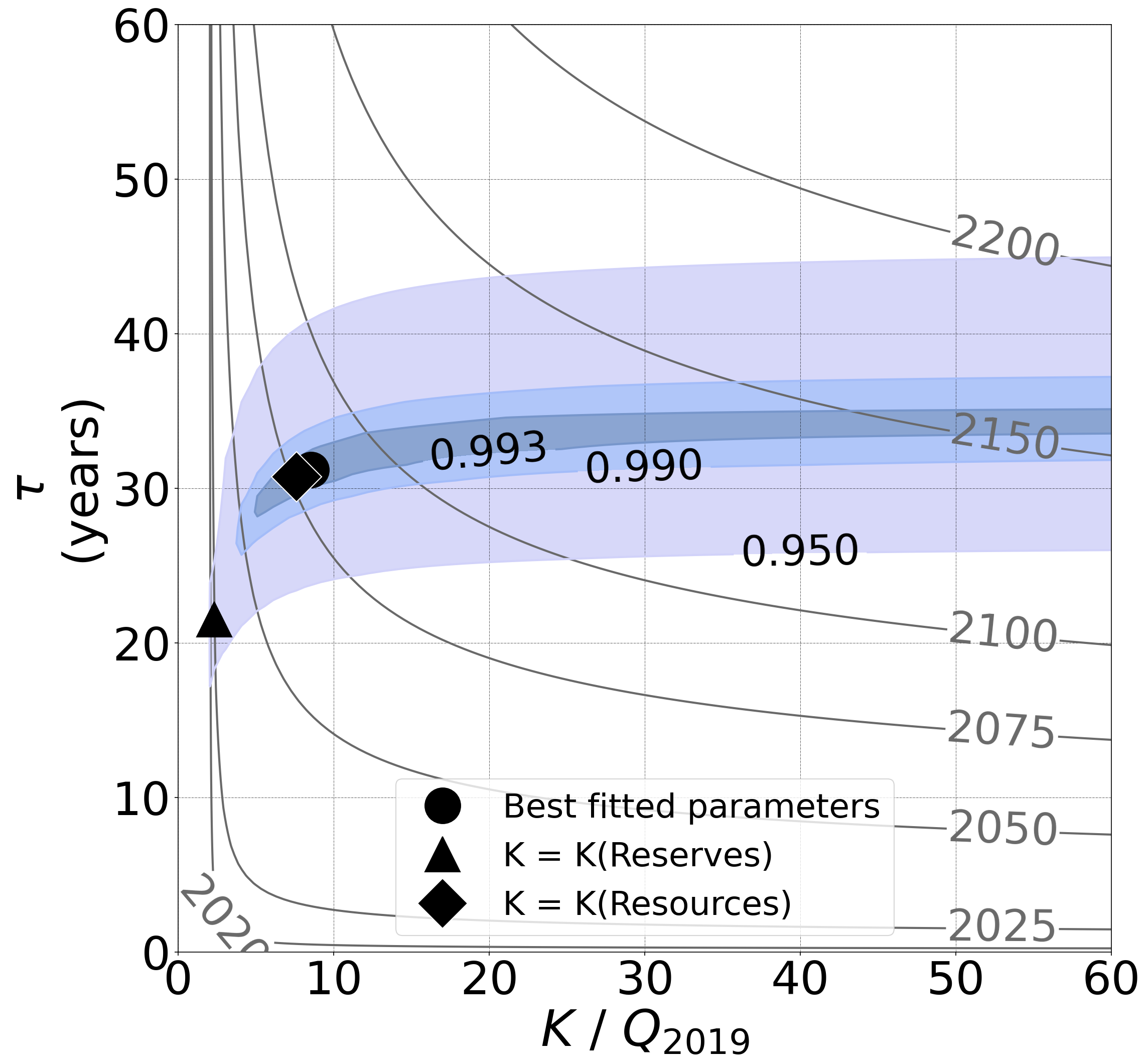}
    \caption{{\bf Placing copper on a Hubbert map.} The three estimated copper peaks of Fig.\ref{fig:Copper_time} are placed on the $\tau$, $K/Q_{2019}$ plane. The shaded blue zones indicate region that share an equal value of $R^2$ (respectively 0.950, 0.990 and 0.993). The fitting place copper peaks (black diamond and dot) before 2100. Taking into account the reserve, estimated by the USGS, gives the most pessimistic peak. }
    \label{fig:Copper_R2zone}
\end{figure}

On the Hubbert map, the peak is represented by a dot and the areas of equivalent correlations are also represented by areas in shades of blue, as shown in Fig \ref{fig:Copper_R2zone}.

Note that the areas of equivalent correlations follow a horizontal asymptotic shape for larger values of $K$. 

\medbreak

This phenomenon traduces the fact that if K has high value relatively to current production, the fit does not recognize an exponential growth but deemed still to come. The resulting tau values is then overestimated, relatively to optimized exponential fit or fit with set K.

\subsection{Application to elements with copper-like extraction evolution}

\begin{figure}[H]
    \centering
  
    \includegraphics[height = 6.5cm]{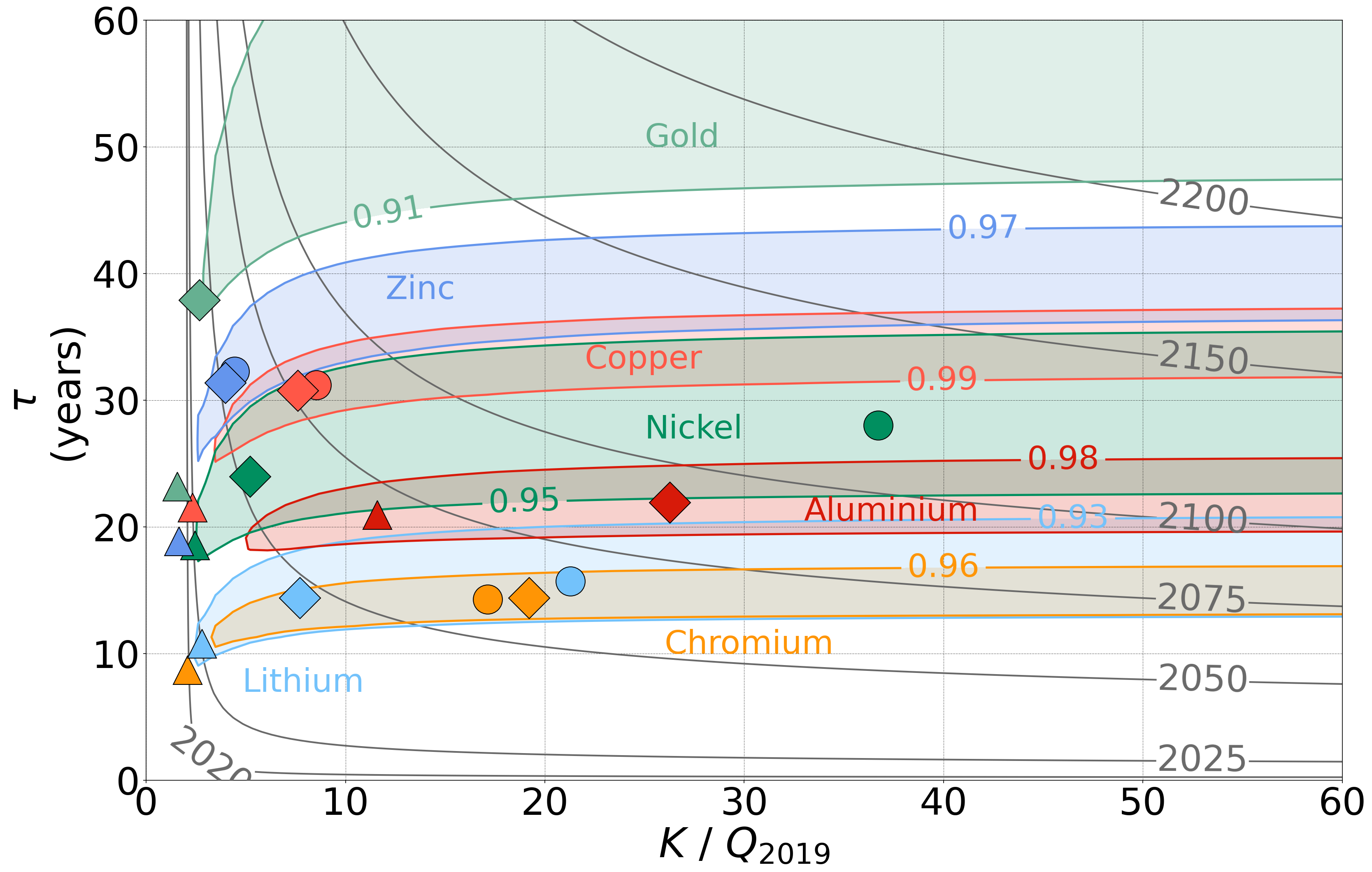}
    \caption{{ \bf Placing the exponentially extracted elements on Hubbert's map.} In addition to copper, eight other chemical elements are showing exponential growth in the amount extracted annually. The dots correspond to a fit of the model with free parameters, the triangles to a fit fixing the value of $K$ to the reserves and the diamonds by fixing the value of $K$ to the resources. All the production peaks based on mining resources occur in the 21st century.}
    \label{fig:Exp_R2zone}
\end{figure}

So far, several other elements show similar extraction dynamics to copper, i.e. their annual production grows exponentially. These chemical elements are aluminum, chromium, gold, lithium, nickel, zinc. As for copper, we fit a Hubbert model to the extraction data for these elements. The estimated peak dates are plotted on a Hubbert's map, see Fig \ref{fig:Exp_R2zone}.

Each of the extraction trends for these elements, with the exception of gold, shows a similar trend to that of copper. Their equivalent areas have the same asymptotic shape. The figure shows the peaks obtained for the model fit by taking for the value of $K$ the value of reserves (triangles) and resources (diamonds) provided by the USGS. The points correspond to a fit with the value of the free parameters. All peak dates obtained fall in the 21st century.

Gold is a particular chemical element that has the lowest ratio of cumulative production to estimated USGS resources. Its annual production is also less well suited to exponential growth because of its greater variability over time. Thus, the peak date obtained by a fit with the value of $K$ fixed to the value of the reserve falls early. Nevertheless, the best overall fitted model of Gold is located out of the range of the map ($\tau$ = 60, $K/Q_{2019}$ = 60).
\medbreak

The underlying assumptions of the model have a priori no reason to hold for ever - unless the stock considered here, the value of $K$, corresponds to the ultimate limit, which we can reasonably doubt. These estimates will be challenged if new mines are found and the exploitable quantity is increased.

Indeed, it appears that several other chemical elements deviate from the simple Hubbert profile when the demand and/or production context changes. However, their dynamics can be divided into different phases. This division makes it possible to apply an updated Hubbert model, with adapted values for $tau$ and $K$. We will name these elements in the following "multi-trend elements".

\subsection{Application to multi-trends elements}

In order to be able to apply a Hubbert model to chemical elements that show several trends in their production, it is necessary to first define categories that describe four different trends that we identify.

\begin{figure}[h!]
    \centering
   \includegraphics[width = 13 cm]{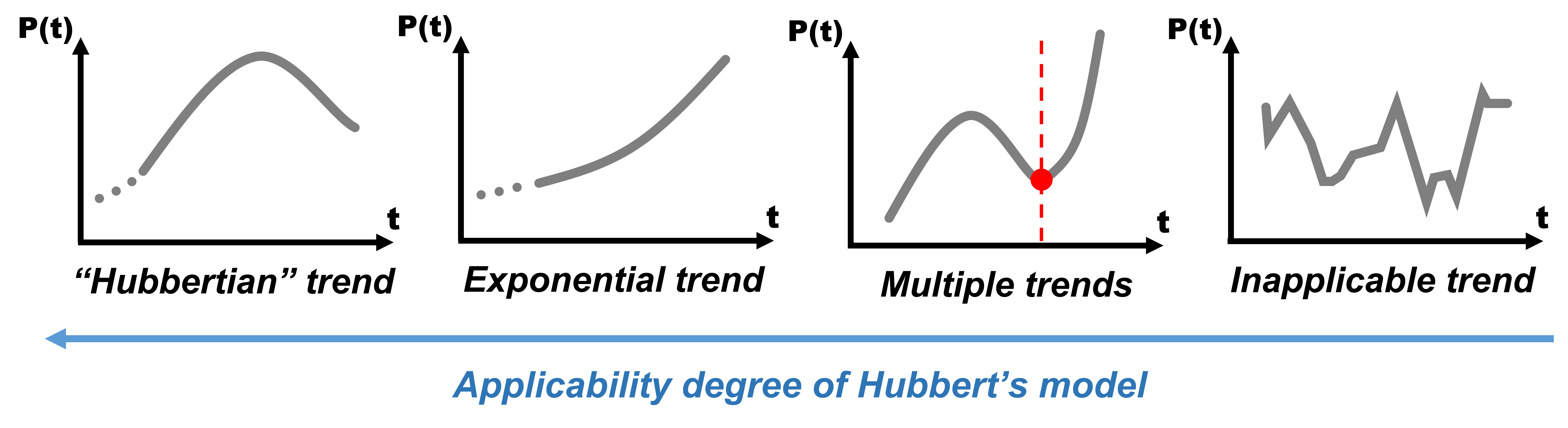}
\caption{{\bf Categorization of the trends in chemical element production based on the applicability of the Hubbert model}. We classify the production trends into four categories. The Hubbertian trend corresponds to an annual extraction in the shape of a bell curve, which therefore presents a peak. The exponential trend corresponds to an exponential growth of the annual production. The multiple trends correspond to evolutions that alternate different growth phases. Erratic trends are those for which the model becomes inapplicable.}
\label{fig:schema}
\end{figure}

Fig \ref{fig:schema} shows the categorization we propose to compare the different temporal evolutions of the production of the elements. Some trends show a peak, a maximum, which is followed by a decreasing annual production. We will call the extraction of these elements a "Hubbertian trend" as these evolutions show the typical bell shape of the yearly production. 

Copper-like elements have already been described as having an exponential growth trend in production for several decades. This exponential growth phase may correspond to the beginning of the dynamics of the Hubbert model, which also present an exponential growth phase. (see Fig \ref{fig:Exp_R2zone}).
The Hubbertian and exponential trend categories better fit the assumptions of a Hubbert model.

The multi-trend elements require considering that the Hubbert model assumption can be applied by dividing the trend into several phases. The first phase shows growth followed by either stagnation or a decrease in production. This first phase is then followed by a new phase of growth in extraction which is often exponential. For this phase, a Hubbert model is applied by not taking into account the data of the first phase but only the data of the second phase. This division is illustrated in Fig \ref{fig:schema}. The first phase, which has its data removed, is on the left of the red dotted line, the second phase which is fitted to a Hubbert model is on the right.

Chemical elements that exhibit an erratic annual extraction trend are not considered for this analysis because these trends do not correspond to Hubbert model dynamics.

\subsubsection{Definition of chemical elements with Hubbertian trend}

\begin{figure}[H]
    \centering
    
    \includegraphics[height = 12 cm]{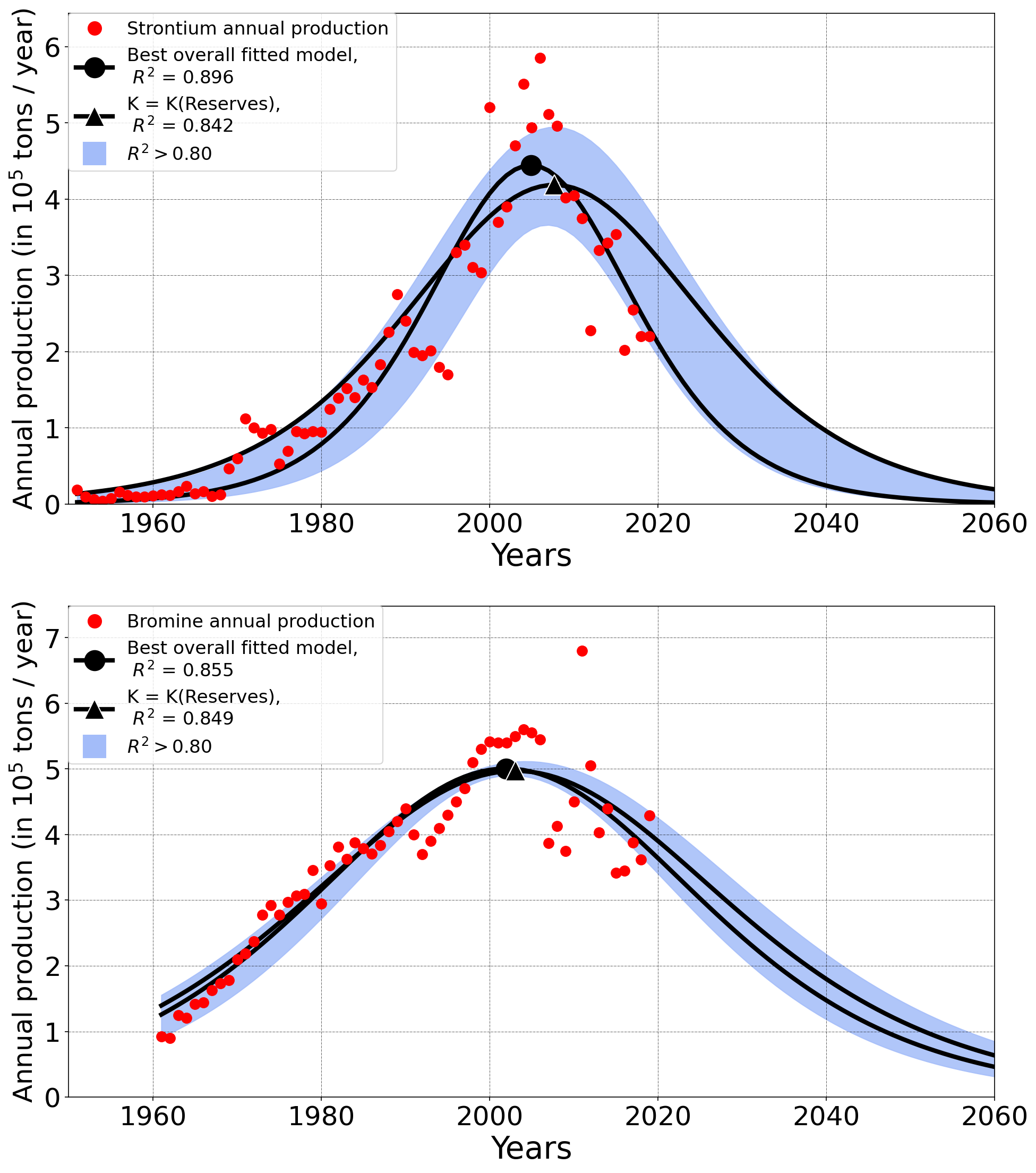}
    \caption{{\bf Fitting a Hubbert model for the case of strontium (top) and bromine (bottom).} Red dots represent USGS data of annual production. The peaks obtained are represented on the time evolution (black line) estimated by the Hubbert model, taking for the $K$ parameter the value of the reserves estimated by the USGS (triangle) or leaving the parameters free (points).}
    \label{fig:Hubb_time}
\end{figure}

As shown in Fig \ref{fig:schema}, Hubbertian-trending elements refer to elements whose production slows down after they have reached a production peak. For example, it is the case for strontium and bromine. We fit a Hubbert model to these chemical elements by taking the value for $K$ either as the USGS reserves (triangles) or by leaving the parameter free (dots). Despite the annual production fluctuations, both types of fit agree to place the Hubbert peak at close values. The peaks therefore seem to have occurred before the year 2020 as shown in Fig \ref{fig:Hubb_time}.

\begin{figure}[H]
    \centering
   
       \includegraphics[height = 6.5 cm]{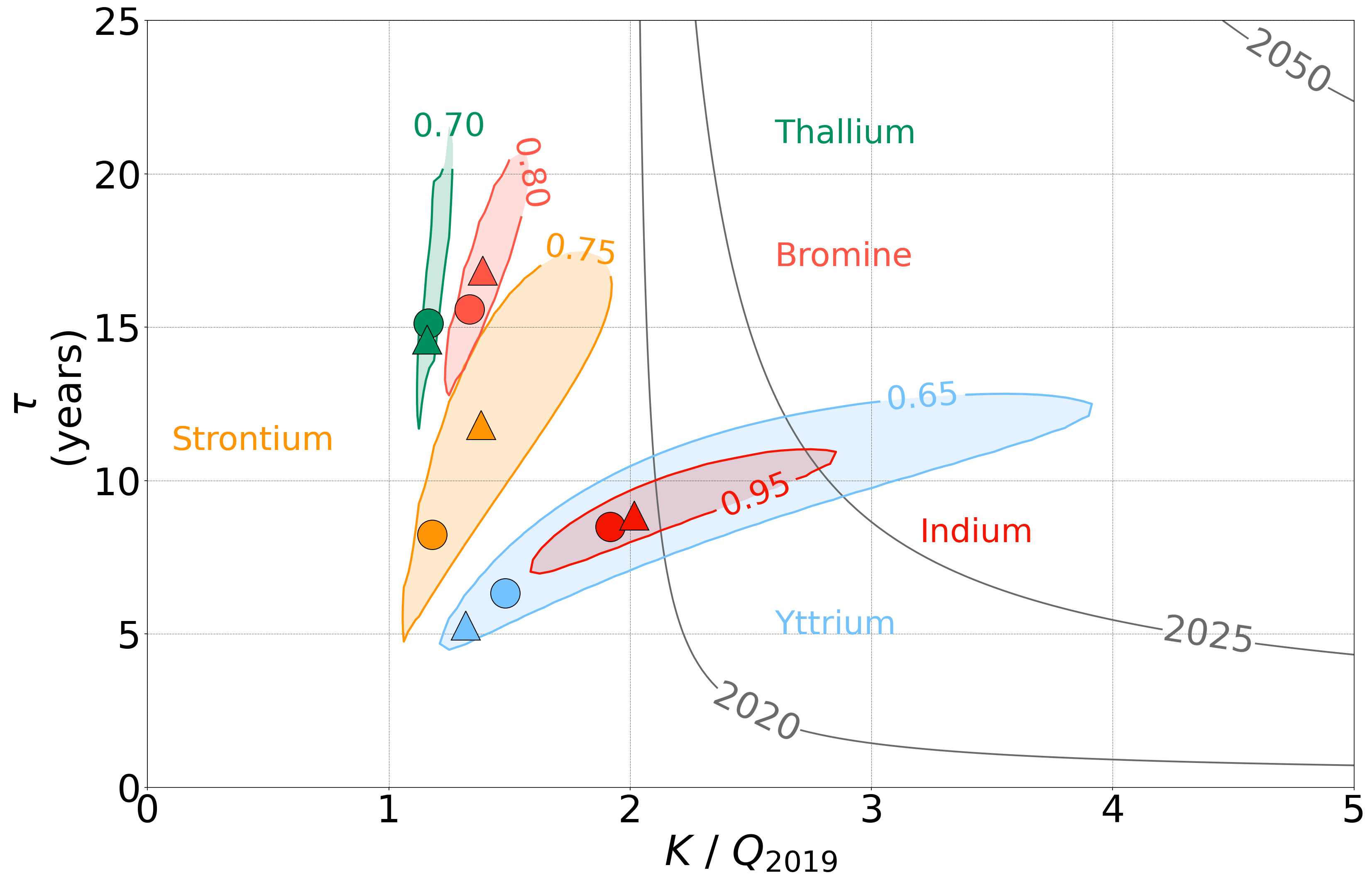}
    \caption{{\bf Placing of Hubbertian chemical elements on a Hubbert map.} The Hubbert model is fitted either by taking for $K$ the value of the reserves estimated by the USGS (triangles) or by leaving the parameters free (dots). Estimated peaks show that they occurred before 2020 and therefore the annual production of these elements is on the declining trend. }
    \label{fig:Hubb_R2zone}
\end{figure}

Then we place this type of elements on a Hubbert map which indicates that the elements thallium, bromine, indium, yttrium and strontium seem to have exceeded their annual production peak and their annual production decreases with time, see Fig \ref{fig:Hubb_R2zone}). 

These mining elements confirm the initial hypothesis of the existence of a production peak. However, it would be wrong to conclude that they only appeared for geological reasons. Strontium, for example, is a material used in cathode ray tubes. Therefore, the rise of this technology until the 2000s, and its subsequent decline, may be a more plausible explanation for the strontium production peak. The fact that there are no \emph{resources} data for these mining elements can be interpreted as a lack of interest in strontium as a raw material.	  






\subsubsection{Chemical elements with multiple annual production trends}

The chemical elements studied in this section show trends in annual production that vary over the study interval. They show an exponential type of growth followed by a stagnation or decrease, and then their production starts to increase again. After an initial maximum in annual production, production increases exponentially again. 

\begin{figure}[H]
    \centering
\includegraphics[height = 12cm]{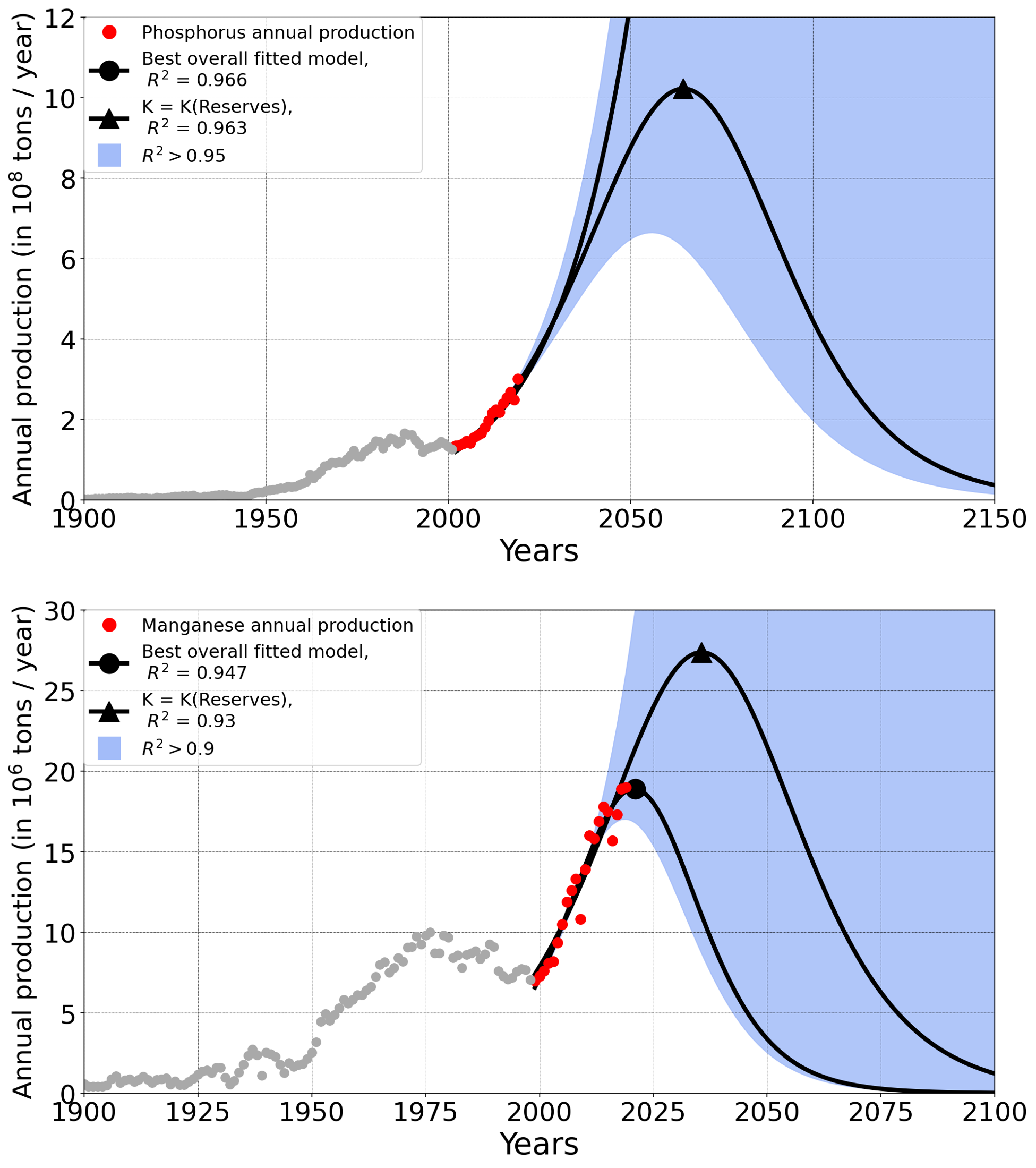}
    
    \caption{{\bf Estimating peak production of phosphorus (top panel) and manganese (bottom panel) that show multiple trends.} Dots represent USGS data of phosphorus and manganese mined per year in tons. Gray dots represent data not used in the model fit. The red dots are the data considered in the model fit. On the basis of the extraction trends after the 2000s, both elements appear to peak before the end of the 21st century.}
    \label{fig:Manganese_time}
\end{figure}

Fig \ref{fig:Manganese_time} shows the annual production of phosphorus (upper panel) and manganese (lower panel). For both elements, it shows a temporary peak or pause in annual production that occurs between 1980 and 1985 before increasing sharply again after 2000.

The production time trends shown in Fig \ref{fig:Manganese_time} indicate that a Hubbert model cannot be applied directly to all data between 1900 and 2019 since only one peak exists for this model. The resulting peak would not correctly characterize the two production trends. A visual inspection of the data shows two exponential phases of extraction separated by a later phase of stagnation or slight decline in production (gray and red dots on Fig \ref{fig:Manganese_time}).
It is possible to consider only the second phase of extraction after the years 2000 and not to take into account the data before this date except in the initial condition. This second phase has an exponential aspect which corresponds to the temporal dynamics of the model. We apply the Hubbert model only to the second phase of exponential extraction, which makes it possible to estimate a production peak resulting from this production revival. Data from the first exponential phase as well as from the stagnation phase are taken into account for the initial condition of the model adjustment considering the total quantity already produced before the 2000s.

\begin{figure}[H]
    \centering
    \includegraphics[height = 6cm]{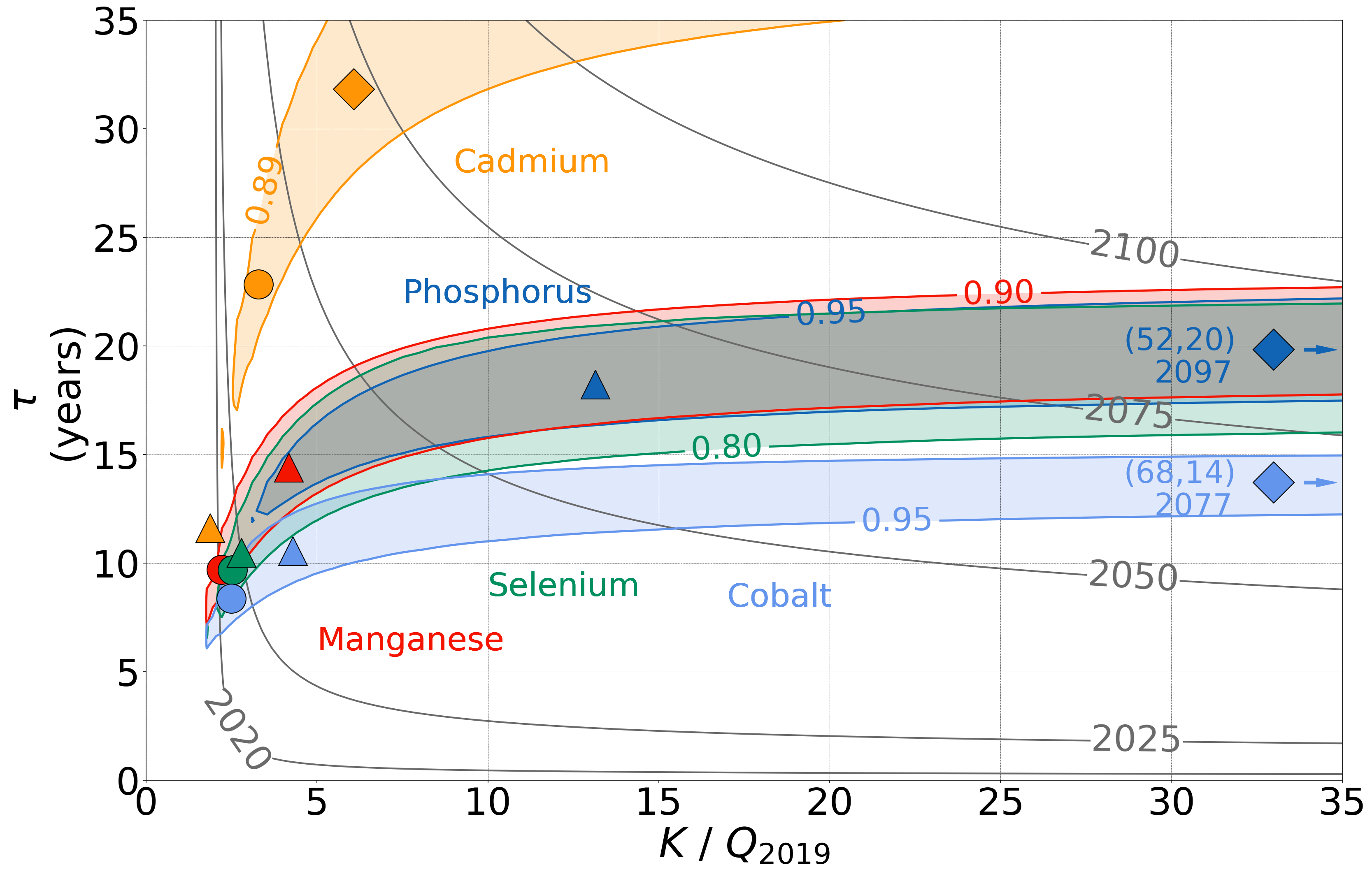}
    \caption{{\bf Placing the multi-trends elements on the Hubbert map.} Following the same method that was used for manganese and phosphorus, a Hubbert model is fitted to the new extraction trends for these elements. The estimated peaks fall during the 21st c. by taking the USGS values for the reserves or ressources (triangles, diamonds) and leaving the parameters free in the fit (dots).}
    \label{fig:Hubb_manganese_R2zone}
\end{figure}

We apply the same method for chemical elements that show the same type of multiple extraction pattern for their annual production. The resulting peak estimates for these elements are shown in Fig \ref{fig:Hubb_manganese_R2zone}. Theses elements also present peaks that fall during the 21st c. by taking into account the values of the reserves (triangles) or ressources (diamands) provided by the USGS, or by letting the parameters free for the fit.

\subsubsection{Elements whose extraction trend cannot be described by a Hubbert model}
	Elements in this category are Arsenic and Beryllium. Arsenic being used in metallurgy processes and in mining, its production may depend on other elements production.
	
	\medbreak
	
	Then, Beryllium is the only element whose \textit{annual production} is of the order of magnitude of hundred tons, which is very few relatively to all other elements. 
	
	\medbreak
	
	Tantalum and Boron are added to this category owing to its noisy data, making impossible to deduce a peak of a general trend.
	
	\medbreak

	A study focused on the mining and purifying processes should enlightening the reasons of these elements specificities.

\pagebreak

\section{Discussion}
While the fit of the Hubbert model to past data can be easily estimated from mathematical indicators, the use and interpretation of the model for the future raises important questions. How much confidence should be placed in the peak date estimates? Is the notion of a peak reliable in the first place? Is the model too simplistic to provide useful insights?
Two main interpretations, forecasting and foresight, must be discussed and distinguished, despite their apparent similarities.

\subsection{A predictive reading of the model}

A forecast approach considers that the future is already determined and tries to estimate this future as precisely as possible through careful measurements of the current situation and a good understanding of fundamentals rules driving the dynamics of the system. From this point of view, it predicts that production will peak and then decline, and can be used to estimate, for example, the date of this peak. Uncertainty regarding the peak position in time results from our imperfect knowledge of the system's parameters, in particular of the ultimately extracted quantity $K$.

Indeed, the maximum estimates are based on two types of data: historical trends and mineral resource estimates. 
\begin{itemize}
    \item Historical data are yearly measurements and present a noise component. The identified trend, embodied by the parameter $\tau$, is consequently sensitive to noise. To ensure the consistency of the $\tau$ value, a sensitivity study must be carried out resulting in confidence interval on the estimated peak. 
    \\Moreover most of these data refer to the \emph{mine production} of an element oxide which excludes recycled compounds. Unfortunately, sometime these data are unavailable and replaced by \emph{smelter} or \emph{refinery} production, comprising a part of recycled matter, leading to a distortion of the mining production trend and then to an inaccurate fit of parameter.         
    \item Global mineral resources data are aggregated states statistics, implying pre-treatments with associated uncertainties. Thus, according to annex B of \textit{Mineral Commodity Summaries 2021} \cite{USGS2021}, reserves and resources data have to be taken with caution:\\
    {
    \footnotesize
    \textit{
    "The USGS collects information about the quantity and quality of mineral resources but does not directly measure reserves, and companies or governments do not directly report reserves to the USGS. Reassessment of reserves is a continuing process, and the intensity of this process differs for mineral commodities, countries, and time period. Some countries have specific definitions for reserves data, and reserves for each country are assessed separately, based on reported data and definitions."}    }
    In addition, every supplementary treatments and hypothesis will then lessen the consistency of the prediction. For instance, aluminium resources data are nonexistent, but built on "Bauxite" resources. Assuming an ordinary conversion mining process, a conversion coefficient of $0.5$ has been applied to bauxite resources and reserves. This factor usually varies from $40\%$ to $60\%$ \cite{Hind1999}.
\end{itemize}
The consistency of the forecasting approach lies on collected data on purpose and its quality. This approach is notably used by Sverdrup \textit{et al.} \cite{Sverdrup2014} and Vidal \textit{et al.} \cite{Vidal2017,Vidal2022} to anticipate mineral resources evolution.

The ability of a forecasting model to accurately predict future events depends on the strength of the assumptions determining the expected behavior. If the evolution of the system is constrained by the fundamental laws of physics, the prediction is more reliable than if it is built on phenomenological rules. The basic justification is that only a finite amount of material can be drawn from a finite stock. However, the practical implementation of the model on real data takes some distance from this basic consideration.

First, the amount of ultimately extracted material $K$ is usually not a purely physical upper-bond (e.g. the number of atoms in the resource under consideration).  This physical limit is generally too far from practical considerations to constitute a significant constraint on the behavior of the system. The actual dynamics will be governed by more limiting constraints (such as Liebig's law of minimum, for example). Instead, in most cases, $K$ is a techno-economic quantity that demarcates deposits that are worth exploiting and materials that are unrecoverable, for example because of excessive dilution. However, this distinction is not necessarily definitive: unlike a law of physics, it is possible to reconsider this line. 

A trade-off thus appears. If the estimation of $K$ is too conservative, considering only quantities competitively available with current technologies (i.e. reserves), the model may tackle the current production dynamics, but has limited forecast capacity because it does not rely on strong physical considerations and can therefore depart from the initial assumptions.

As mentioned in the part "Evaluation of K",  both \emph{Reserves} and \emph{Resources} change over time and depend on mining exploration which is itself driven by the economic context including material demand. These mining concepts can be compared to the headlights of a vehicle moving on a road. If the need is pressing, means (economic and technical) will be allocated to increase the range of the beam. However, the distance made visible will tell more about the trend and concerns of the immediate future than the remaining length of the road, which would be the true "K". Hence, mining \emph{resources} are of a different nature than the "Ultimate resources" of the original article of Hubbert \cite{Hubbert1956}, mainly because it is not documented for this purpose. However using a resources-based "K" can be justified by order of magnitude considerations. Indeed, \emph{resources} and \emph{reserves} give order of magnitude of the scale of mining industry. This tends to introduce a "realistic" industrial reference or at least a physical meaning compared to curves only fitted with mathematical optimization.  
Notice that Fig \ref{fig:Exp_R2zone} shows for zinc copper and chromium a resources-based peak (diamond) close to the best fitted peak (dot). This could be a clue that for those mining elements resources are estimated with models similar to the Hubbert's one. This consequently mitigates the consistency of the previous argument for zinc, copper and chromium.




Similar considerations can be applied to the other assumptions of the model. There is no physical law that forces the typical time $\tau$ to remain constant, nor the production to remain proportional to the extracted and remaining quantities. This parameter is strongly influenced by economic considerations such as the demand for materials.

Thus, it appears that the predictive ability of a Hubbert model is limited. While the extraction of a finite resource will certainly come to an end at some point, this consideration may not be the actual driving force behind the observed behavior, even if the model provides a satisfactory phenomenological description. From this point of view, the estimated position and value of the peak must be treated with caution. However, the very existence of the peak gives an important predictive capacity. When the peak is reached, either the production will actually decrease or the system will deviate from one of the basic assumptions of the model. From this perspective, the Hubbert peak provides a forecast of the range of validity of the business-as-usual dynamics.

The classic case of this consideration is certainly the historical application of the model to oil production in the United States. National production followed the trend predicted by Hubbert until after the peak, but then increased again to unprecedented values. This "multiple trend" behavior is explained by the exploitation of unconventional resources, which were excluded from the initial $K$ estimate that was based solely on conventional oil.

Considering the Hubbert model estimates for the resources of the 12 elements reported in Figures \ref{fig:Exp_R2zone} and \ref{fig:Hubb_R2zone}, a striking feature is that most peaks are estimated before the end of the century. Following the previous discussion, we conclude that either these materials will become critical or something will change from the current dynamics.

While the model can be used to estimate \emph{when} the system will deviate from its previous dynamics, it does not contain predictive information about which assumption will actually be broken. A complementary approach based on foresight can help to estimate which evolution would be most desirable, or most disadvantageous.

\subsection{A foresight reading of the model}

As opposed to forecasting, foresight considers that the future is open and to be built and aims at providing tools to help in the decision making process \cite{Berger2008}. As a foresight tool, it provides insight into strategies that can be considered to circumvent the problem of resource depletion. Previously, we used the model to make an educated guess at possible time scales before reaching a production peak that would be due solely to geological constraints.

Here we use the model to discuss the effectiveness of strategies to delay peak production as late as possible. Which parameter of the model's dynamics should be changed in order to most efficiently push back the production peak? Is it more efficient to try to find new resources and thus increase the value of $K$, or to reduce demand which corresponds to increasing the value of $\tau$?

\begin{figure}[H]
    \centering
     \includegraphics[height = 6cm]{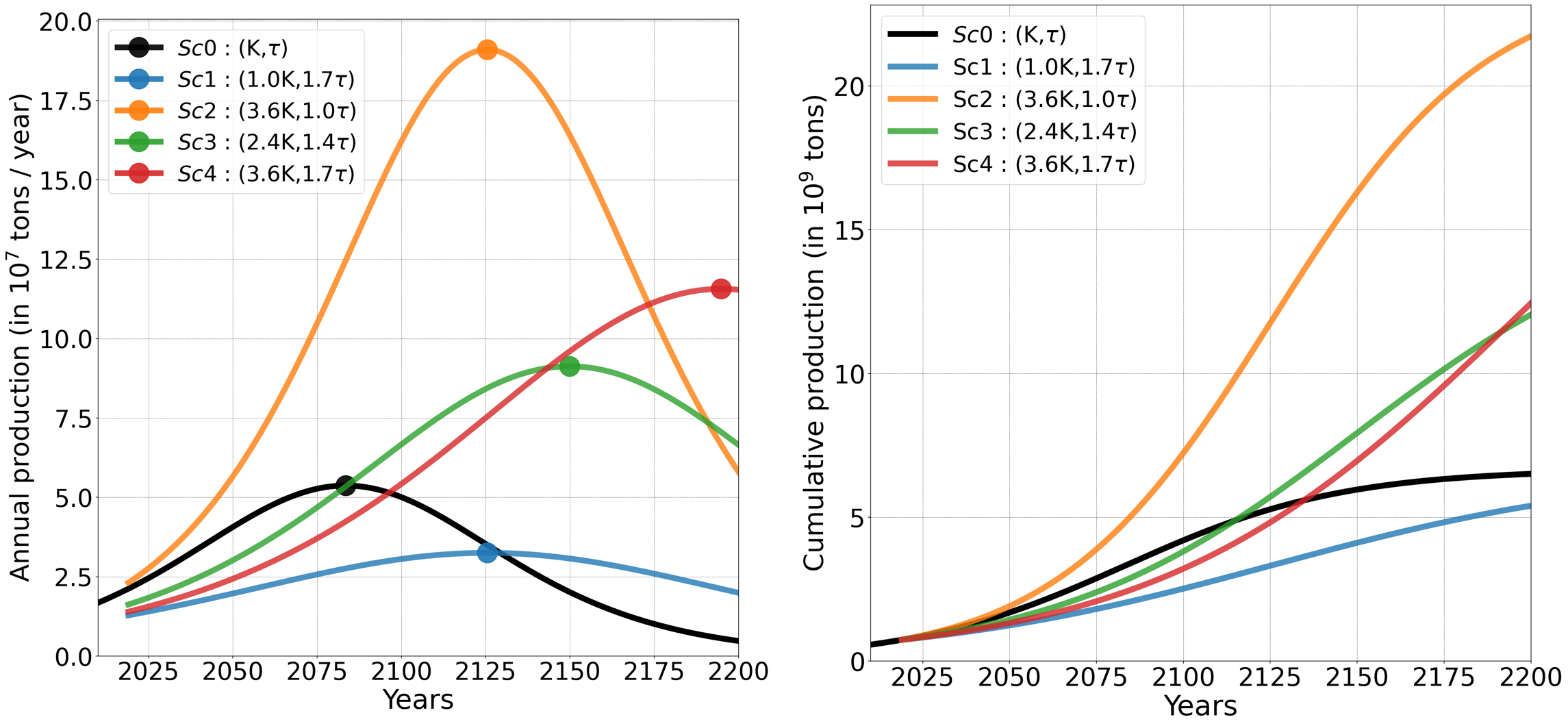}

    \caption{{\bf Time evolution of foresight scenarios}. Annual production (left) and cumulative production resulting from delaying strategies of the peak of the scenario 0 based on a Copper scenario and defined by the triplet ($Q(2019) = 7.4 \ 10^{8} \ tons, \ K = 9 \cdot Q(2019), \ \tau = 31 \ yrs$). }
    \label{fig:prospective_time}
\end{figure}

As an example, we consider delaying peak copper production to the year 2125 from the 2083 estimate in the business-as-usual scenario based on currently identified resources, see Fig \ref{fig:prospective_time} denoted Scenario 0 (Sc0), black line. To achieve this result, a first option would be to reduce or stabilize the demand for raw materials, in order to slow down the extraction rate $\tau \rightarrow \tau ' = 1.7 \tau$ (Scenario 1, blue line on Fig.\ref{fig:prospective_time}). The numerical factor can be estimated from:

\begin{align}
    \Delta t_{{\rm peak}}	&=\tau'\ln\left(\frac{K}{Q_{0}}-1\right)-\tau\ln\left(\frac{K}{Q_{0}}-1\right)\\
\Rightarrow\frac{\tau'}{\tau}	&=\frac{\Delta t_{{\rm peak}}}{\ln\left(\frac{K}{Q_{0}}-1\right)}\simeq 1.7
\end{align}

The peak can be delayed by the same amount of time by increasing mineral exploration to discover new resources and increase reserves (economically extractable quantity) before the end of the exponential phase of scenario 0. The second scenario leads to a change in $K\rightarrow K' = 3.6K$ (Scenario 2, orange line on Fig.~\ref{fig:prospective_time}) since:

\begin{align}
\Delta t_{{\rm peak}}	&=\tau\ln\left(\frac{K'}{Q_{0}}-1\right)-\tau\ln\left(\frac{K}{Q_{0}}-1\right)\\
\Rightarrow\frac{K'}{K}	&=\left(1-\frac{Q_{0}}{K}\right)\exp\left(\frac{\Delta t_{{\rm peak}}}{\tau}\right)+\frac{Q_{0}}{K} \simeq 3.6     
\end{align}

The comparison between these two options highlights that decreasing $\tau$ (by reducing or stabilizing demand) is globally more efficient than increasing $K$ through the discovery of new resources for example. The height of the production peak is however much higher in the case of an increase in $K$. Note that to postpone the peak by only 42 years, keeping the same exponential growth dynamics, it is necessary to find 3.6 times the currently known resources.

This difference can be explained by the dynamics of the model at the beginning which corresponds to an exponential growth. This results in a logarithmic dependence of the position of the peak with respect to the total amount of available resource $K$.

It is also worth considering changing both parameters at the same time. Applying both of the above changes simultaneously shifts the peak by about 100 years, more than twice their individual contributions (Scenario 4, red line in Fig.~\ref{fig:prospective_time}). The non-linearity of this behavior leaves room for a finer-grained analysis of an optimal strategy - that is, a path that maximizes peak carryover with minimal changes from the current situation.

\begin{figure}[H]
    \centering
    \includegraphics[height = 6.5cm]{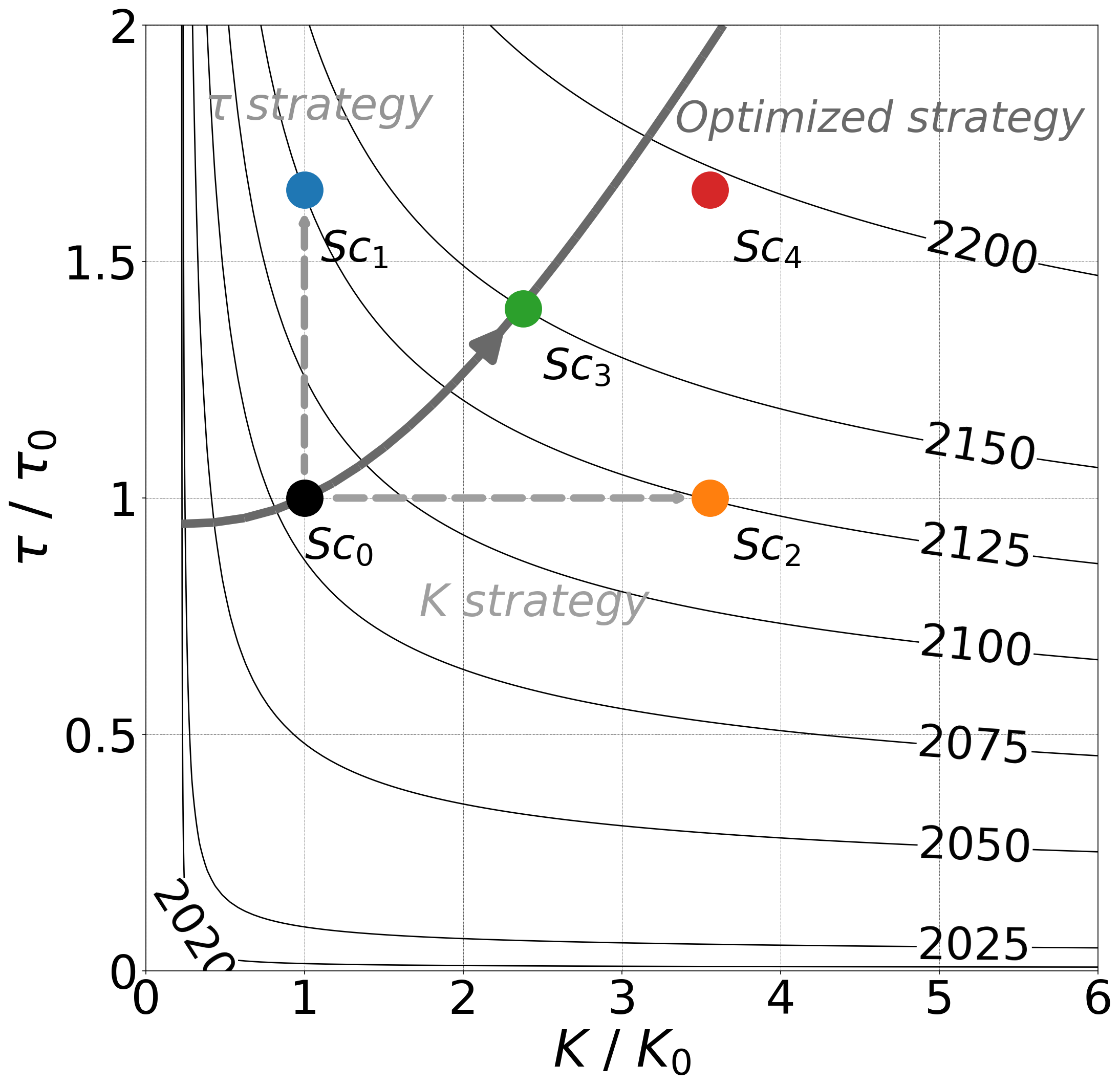}
    \caption{{\bf Foresight scenario based on the Hubbert's map.} The curve illustrates the most effective strategy for delaying the peak by acting on the two parameters $\tau$ and $K$. This strategy requires both slowing down the mining activities (by increasing $\tau$) and finding more resources (by increasing $K$). It corresponds to the gradient in parameter space.}
    \label{fig:optimal_scenario}
\end{figure}

The Hubbert map presented earlier provides a simple trajectory of such a strategy optimized in the two-parameter space (see Fig.~\ref{fig:optimal_scenario}). On this Hubbert map, each pair of parameters corresponds to a point of the plane $(\tau, \frac{K}{Q_{0}})$. This point defines a possible scenario that defines a choice of the time scale of mining and potential discovery of new resources.

The optimal strategy corresponds to the line orthogonal to the iso-date gradient. To delay the peak by 42 years, it is sufficient to increase reserves by a factor 1.9 and the extraction time by a factor 1.2 (scenario 3, green line in Fig.~\ref{fig:prospective_time}). This change may be easier than trying to get the same result by changing $K$ or $\tau$ alone.

In this example, the effort for changing $K$ by a given factor is considered as equivalent to the effort for changing $\tau$ by the same factor. In reality, we can consider that increasing reserves by a factor 2 is significantly easier, or harder, than decreasing the extraction rate by a similar factor. Such considerations can be readily included in the previous analysis, by rescaling the axes of the Hubbert map according to the relevant weighting.

\medbreak
The five scenarios in Fig \ref{fig:optimal_scenario} reflect possible dynamics expressed by different sets of model parameters. Each of them can thus be considered in part as a global commodity management objective. This work does not analyze the extent to which they would result from constraints suffered or chosen.

 \subsection{Phenomena and constraints affecting the desirability of scenarios}

\medbreak
The phenomena and constraints affecting one or both of the model's parameters are numerous and dealing with their interactions would require ambitious systemic evaluations that go far beyond this study, but we propose to illustrate some of them in the following section. 

\subsubsection{Materials demand for the energy and industrial transitions}

Energy transitions in the current socio-economic context raise concerns about increasing raw material needs in strategic industries (e.g., infrastructure and the energy sector).
Like the copper case, well documented by Northey \textit{et al.} \cite{Northey2014}, the solar industry highlights the pressure of demand on the criticality of metals. The global metal demand of the photovoltaic industry to mitigate climate change is estimated based on a scenario developed by P. Verlinden \cite{Verlinden2020}. To reach the target of 1 TeraWatt (TW) of installed capacity with photovoltaic technologies (Mono PERC with fixed tilt), it would require 29 000 tons of Silver which corresponds to 94 \% of the worldwide production in 2018. Lithium is another example of element associated with a group of technologies which is increasingly stressing the demand by comparison with current mining production \cite{Greim2020}).

 

Based on these two examples, one could conclude that the mining industry should aim to drastically increase reserves, i.e. significantly increase $K$.  

Furthermore, with the intention of acting on both parameters, policies could consider minimizing the consumption of these critical materials. This reduction could be achieved by designers and producers of renewable energy technologies. In other words, it is a matter of reducing the demand for materials through technological efficiency. This is reflected in practice by the fact that the European Commission and other national institutions publish reports on the criticality of raw materials \cite{EC2020}, \cite{EC2017} and call on policy makers to reduce consumption based on resource efficiency.

However, the effects of efficiency gains are uncertain and sometimes even counterproductive in the current socio-economic context, as this may favor rebound effects. \cite{Gossart2015,wallenborn2018,Parrique2019}. On the other hand, energy needs being inversely proportional to the mineral content \cite{Northey2014}, the coupling between energy and raw material needs generates a positive feedback on the raw material demand. Under these conditions, seeking to increase the $K$ parameter could conflict with the initial objective, namely the fight against climate change. 


\subsubsection{The planetary boundaries framework and the phosphorus case}


The case of phosphorus, as a component of fertilizer, illustrates a link between mining, agriculture and the phosphorus cycle, a global limitation identified in 2009 by Rockstrom \textit{et al.} \cite{Rockstrom2009} and updated in 2015 by Steffen \textit{et al.}  \cite{Steffen2015} then in 2022 by Persson \textit{et al.} \cite{Persson2022}. This case points to an environmental limit to human activities.       

In this regard, an increase in phosphorus extraction from rock can lead to an increase in the flux of phosphorus dissipated into freshwater rivers and then into the oceans. It is currently estimated to $22 \ Tg \ P/yr$ \cite{Carpenter2011}, \cite{Liu2008}), which would at the end push more humanity into the buffer zone of $[11–100] \ Tg \ P/yr$ and possibly cross a threshold of no return \cite{Steffen2015}.

Consequently, and within this sustainability framework, scenarios that include a trend toward increased phosphorus production are undesirable and, to some extent, unrealistic given the response of the planetary system to such perturbations. In other words, while the business-as-usual scenario implies that material demand determines material supply as long as supply is considered unlimited, the planetary boundaries framework calls for a reversal of this paradigm by accepting/establishing a limitation of supply and adapting society accordingly.

Indeed, current and future mining activities raise environmental concerns (e.g., \cite{Lapcik2018,Kaunda2020}). These environmental impacts must be integrated as ecological boundaries into resource management strategies regarding the needs of society to be met.

\subsubsection{Social pressure and (mid and) long-term planning}

Social, economic and political phenomena could also limit the interest of increasing trend scenarios for local and global populations. The 2020 report of the European Environment Agency’s (EEA) called “State of the Environment” states that \textit{"Europe will not achieve its 2030 goals without urgent action during the next 10 years to address the alarming rate of biodiversity loss, increasing impacts of climate change and the over-consumption of natural resources."} \cite{EEA2020}. However, policies that delineate what can be produced (i.e., infrastructure or consumer goods) with the raw material extracted and how, for example, through incentives and restrictions, are rare and sometimes ineffective, as illustrated by plastics pollution prevention \cite{daCosta2021}. The framework for the continuation of the current socio-economic paradigm in which the current criticality of raw materials is defined, as a decision support tool, could be strongly affected by ambitious resource governance strategies \cite{Ali2017} and growing social pressures from local populations \cite{Engels2018,Crost2020}.


Thus, these few elements reflect the systemic nature of the problem of raw material extraction. Therefore, given the ecological constraints and social needs, the question is: "Which scenario should be considered as a goal to be achieved? ". Extensive interdisciplinary work is still needed to assess the plausibility and relevance of scenarios to support large-scale sustainability strategies with time horizons beyond a century.

\section{Conclusion}
The Hubbert model offers a phenomenological good description for several mining elements, most of which are currently before the theoretical peak, i.e. in the exponential growth phase. For the time being, this phase remains valid and no reduction in demand for raw materials is envisaged in the scenarios and projections for the 2050 or 2060 horizon. A key strength of the model is that the location of the peak depends only weakly (logarithmically) on the stock size. Significant uncertainties on the stock size thus have little influence on the peak estimates. Moreover, some elements have already reached a peak such as Bromine and Strontium.  


Nevertheless, the underlying assumptions of the model, i.e. continuous growth in demand and supply of raw materials and the constraint of geological depletion, have no reason a priori to hold forever, except if the considered stock corresponds to the ultimate limit. It appears indeed that several elements depart from the simple Hubbert profile when the demand and/or production context evolves. However, their new dynamics can in turn be described by an updated Hubert model, with adapted values for $\tau$ and $K$.

The Hubbert model, presented here in its simplest form, is thus not a forecasting model on a global scale. First, because it is only able to estimate what will happen as long as current conditions remain unchanged. Second, and to a lesser extent, because of the problematic nature of the data usually computed to obtain predictions.

It offers nonetheless fruitful foresights. It sets boundaries on the business-as-usual trajectories, highlighting the need of departing from current conditions to avoid production peaks. It also allows us to estimate the influence on the production peak of deviations from current conditions. The Hubbert plane presented in this article makes this reading explicit. In particular, it shows that the increase in available stock for the mining industry delays the peak, but only in a limited way as compared to the influence of altering $\tau$ the time scale of mining and by extension the supply rate.

The discussion emphasizes the necessity of additional, systemic and interdisciplinary studies to assess the sustainability and desirability of foresight scenarios in general, including those for the mining industry. All the more so as other constraints, which are not taken into account here, will be stronger in the medium term such as limited access to fossil fuels and geopolitical considerations that limit access to raw materials. In this respect, the group of business-as-usual scenarios presented in this article is probably not sustainable in the second half of this century. 

\section{Declaration of competing interest}
We wish to confirm that there are no known conflicts of interest associated with this publication and that no specific grant for this work has been received that could have influenced its outcome. 

\section{Credit authorship contribution statement}

\textbf{Lucas Riondet:} Conceptualization, Data Curation, Formal Analysis, Investigation, Software, Visualization, Writing, Validation, Original Draft Preparation. \textbf{Daniel Suchet:} Conceptualization, Methodology, Formal Analysis, Validation, Writing original Draft Preparation. \textbf{Olivier Vidal:} Writing - Review \& Editing. \textbf{José Halloy:} Supervision, Conceptualization, Project Administration, Writing - review \& Editing.


\pagebreak

\appendix

\pagebreak

\section{Supporting information}

\textbf{Applicability of Hubbert model to global mining industry: Interpretations and insights}

\medbreak
Lucas Riondet, Daniel Suchet, Olivier Vidal, José Halloy

\pagebreak
\section{Appendix: Raw data} \label{appendix:rawdata}

\begin{table}[h!]
	\begin{center}
		\caption{ Database. collected from USGS (2021) and resulting K values}
		\label{tab:rawdata}
		\renewcommand\arraystretch{1.3}
		\begin{adjustbox}{angle=90}
		\begin{tabular}{|c|c|c|c|c||c|c|} 
			\hline
			\multicolumn{2}{|c|}{ \multirow{2}{*}{\textbf{Element}}} &\textbf{Reserves} & \textbf{Resources} & $ \mathbf{\sum_{t=1900}^{2019} P(t)}$ & \textbf{K(Reserves)} & \textbf{K(Resources)} \\
			\multicolumn{2}{|c|}{}  & tons & tons & tons & tons & tons \\
			\hline

			\multirow{8}{*}{\rotatebox{90}{Exponential}} & Aluminium & $ 1.50 \cdot 10^{10} \ (*) $ &  $ 3.75 \cdot 10^{10} \ (*) $  &  $1.44 \cdot 10^{9} $  & $ 1.64 \cdot 10^{10} $ & $ 3.89 \cdot 10^{10} $ \\
			\cline{2-7}
		
			& Chromium &  $ 5.70 \cdot 10^{8} $ &  $ 1.20 \cdot 10^{10}$  &   	$6.64 \cdot 10^{8} \ (**) $ & $ 1.23 \cdot 10^{9} $ & $ 1.27 \cdot 10^{10} $ \\
			\cline{2-7}
			
			& Copper &  $ 8.70 \cdot 10^{8} $ &  $ 5.60 \cdot 10^{9} $  &  $7.28 \cdot 10^{8} $ &  $ 1.60 \cdot 10^{9} $ & $ 6.33 \cdot 10^{9} $ \\
			\cline{2-7}
			
			& Gold &  $ 5.3 \cdot 10^{4} $ &  $ 3.30 \cdot 10^{5} $  &  $1.66 \cdot 10^{5} $ & $ 2.19 \cdot 10^{5} $ & $ 4.96 \cdot 10^{5} $ \\
			\cline{2-7}
			
			& Lithium &  $ 2.10 \cdot 10^{7}$ &  $ 7.98 \cdot 10^{7}$  &   $1.43 \cdot 10^{7} \ (**) $ & $ 3.53 \cdot 10^{7} $ & $ 9.41 \cdot 10^{7} $ \\
			\cline{2-7}
			
			& Nickel &  $ 9.40 \cdot 10^{7}$ &  $ 3.00 \cdot 10^{8}$  & $7.15 \cdot 10^{7} $ & $ 1.66 \cdot 10^{8} $ & $ 3.72 \cdot 10^{8} $ \\
			\cline{2-7}
			
			
			& Zinc &  $ 2.50 \cdot 10^{8}$ & $ 1.90 \cdot 10^{9} $  &  $5.45 \cdot 10^{8} $ & $ 7.95 \cdot 10^{8} $ & $ 2.44 \cdot 10^{9} $ \\
			\hline
			\hline
			
			\multirow{5}{*}{\rotatebox{90}{Multiple trends}} & Cadmium &  $ 7.50 \cdot 10^{5} $ &  $ 5.70 \cdot 10^{6} $  &  $1.27 \cdot 10^{6} $ & $ 2.02 \cdot 10^{6} $ & $ 6.97 \cdot 10^{6} $ \\
			\cline{2-7}
			
			& Cobalt &  $ 7.10 \cdot 10^{6}$ &  $ 1.45 \cdot 10^{8}$  &   $2.50 \cdot 10^{6} $ & $ 9.60 \cdot 10^{6} $ & $ 1.47 \cdot 10^{8} $ \\
			\cline{2-7}
			
			& Manganese &  $ 1.30 \cdot 10^{9}$ &  $ - $  &  $7.03 \cdot 10^{8} $ & $ 2.00 \cdot 10^{9} $ & $ - $\\
			\cline{2-7}
			
			& Phosphorus &  $ 7.10 \cdot 10^{10}$ &  $ 3.00 \cdot 10^{11}$  & $9.21 \cdot 10^{9} $ & $ 8.02 \cdot 10^{10} $ & $ 3.09 \cdot 10^{11} $ \\
			\cline{2-7}
			
			& Selenium &  $ 1.00 \cdot 10^{5}$ &  $ - $  &   $1.06 \cdot 10^{5} $ & $ 2.06 \cdot 10^{5} $ & $ - $ \\
			\hline  
			\hline               
			
			\multirow{5}{*}{\rotatebox{90}{Hubbertian}} & Bromine &  $ 1.20 \cdot 10^{7}$ &  $ 1.00 \cdot 10^{9}$  &    $ 2.15 \cdot 10^{7} $ & $ 3.35 \cdot 10^{7} $ & $ 1.02 \cdot 10^{9} $ \\
			\cline{2-7}
			
			& Indium &  $ 1.60 \cdot 10^{4}$ & $ - $  &  $ 1.48 \cdot 10^{4} $ & $ 3.08 \cdot 10^{4} $ & $ - $  \\
			\cline{2-7}
			
			& Strontium &  $ 6.80 \cdot 10^{6} $ &  $ 1.00 \cdot 10^{9} $ &  $ 1.30 \cdot 10^{7} $  & $ 1.98 \cdot 10^{7} $ &  $ 1.01 \cdot 10^{9} $\\
			\cline{2-7}
			
			& Thallium &  $ 3.80 \cdot 10^{2}$ &  $ 6.47 \cdot 10^{5} $  &  $ 5.10 \cdot 10^{2} $ & $ 8.90 \cdot 10^{2} $ & $ 6.48 \cdot 10^{5} $ \\
			\cline{2-7}
			
			& Yttrium &  $ 5.00 \cdot 10^{4}$ & $ - $  &  $ 1.47 \cdot 10^{5} $ & $ 1.97 \cdot 10^{5} $ & $ - $ \\
			\hline                   
		\end{tabular}
	\end{adjustbox}

	\end{center}
\end{table}

(*) Assuming a conversion coefficient between bauxite and Aluminium oxyde reserves/resources of 1/2.

\medbreak

(**) Element's production with discontinuity between the two types of USGS's data: long time series data (from 1900 to 2000s) and The Mineral Commodity Summary (since 1994).

\pagebreak

\section{Appendix: Parameter K definition in multiple trends case} \label{appendix:Koverestimation}

\begin{figure}[H]
    \centering
    \includegraphics[height = 6cm]{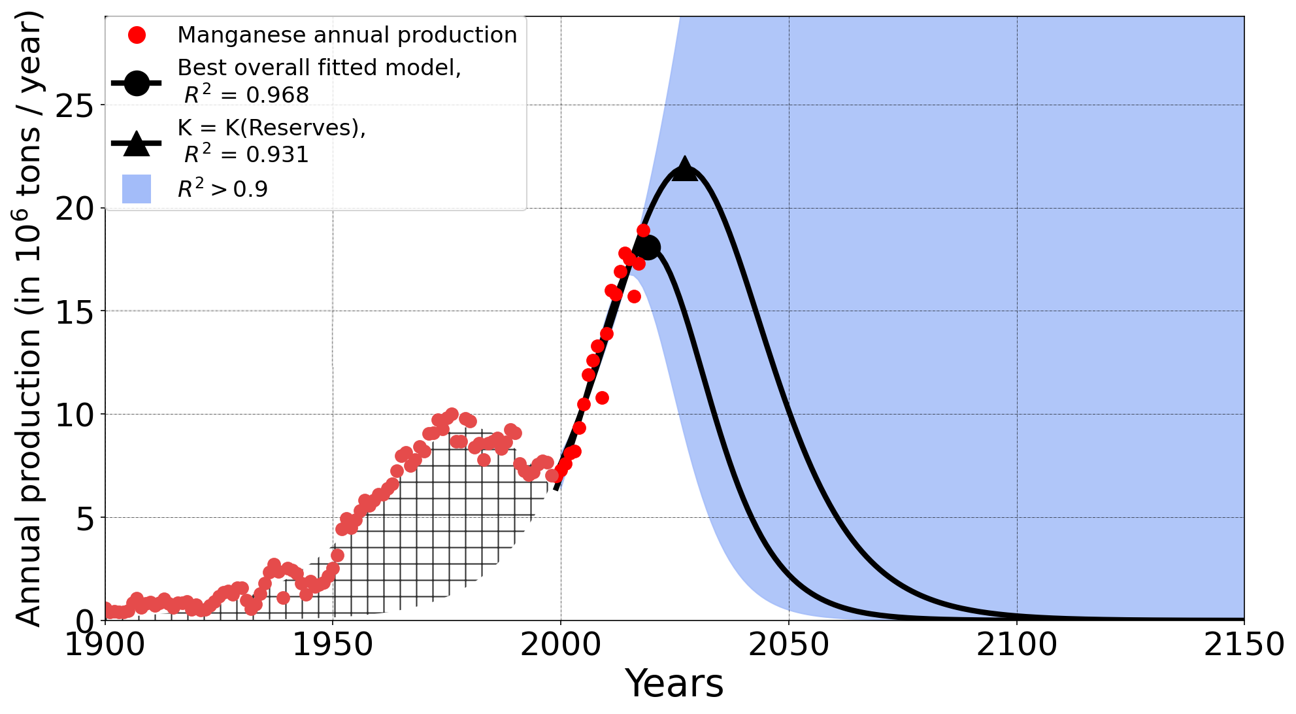}
    \caption*{Hatched area represents the part of the cumulative production artificially included in the ultimate quantity to be extracted because of the computation convention when applied to \emph{multiple trends} elements. It leads to overestimating the \emph{Ultimately Recoverable Resources} (URR).     
    } \label{fig:Koverstimation} 
\end{figure}

\pagebreak

\section{Appendix: Hubbertian trend elements} \label{appendix:Hubbertian}
{\bf Elements:} Thallium, Bromine, Strontium, Indium, Yttrium.
\begin{figure}[H]
    \centering
    \includegraphics[height = 3cm]{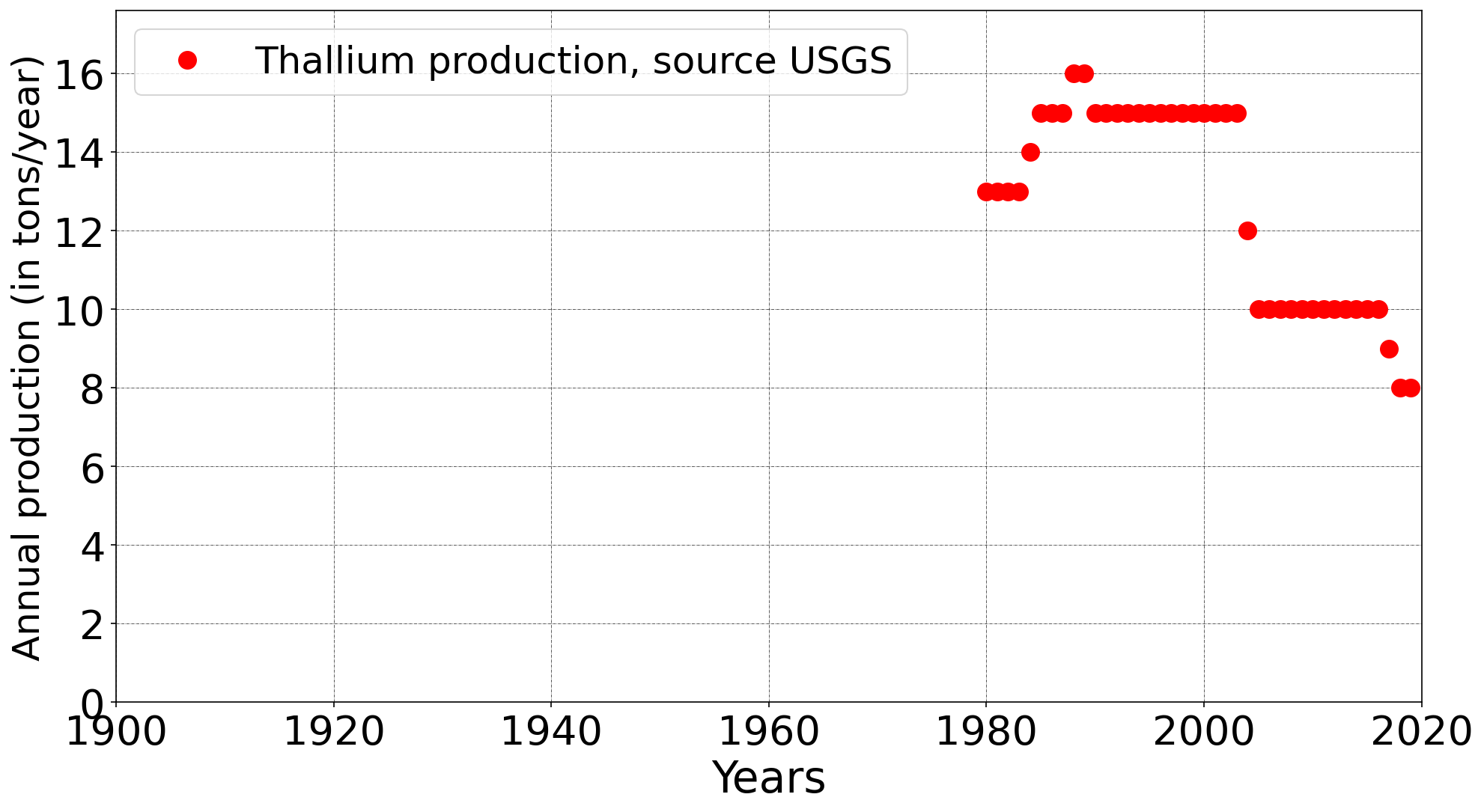}
    \includegraphics[height = 3cm]{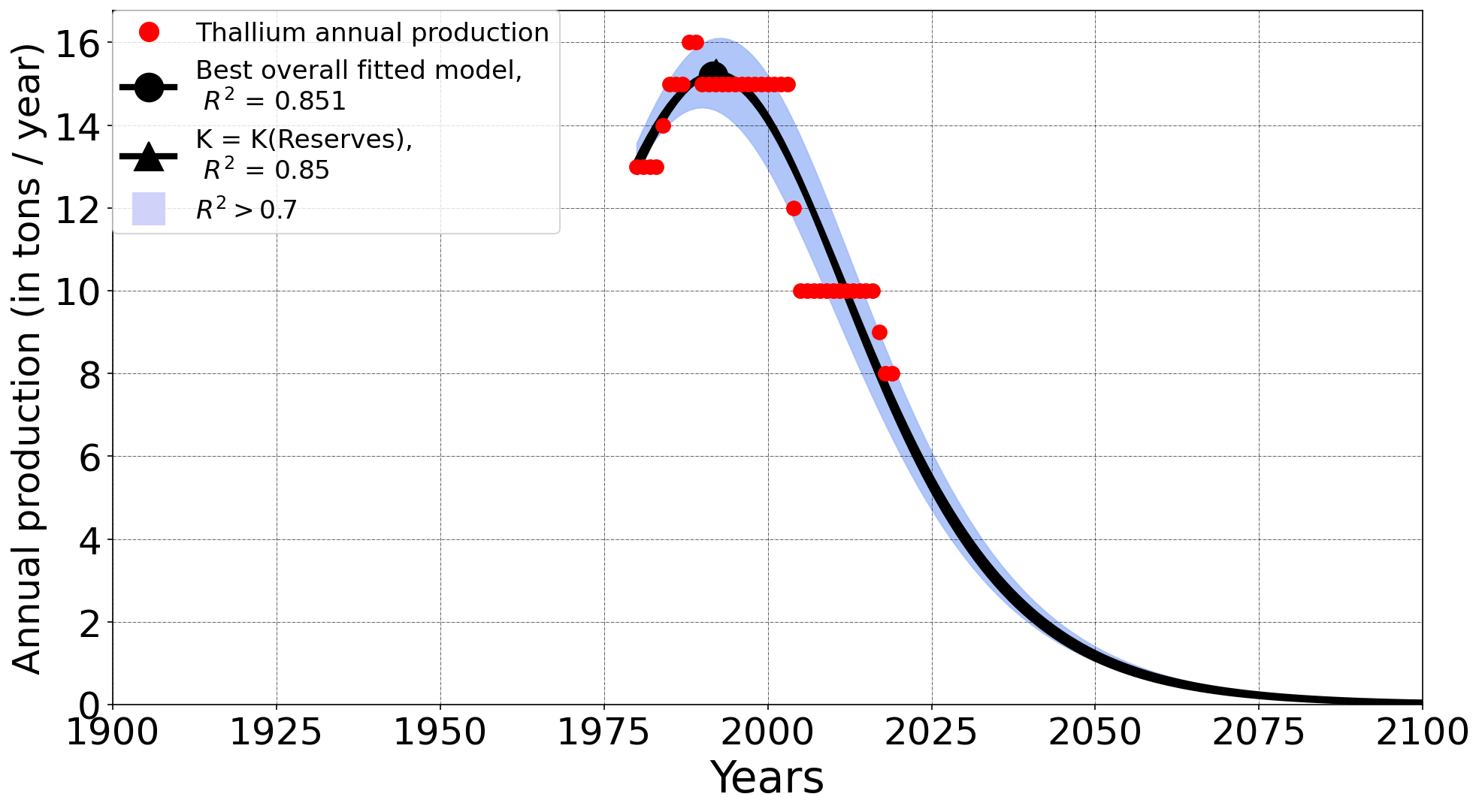}

    \includegraphics[height = 3cm]{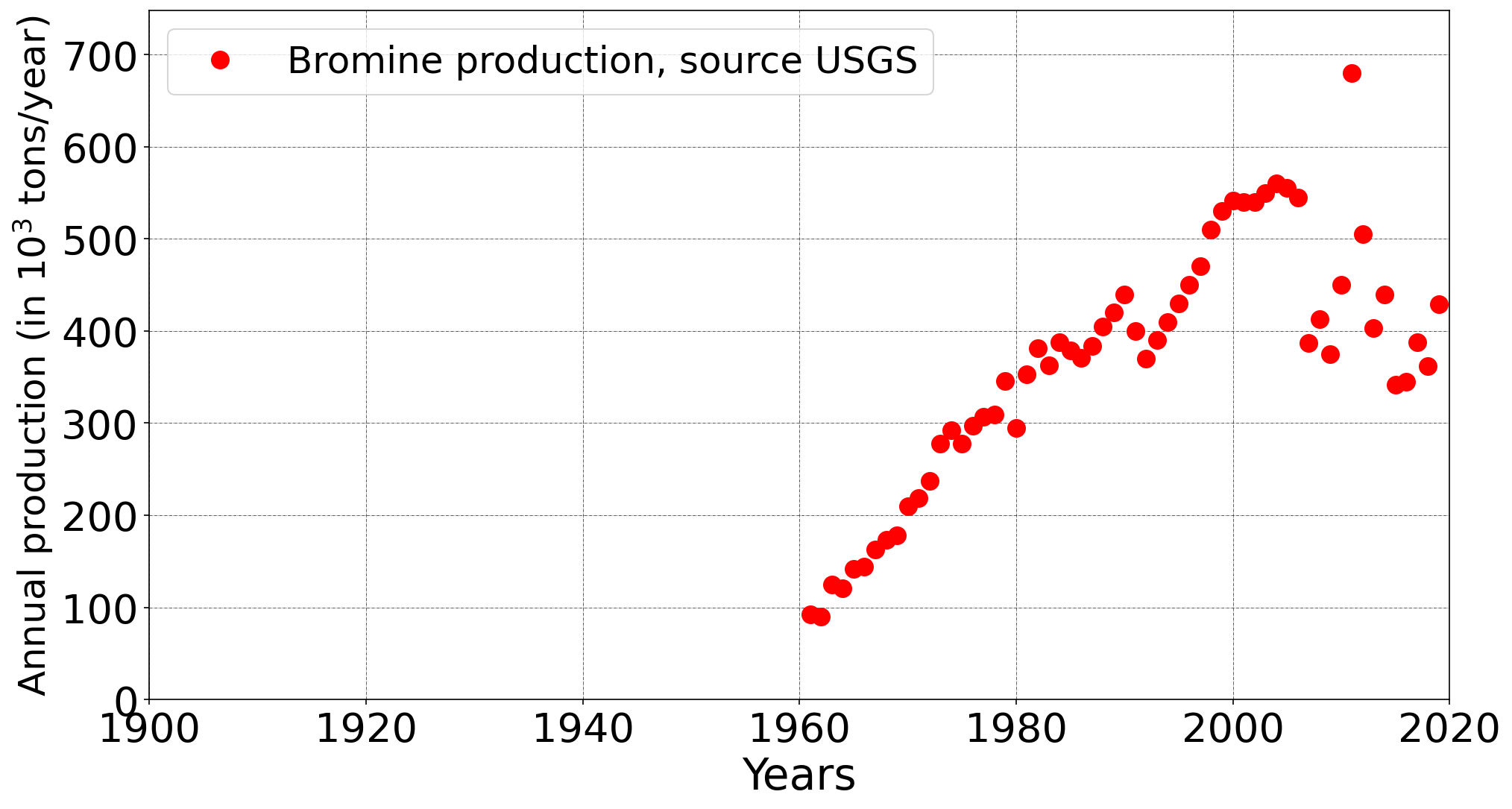}
     \includegraphics[height = 3cm]{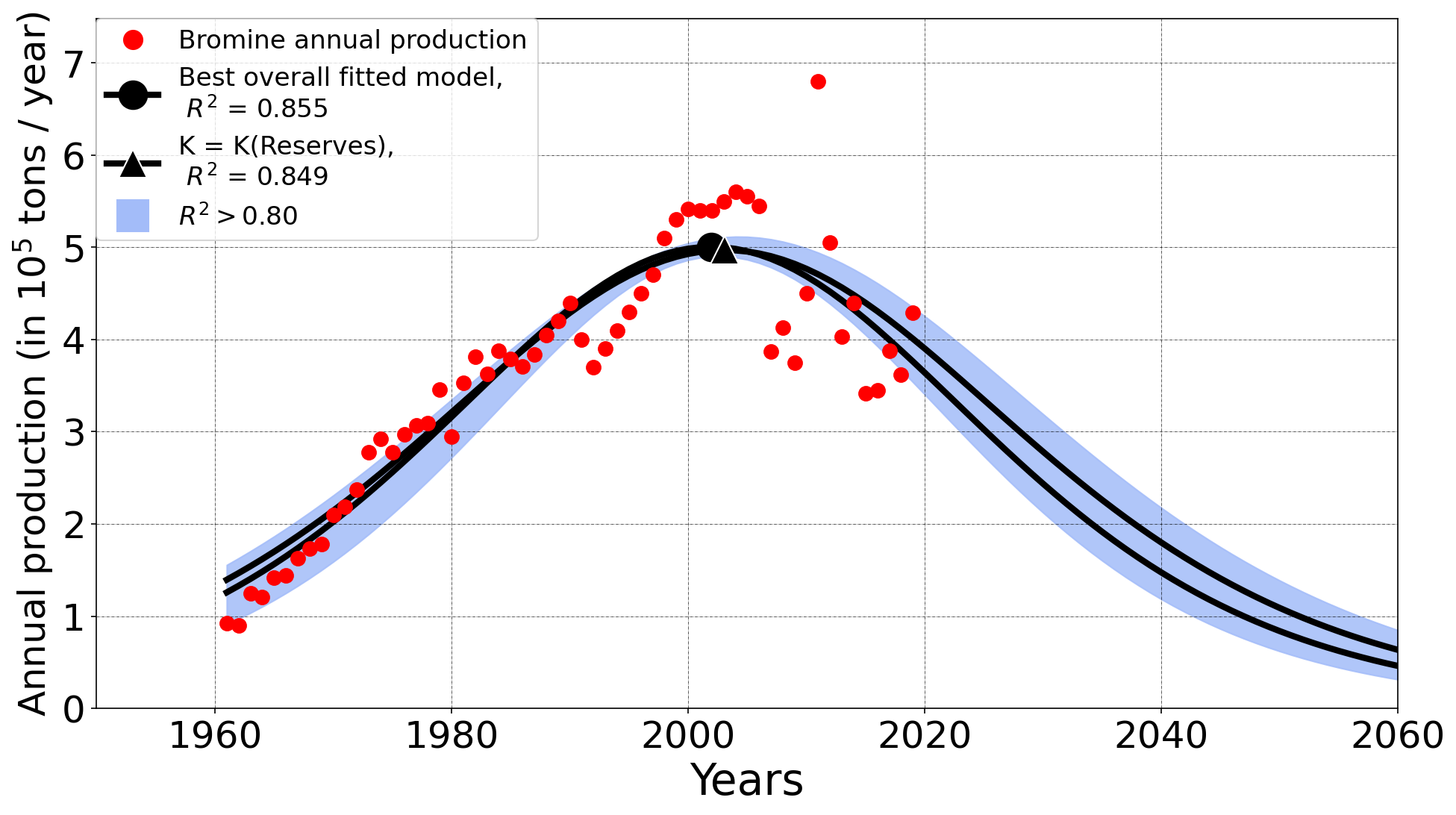}
    
    \includegraphics[height = 3cm]{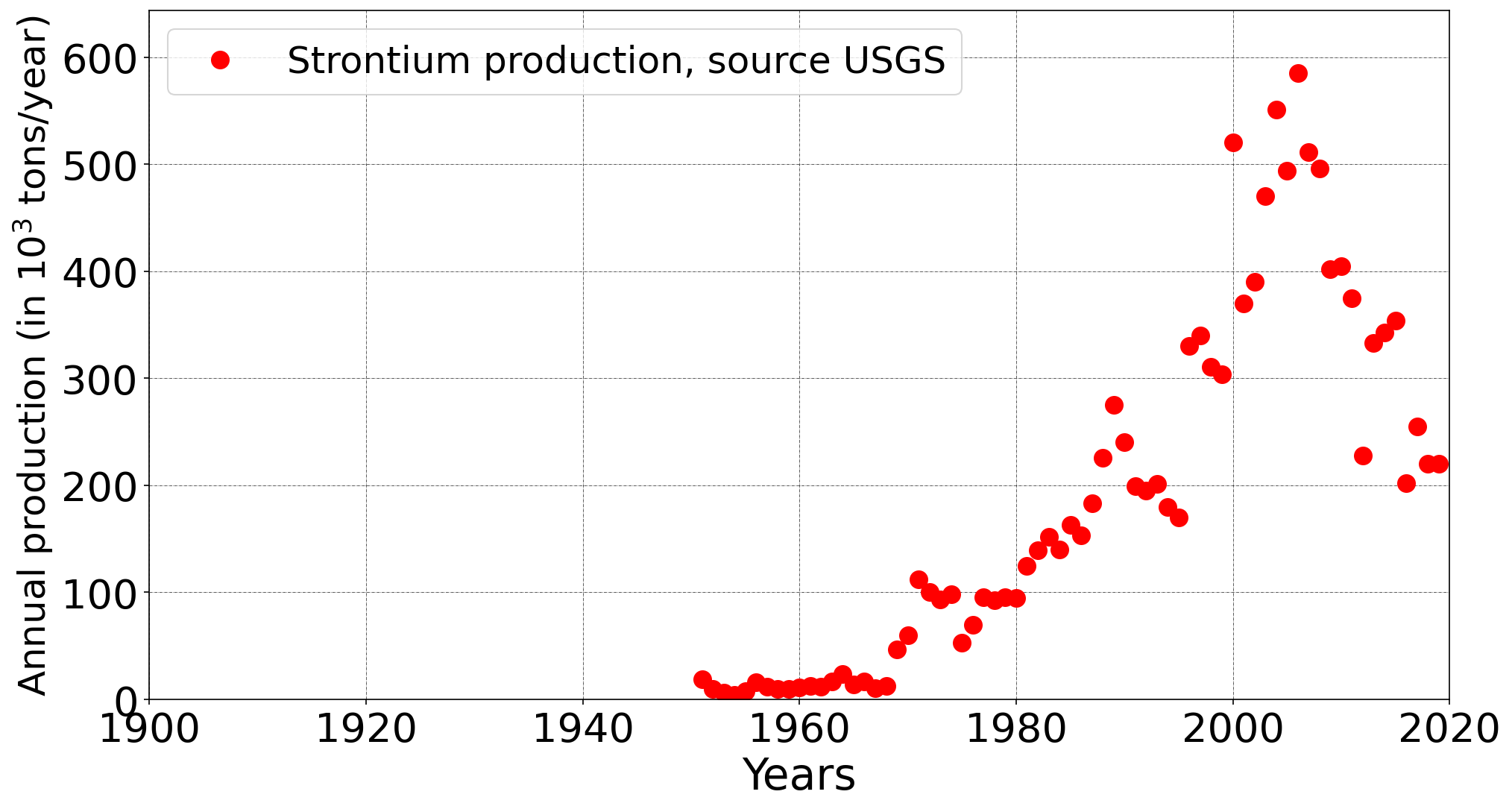}
     \includegraphics[height = 3cm]{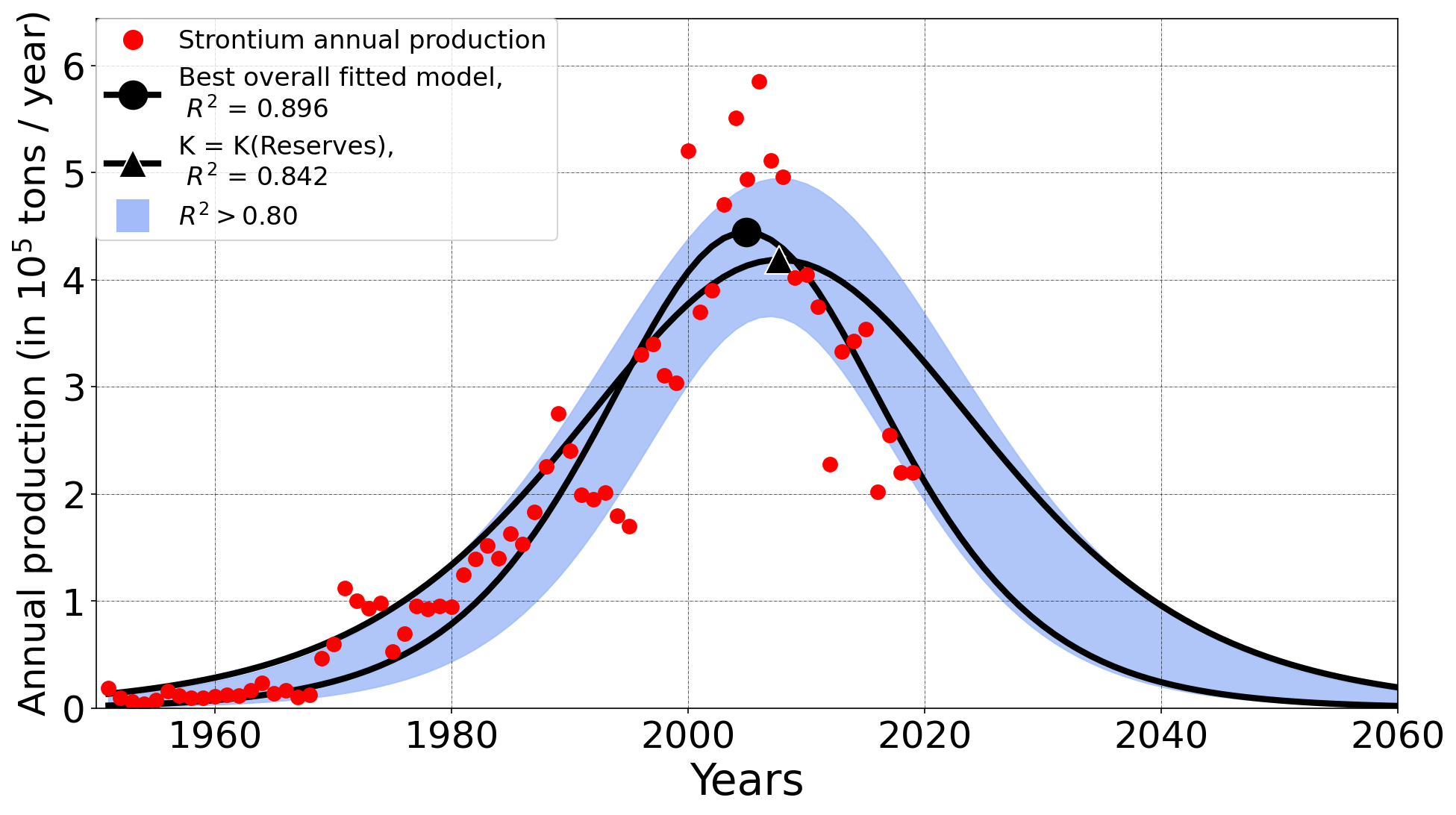}
     
    \includegraphics[height = 3cm]{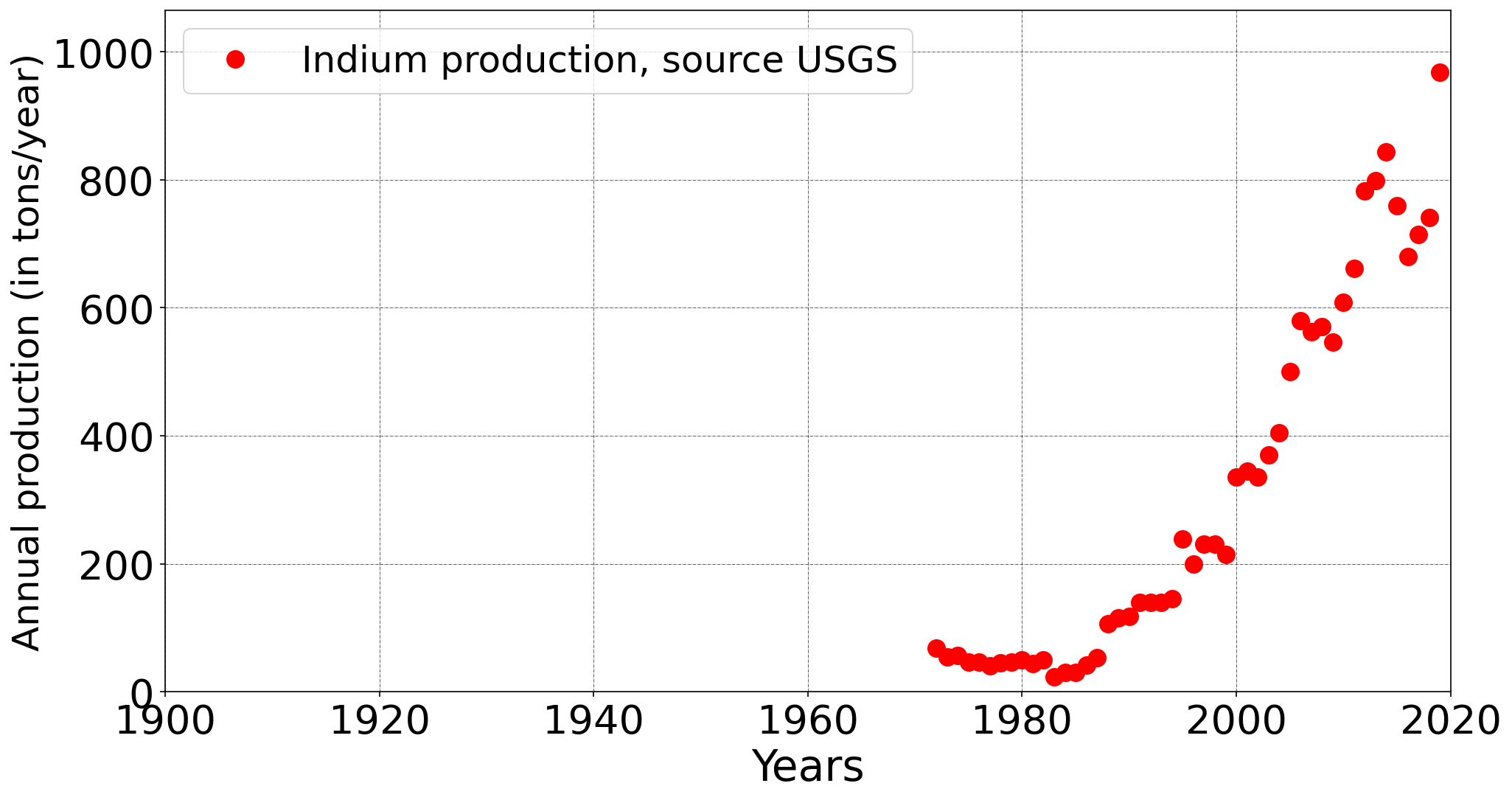}
    \includegraphics[height = 3cm]{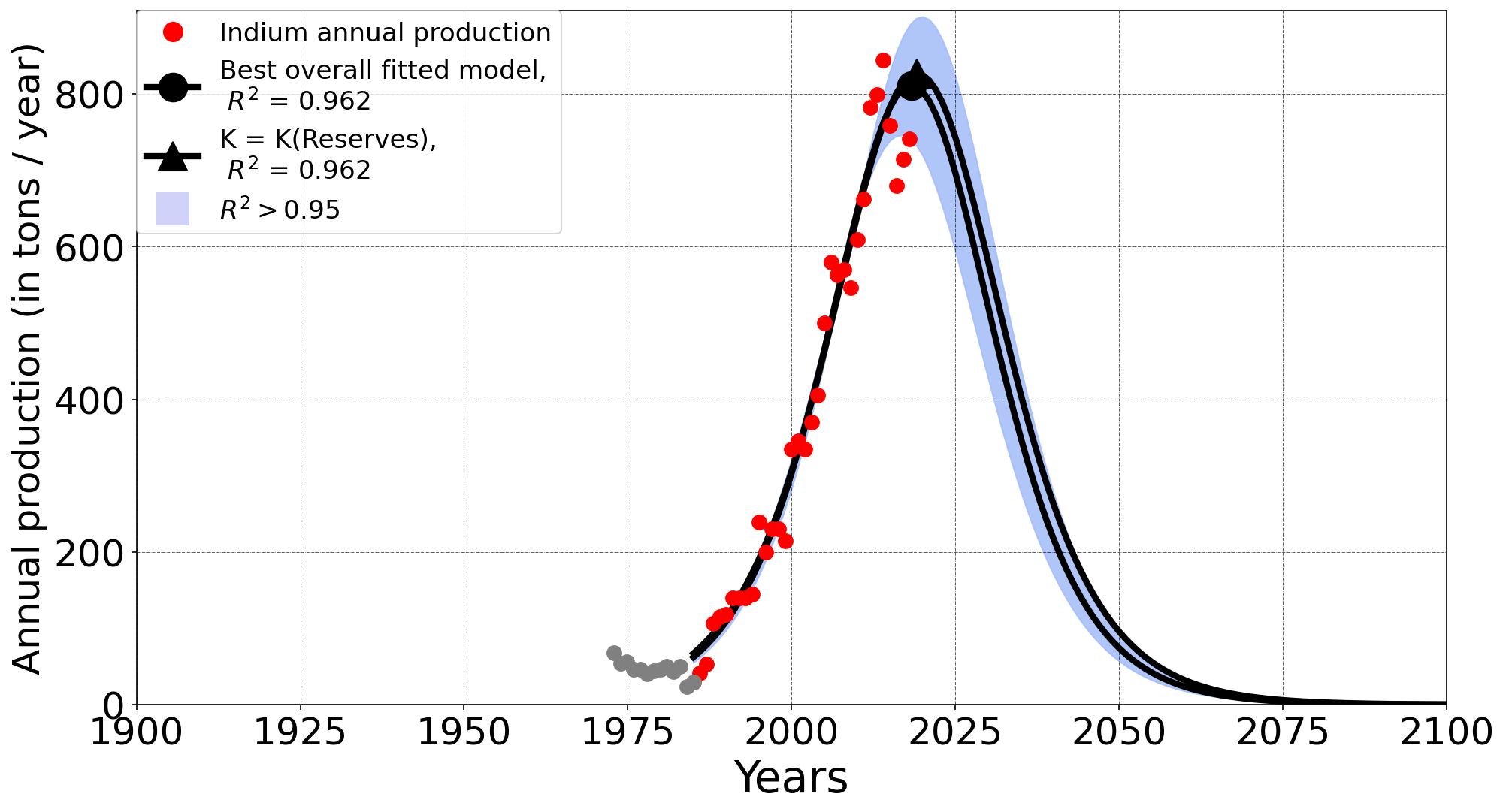}

    \includegraphics[height = 3cm]{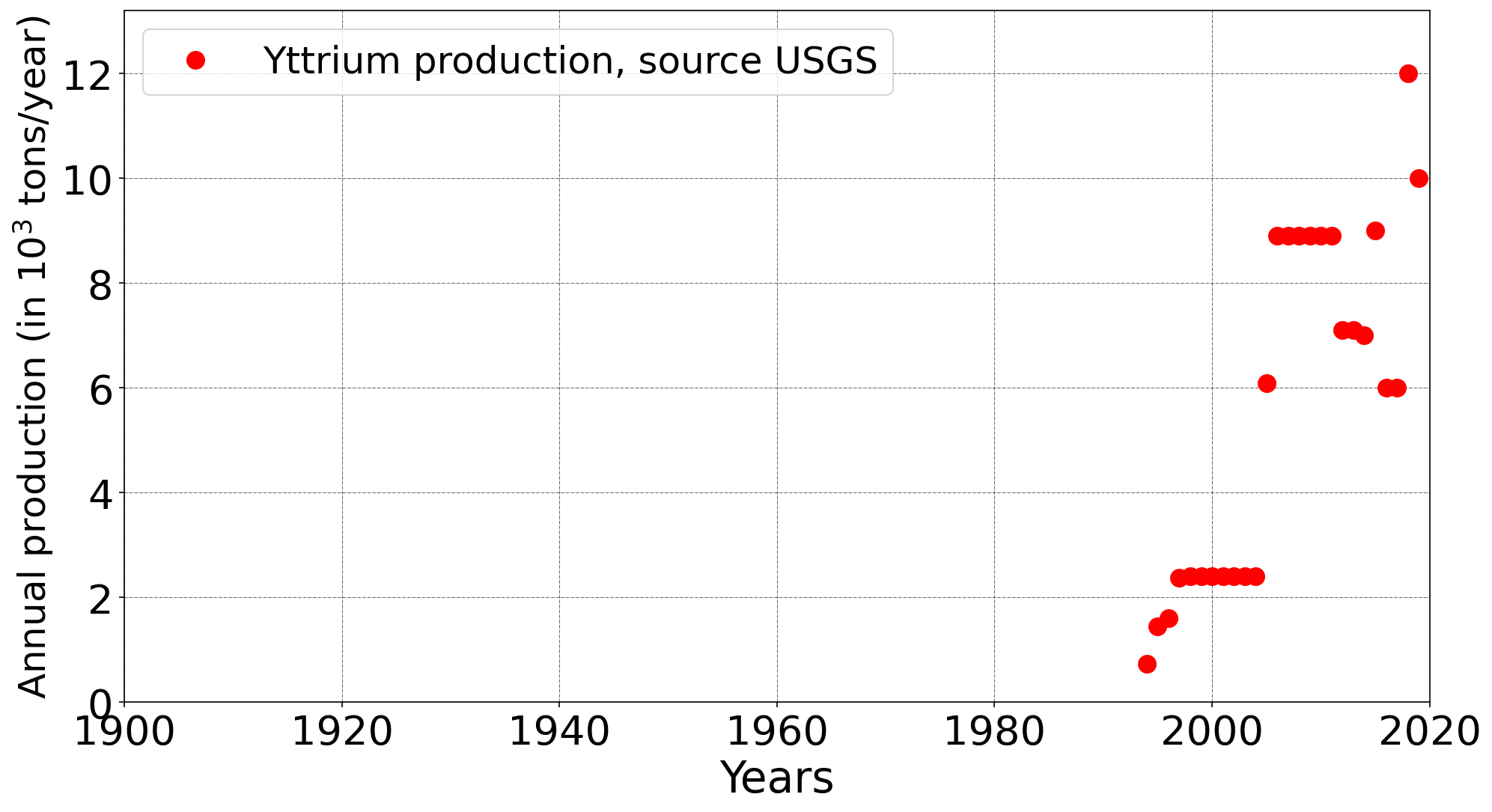}
    \includegraphics[height = 3cm]{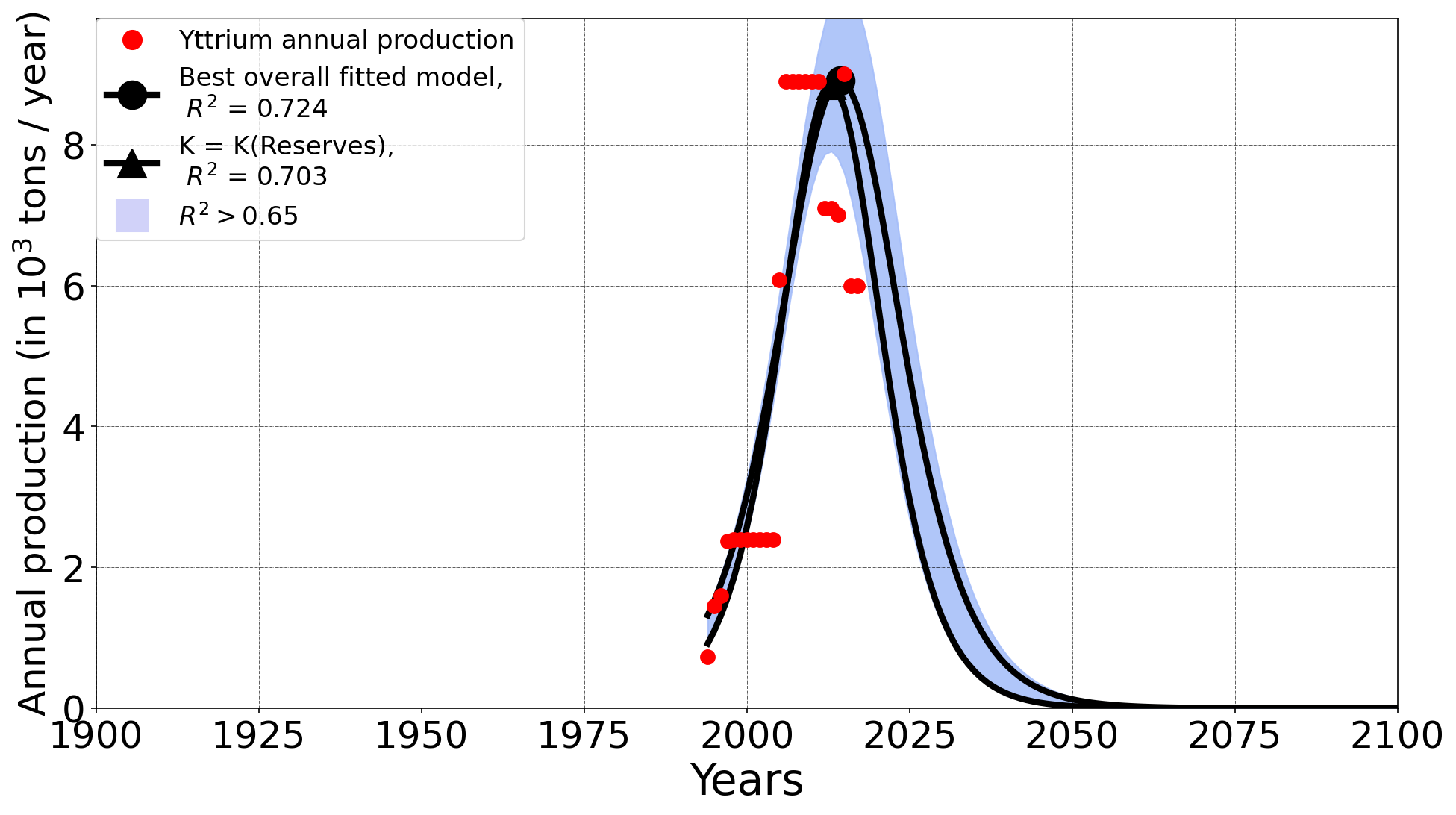}

    \caption*{} \label{fig:Hubbertian}
\end{figure}

\pagebreak

\section{Appendix: Exponential trend elements} \label{appendix:Exponential}
{\bf Elements:} Gold, Zinc, Copper, Nickel, Aluminium, Lithium, Chromium. 
\begin{figure}[H]
    \centering
    \includegraphics[height = 3cm]{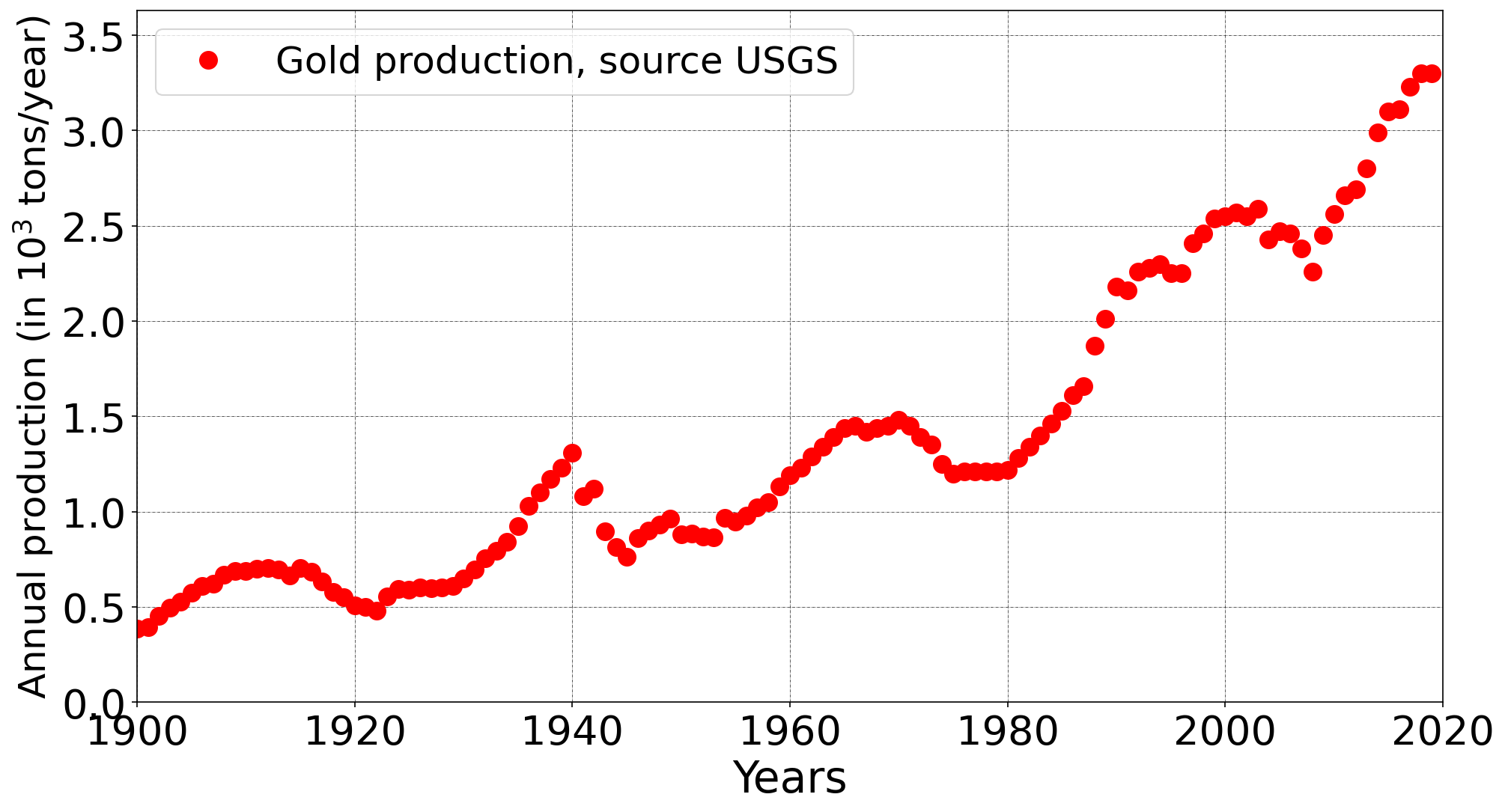}
    \includegraphics[height = 3cm]{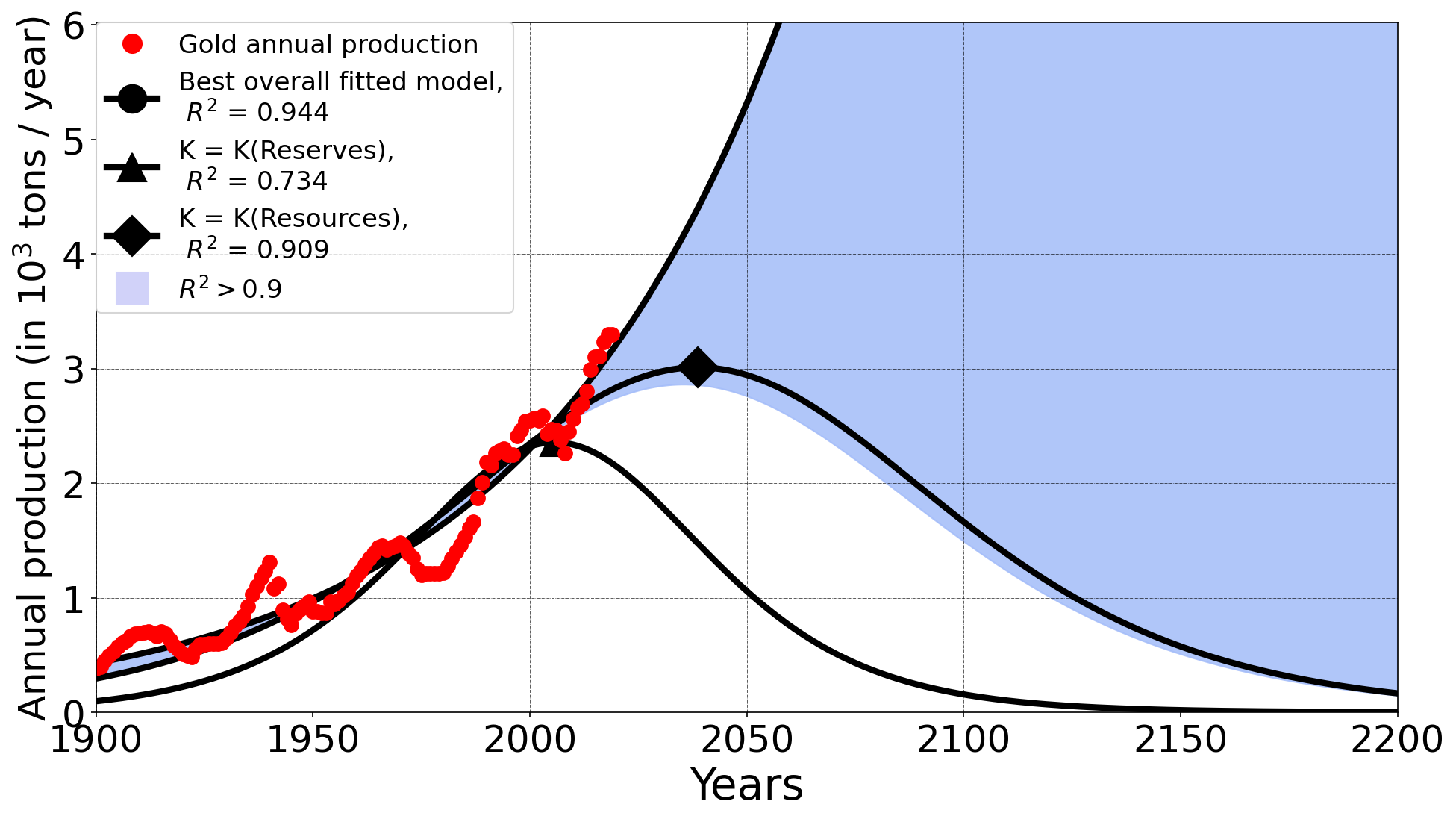}

    \includegraphics[height = 3cm]{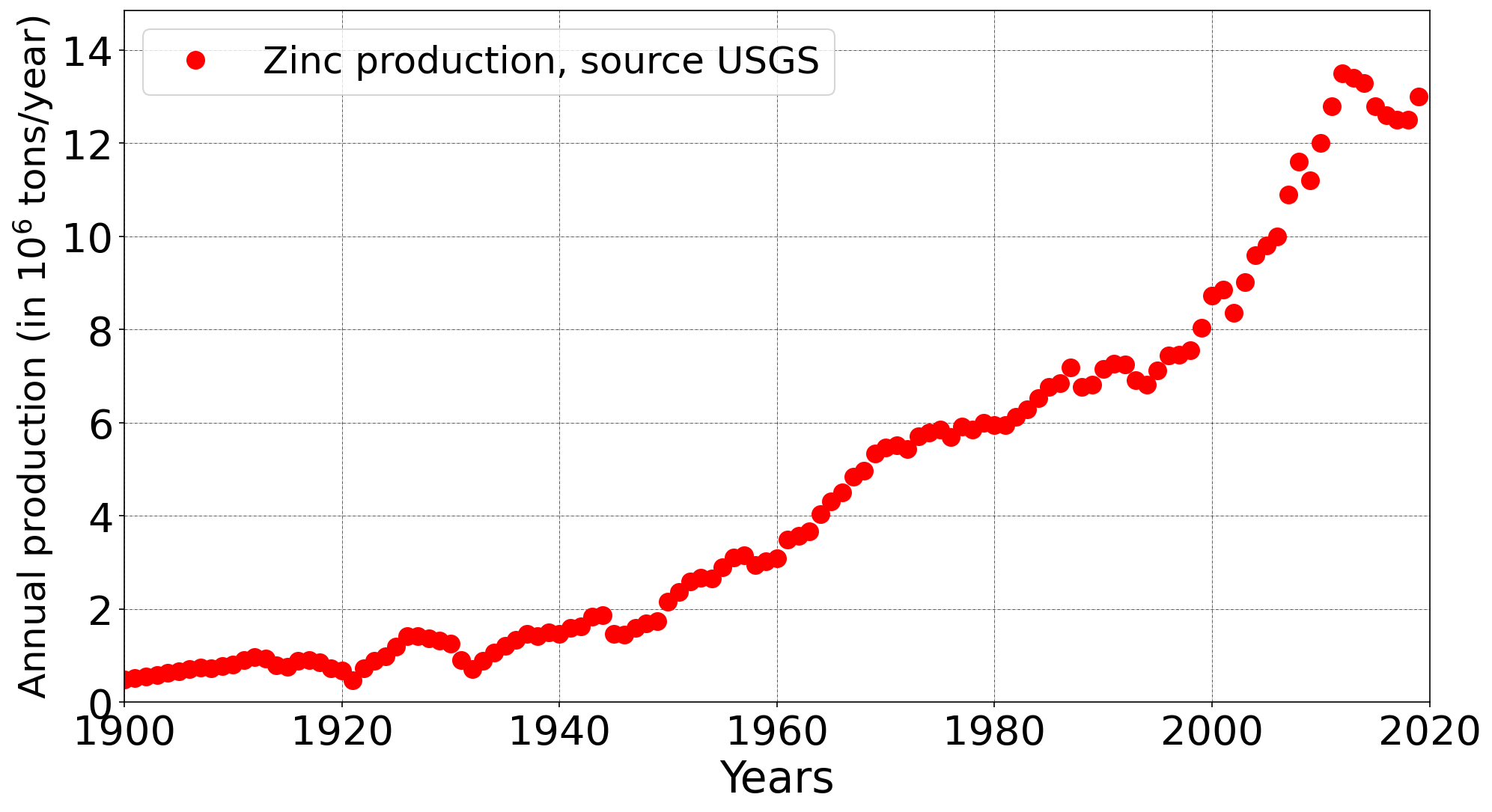}
    \includegraphics[height = 3cm]{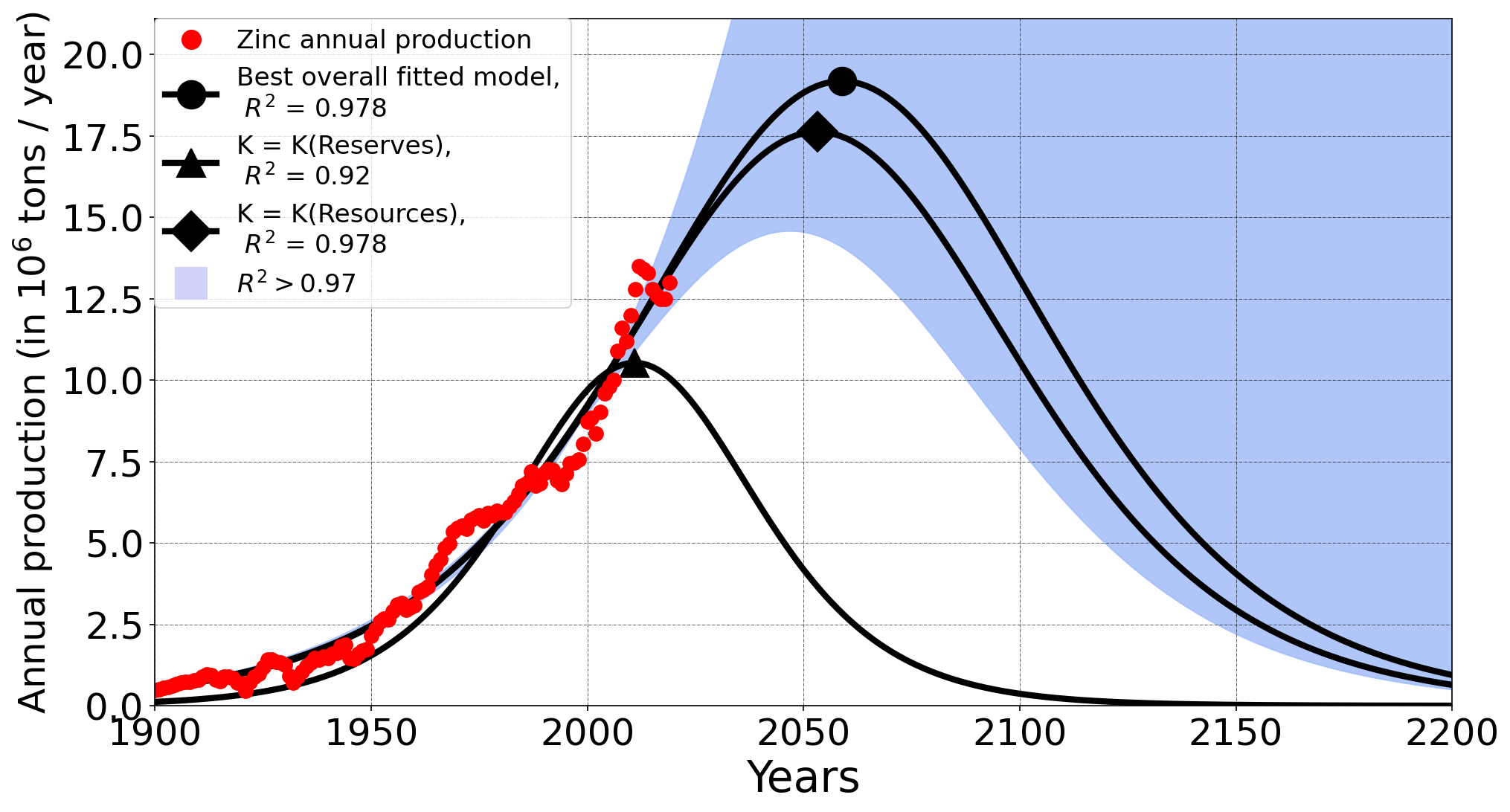}

    \includegraphics[height = 3cm]{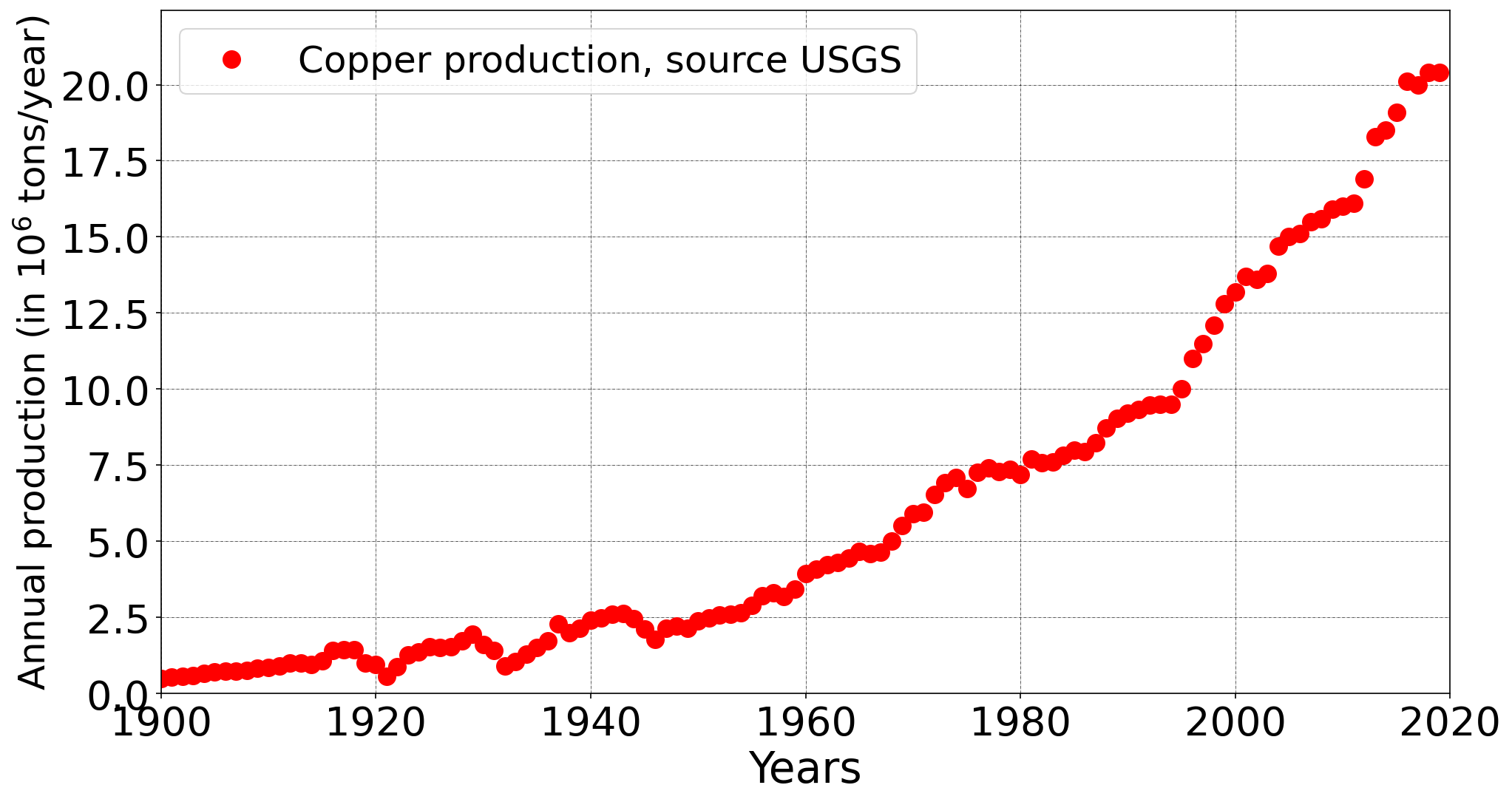}
    \includegraphics[height = 3cm]{Figure_5_Copper_Hubbert_model_time_representation_2019.png} 

    \includegraphics[height = 3cm]{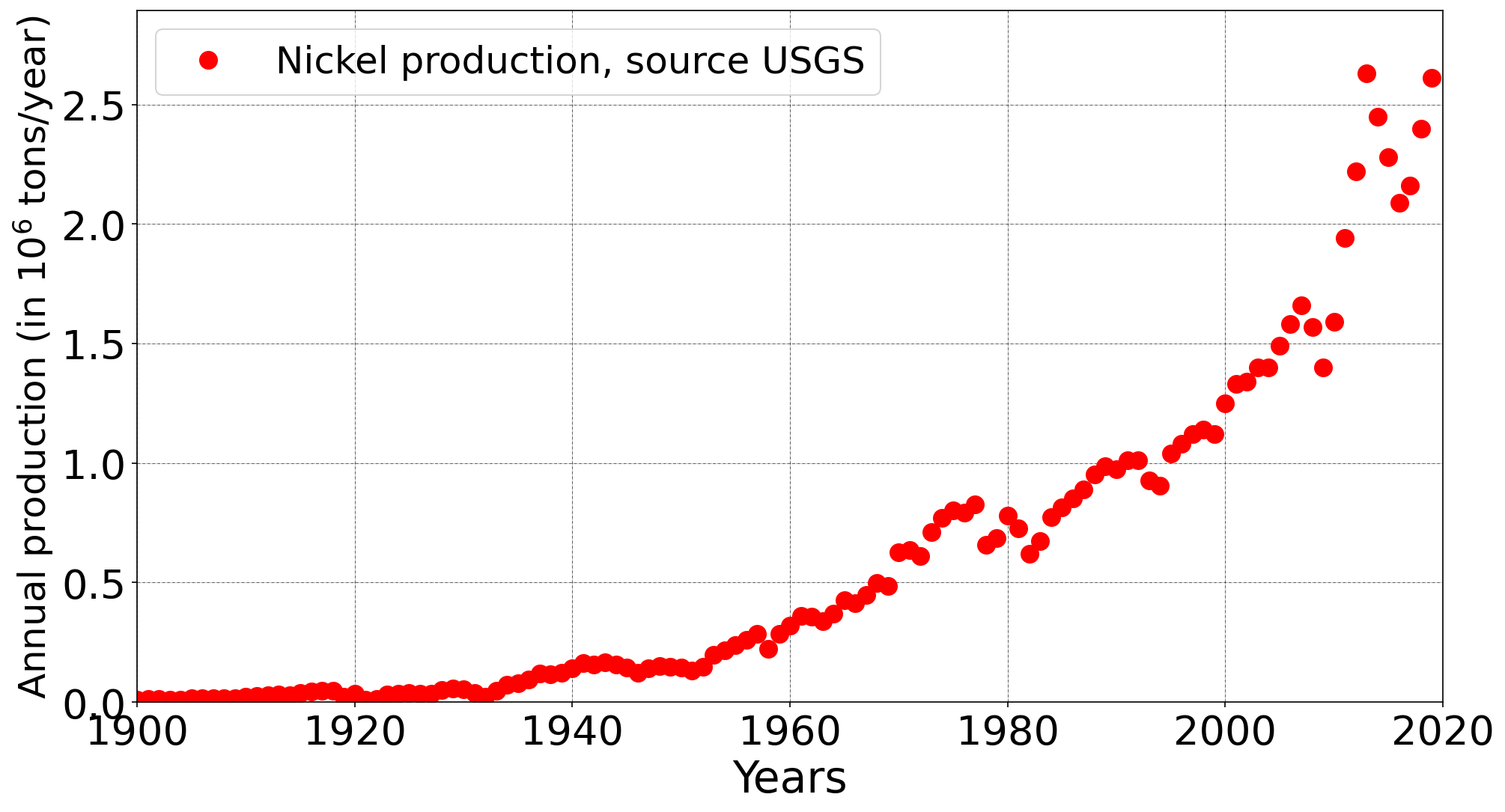}
    \includegraphics[height = 3cm]{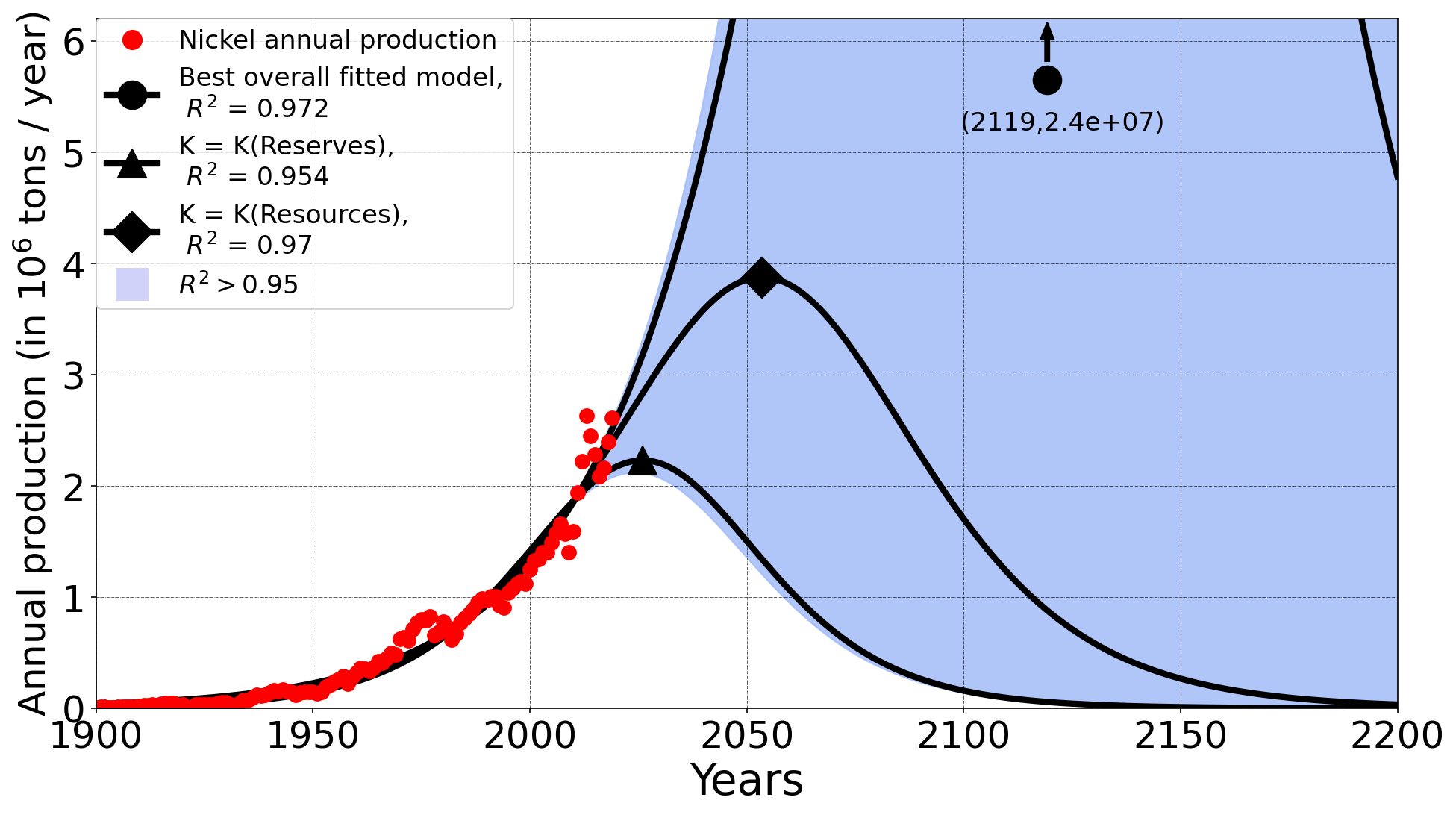}

    \includegraphics[height = 3cm]{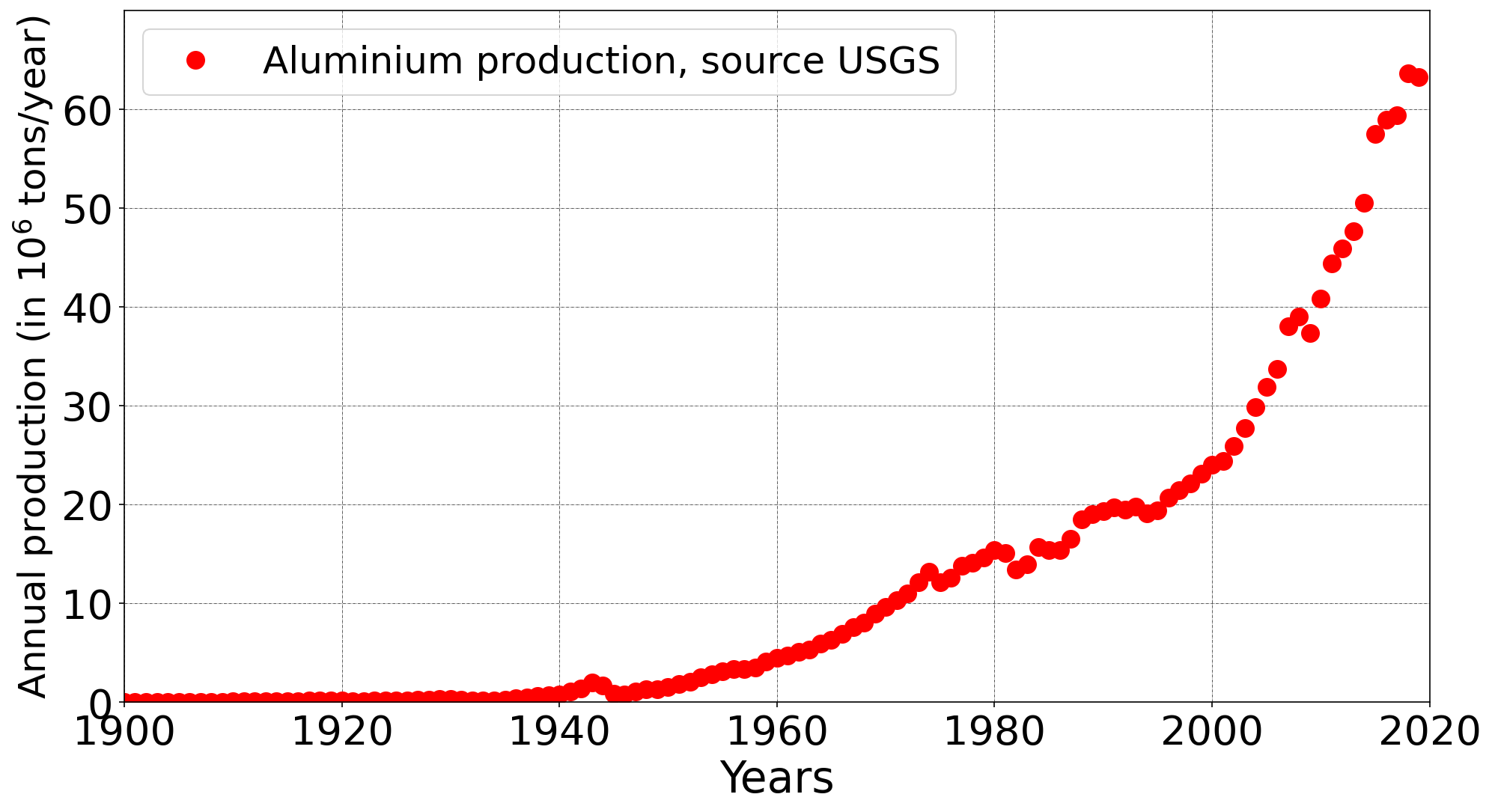}
    \includegraphics[height = 3cm]{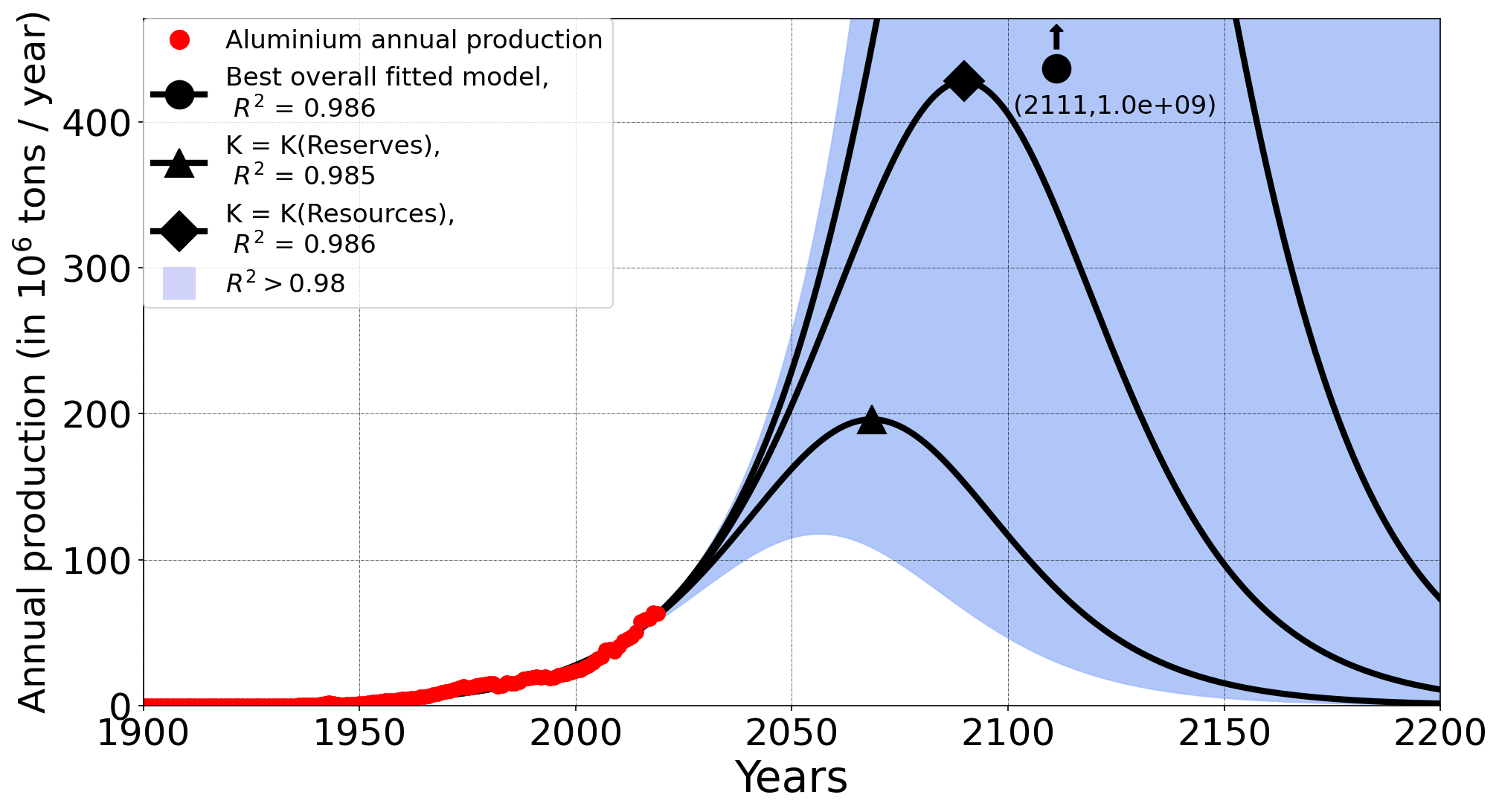}

    \includegraphics[height = 3cm]{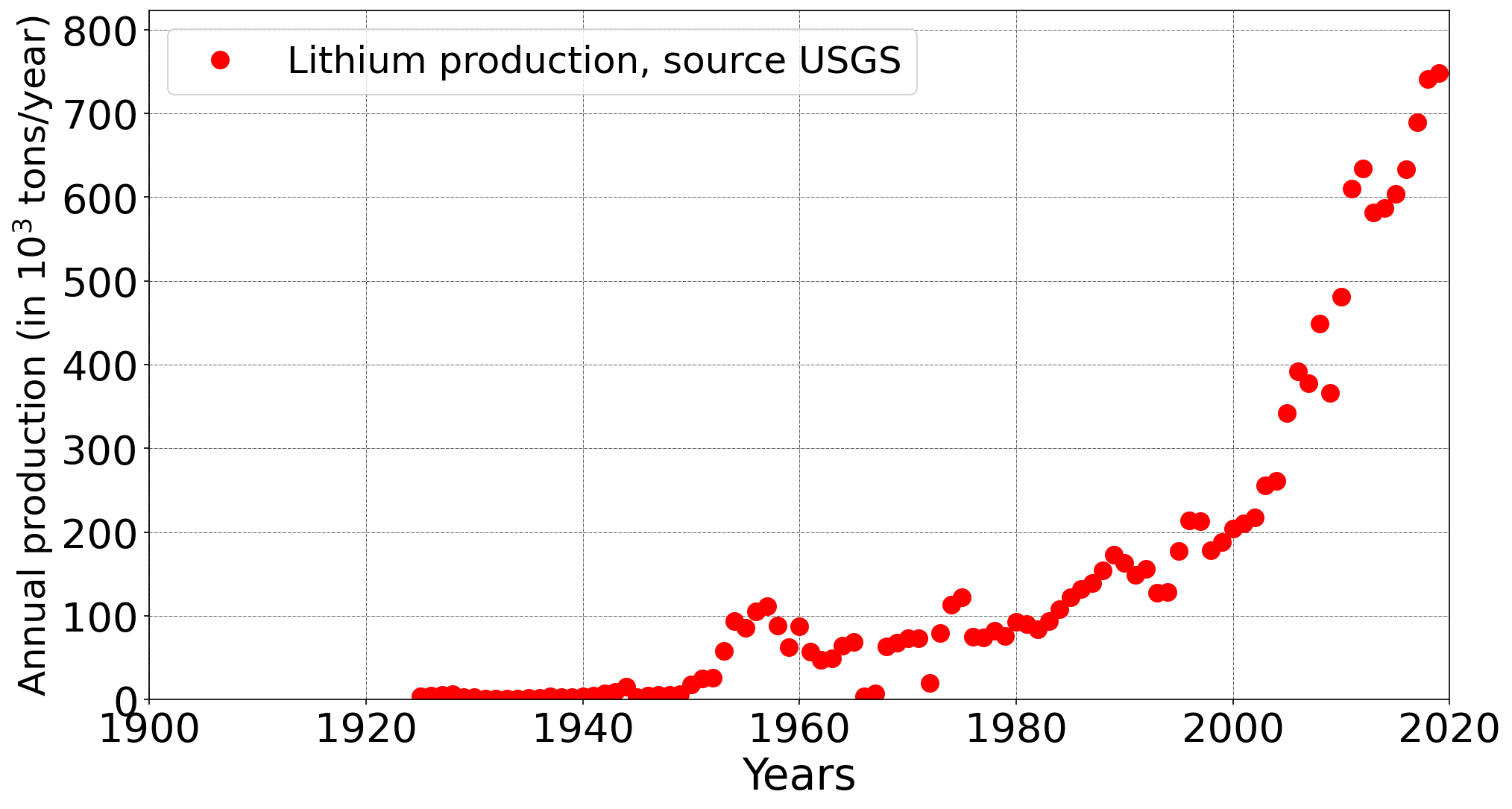}
    \includegraphics[height = 3cm]{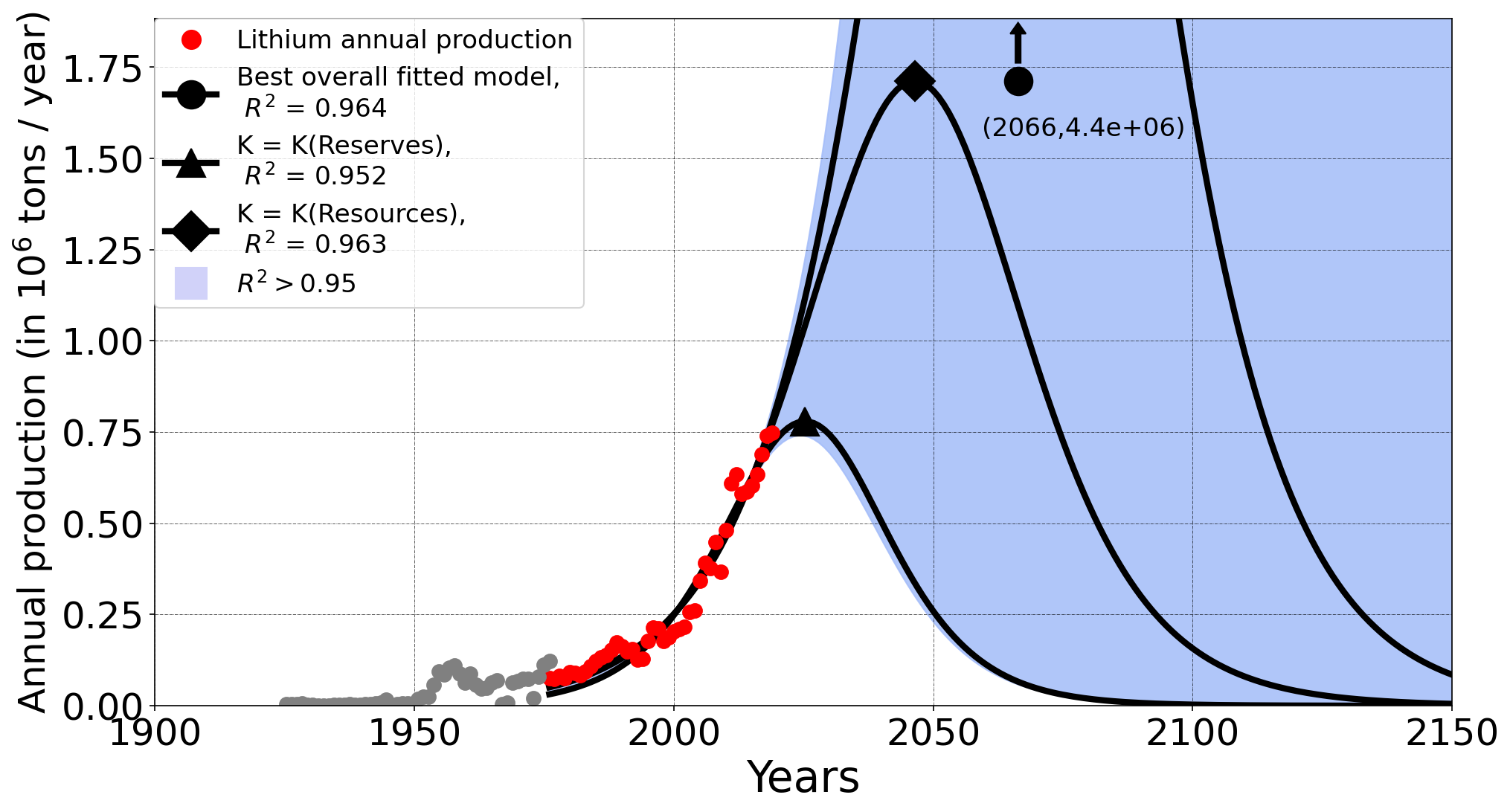}
    \caption*{}  \label{fig:Exponential} 
\end{figure}

\begin{figure}[H]
    \centering
    \includegraphics[height = 3cm]{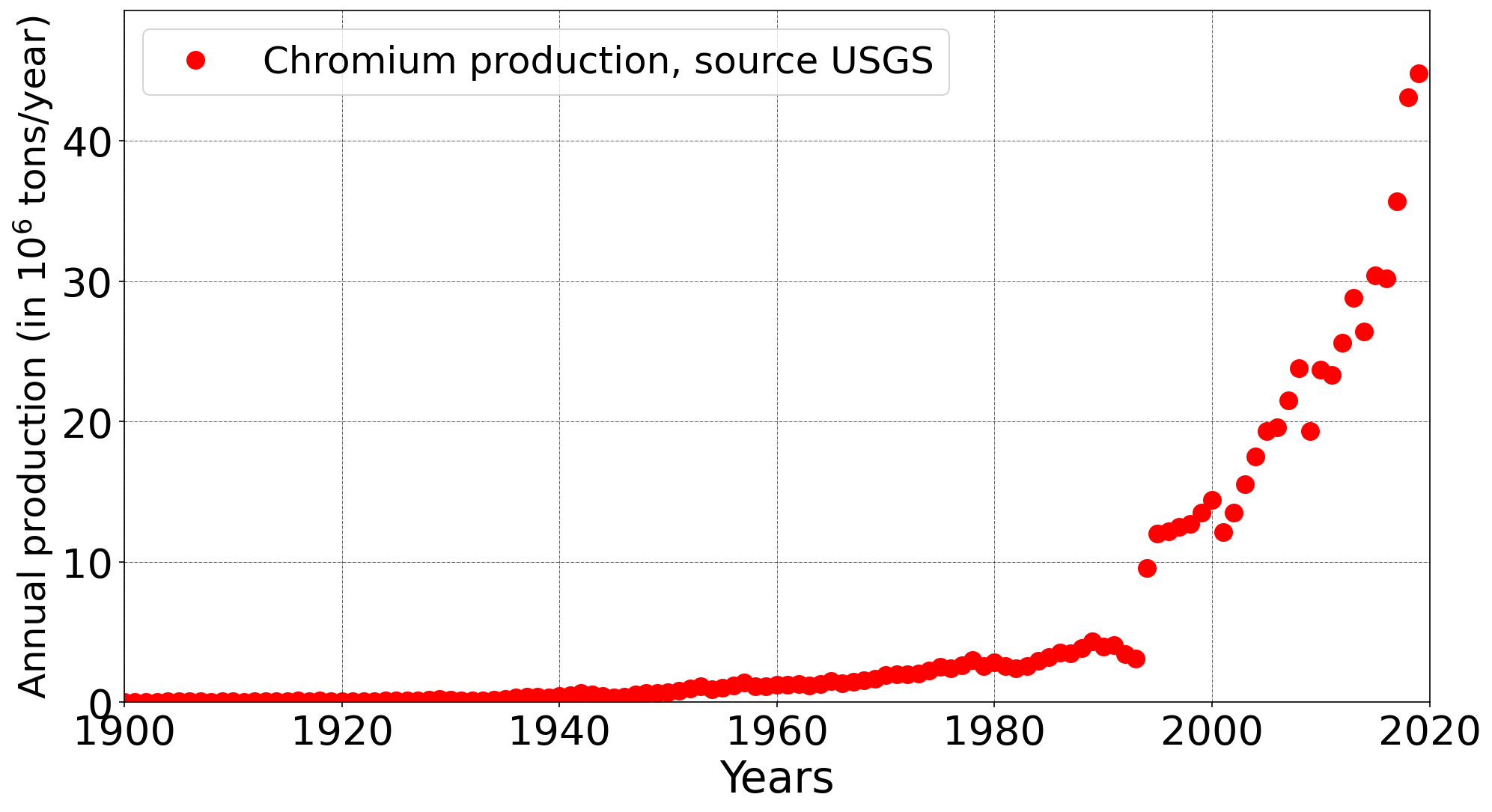}
    \includegraphics[height = 3cm]{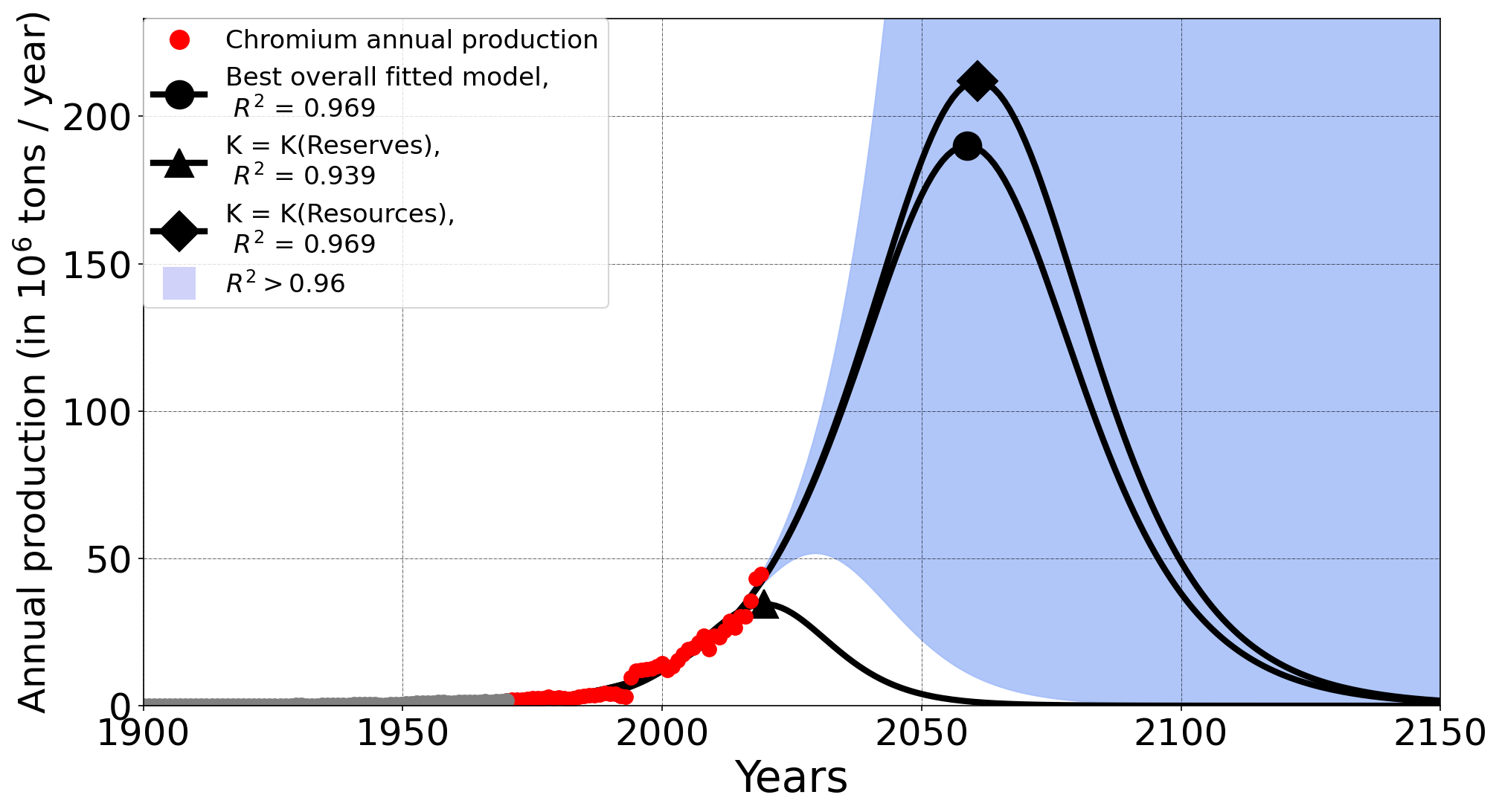}    \caption*{} 
\end{figure}

\pagebreak

\section{Appendix: Multiple trends elements} \label{appendix:Multiple}

{\bf Elements:} Cadmium, Phosphorus, Selenium, Manganese, Cobalt.
\begin{figure}[H]
    \centering
    \includegraphics[height = 3cm]{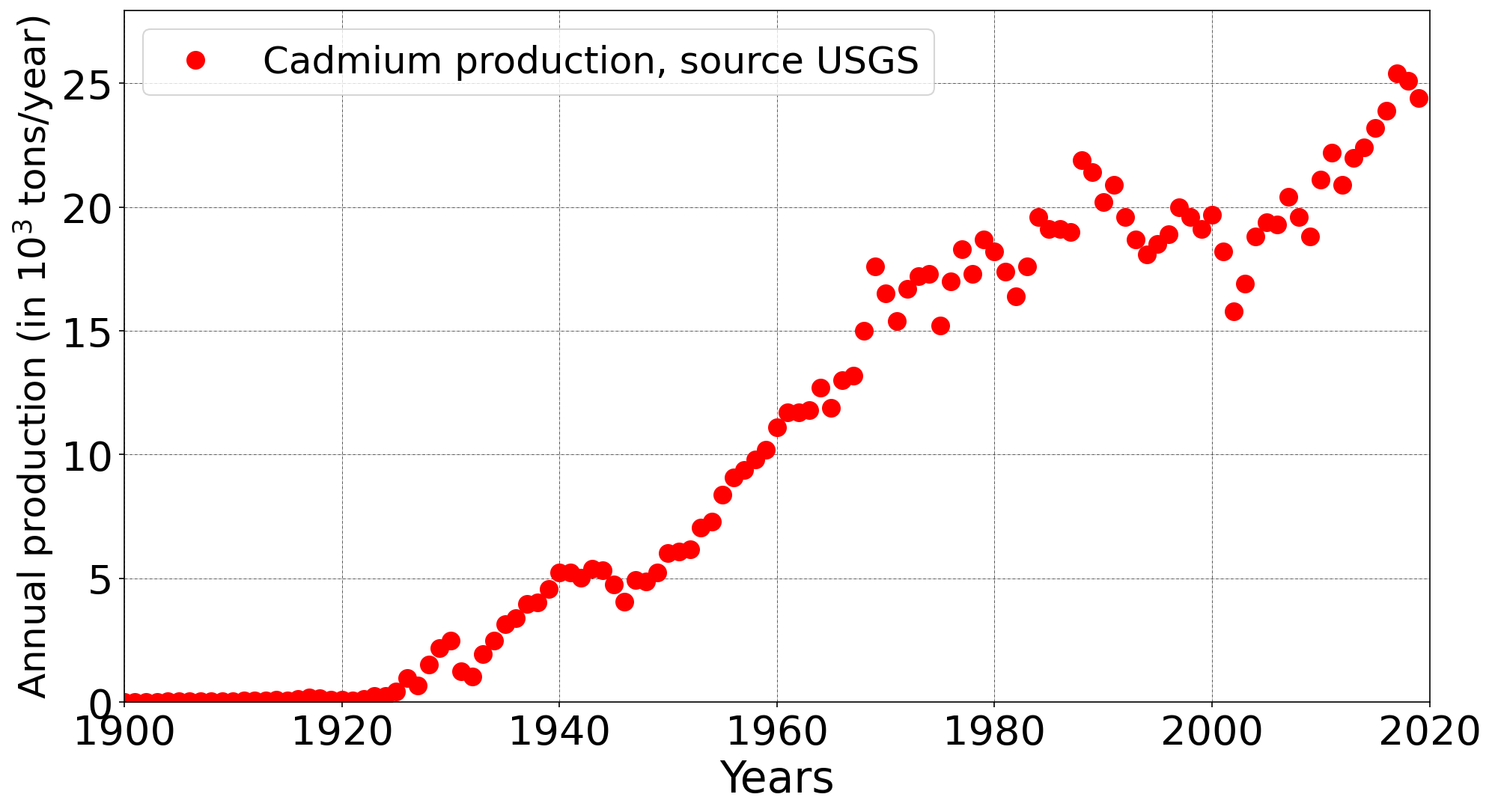}
    \includegraphics[height = 3cm]{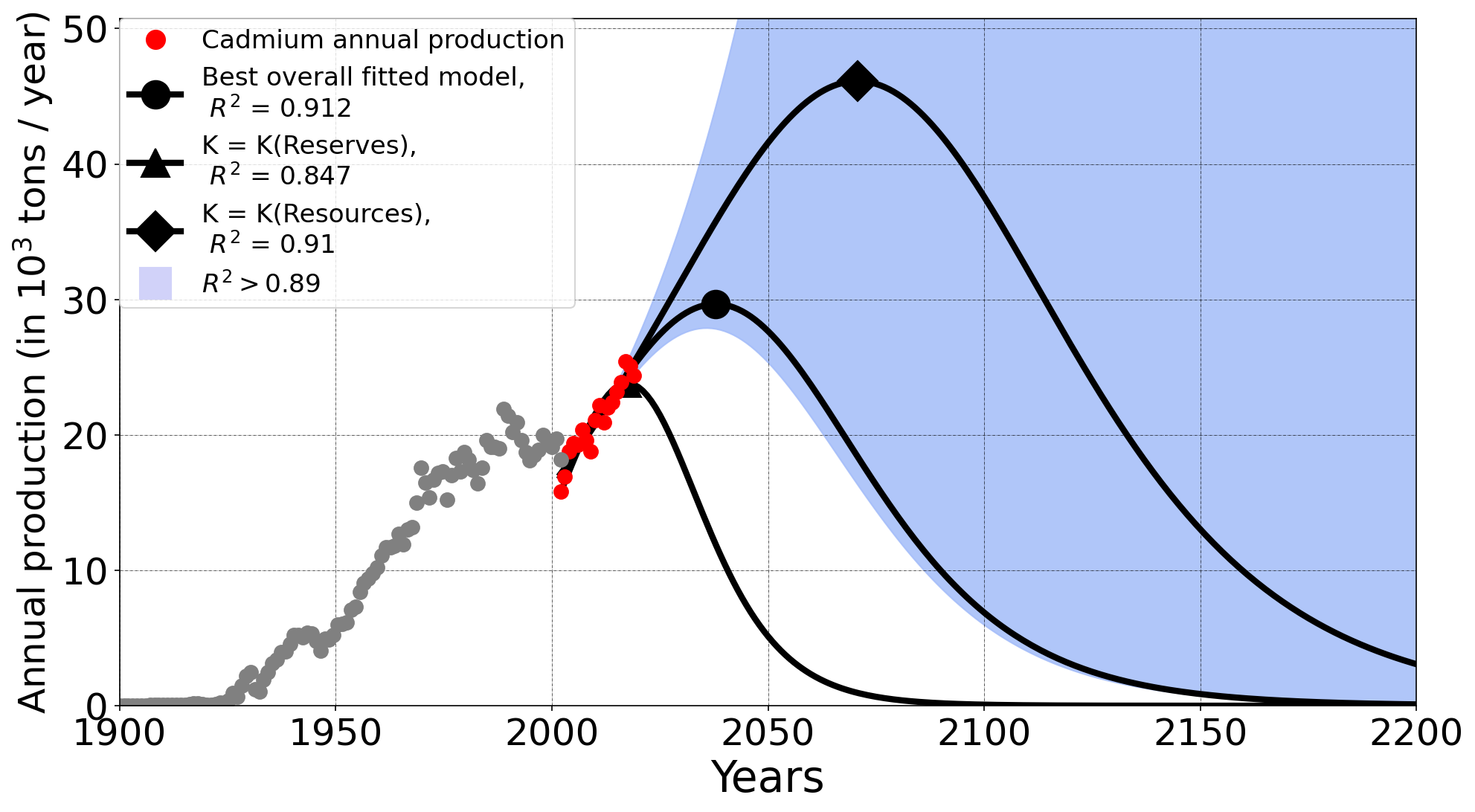}

    \includegraphics[height = 3cm]{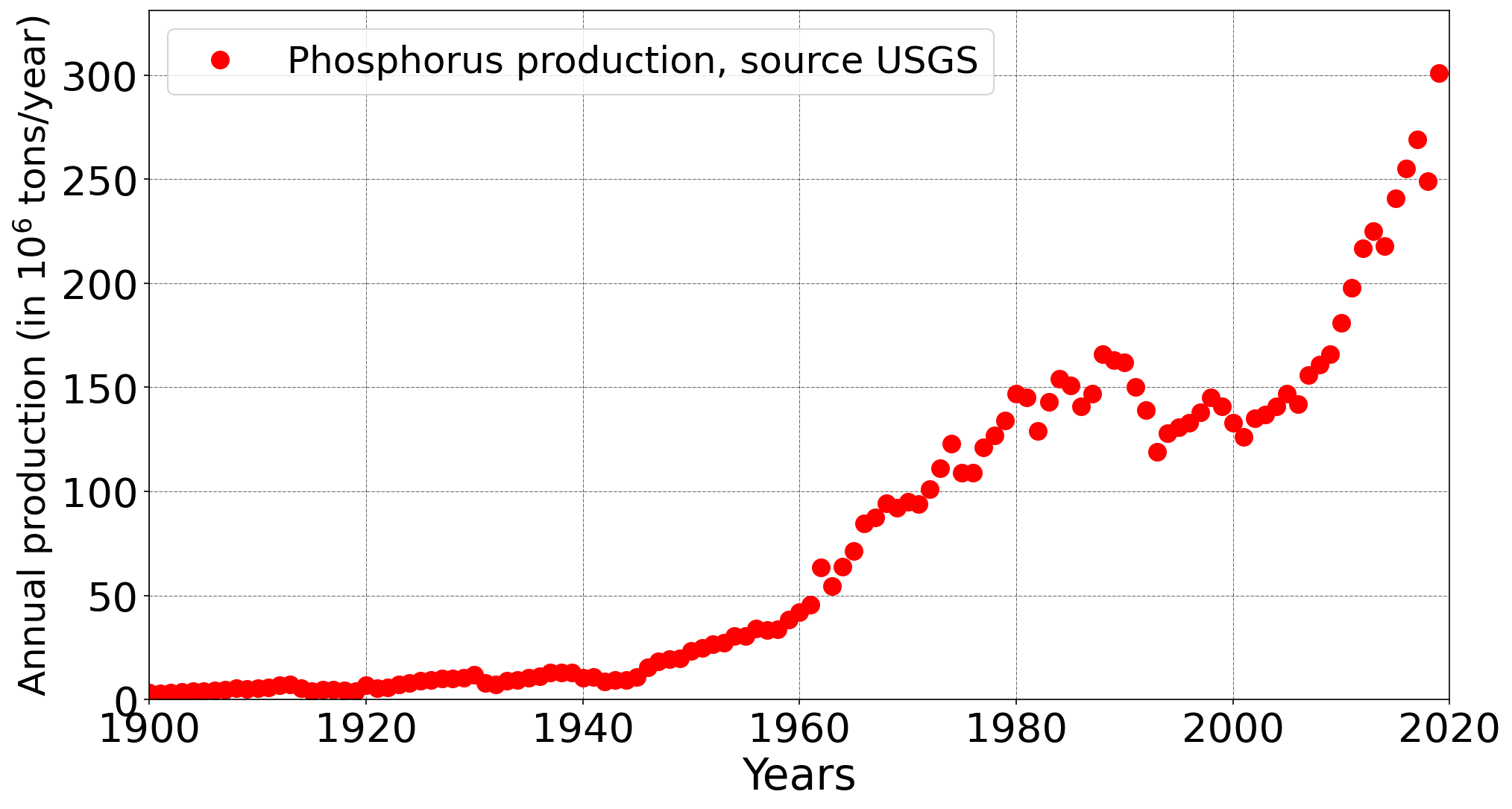}
    \includegraphics[height = 3cm]{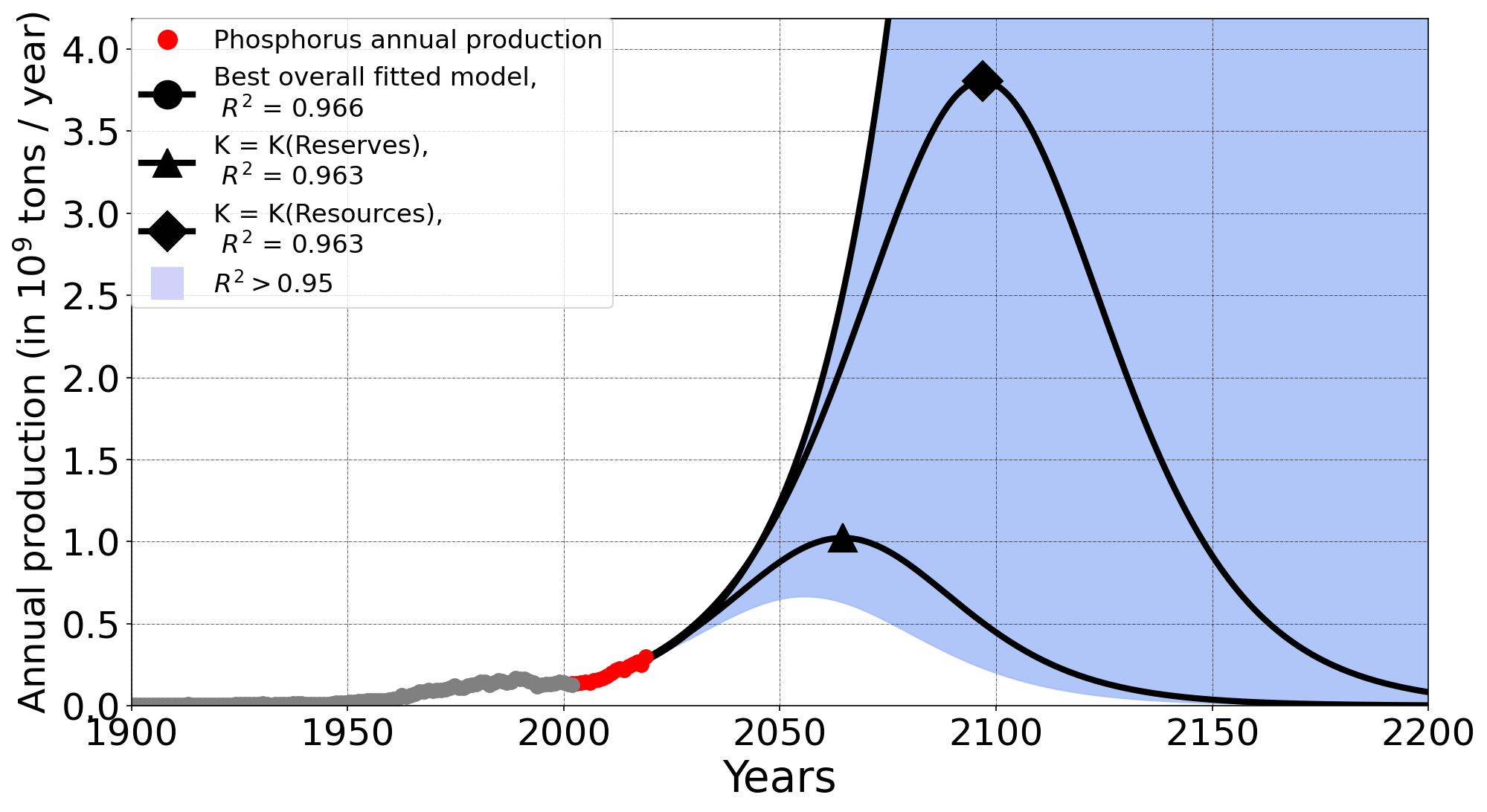}

    \includegraphics[height = 3cm]{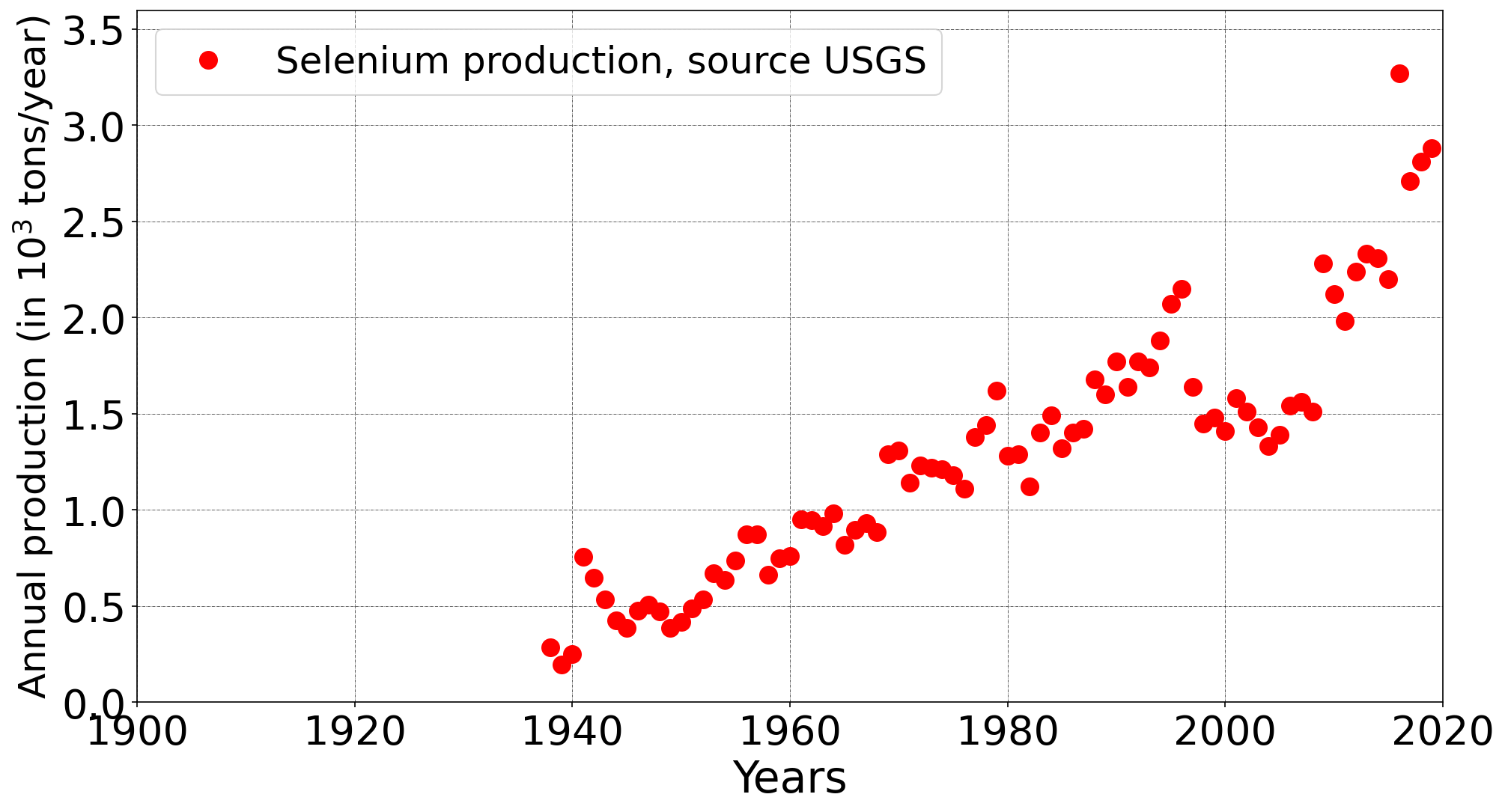}
    \includegraphics[height = 3cm]{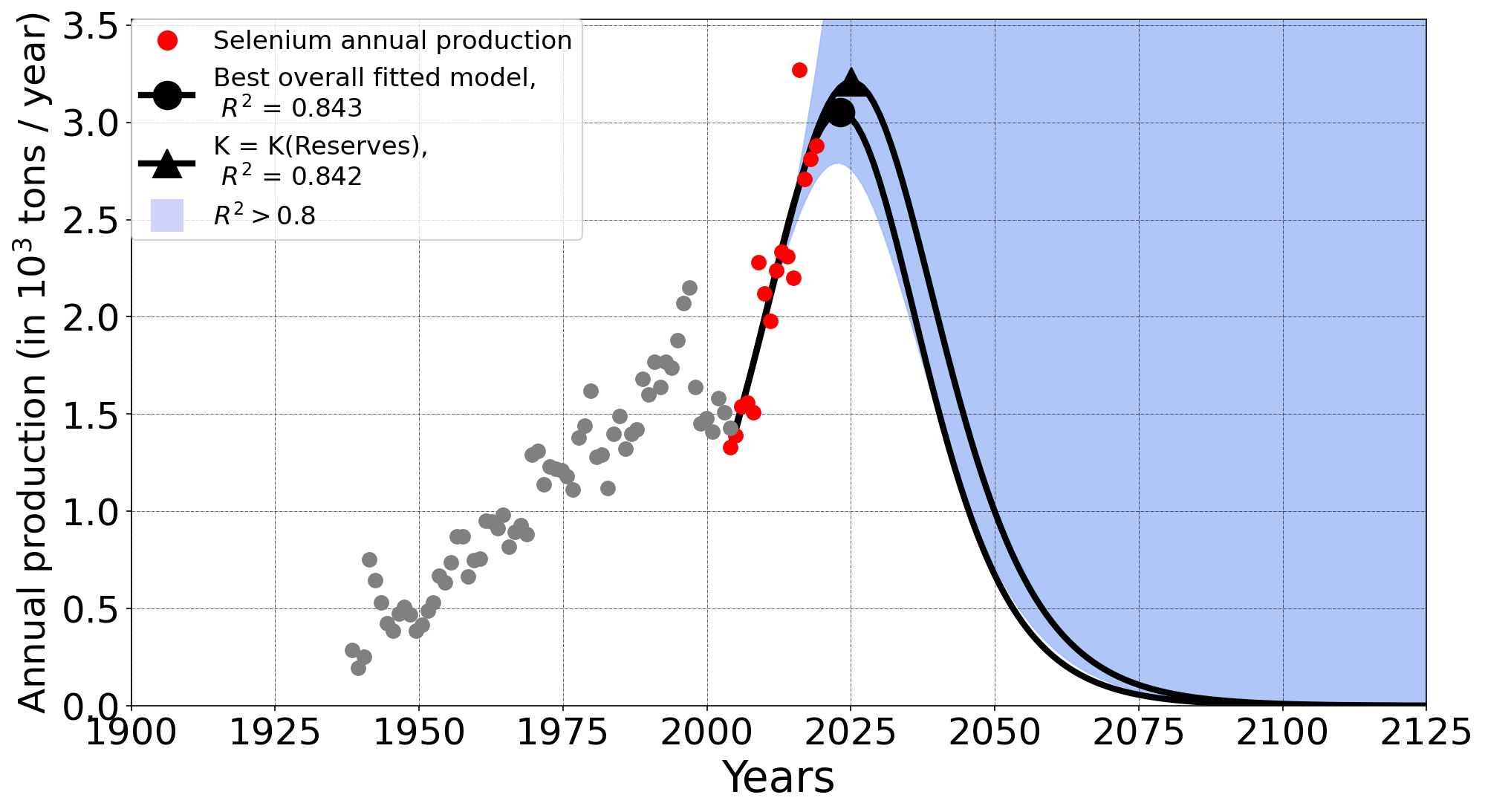}

    \includegraphics[height = 3cm]{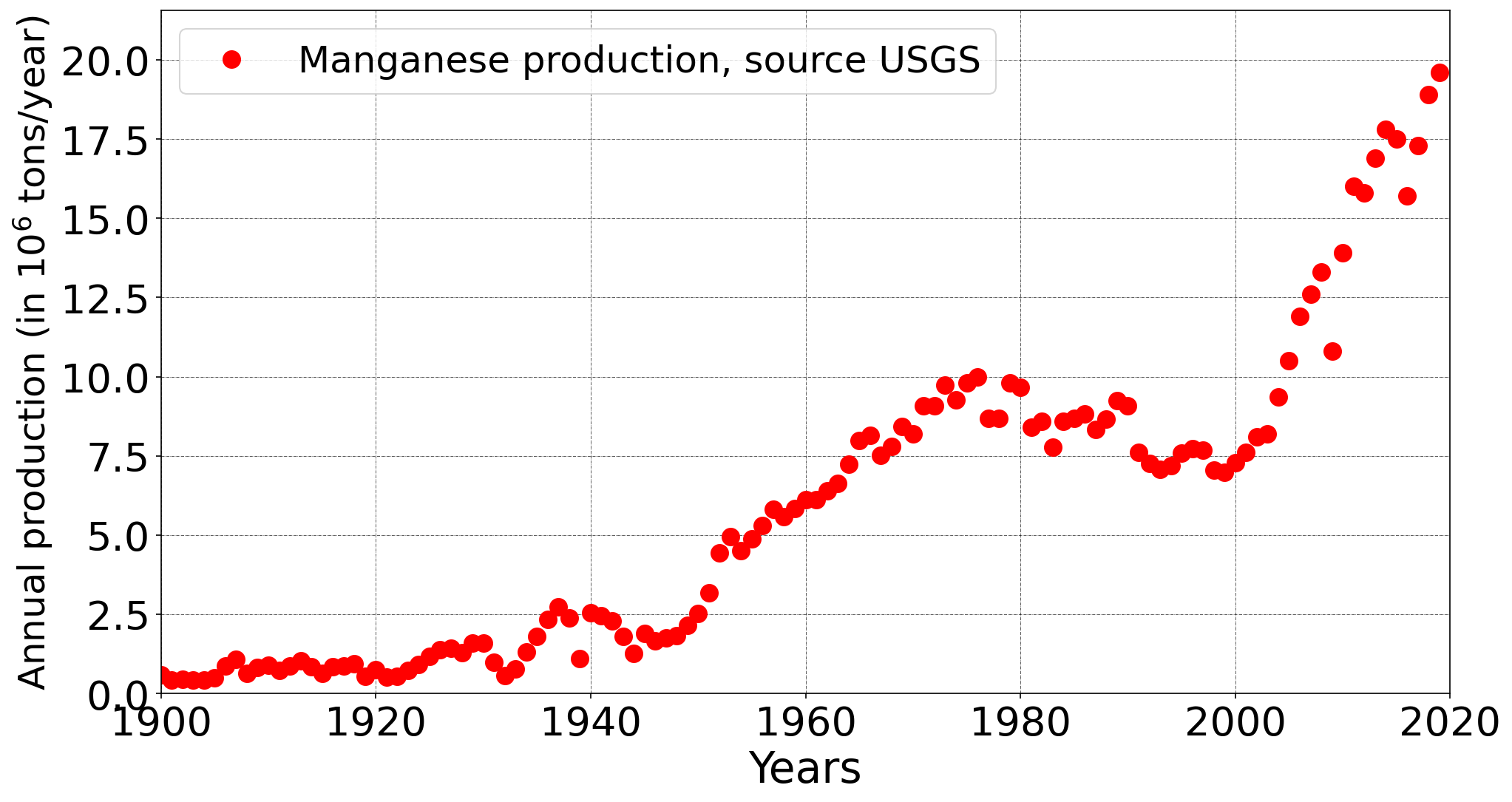}
    \includegraphics[height = 3cm]{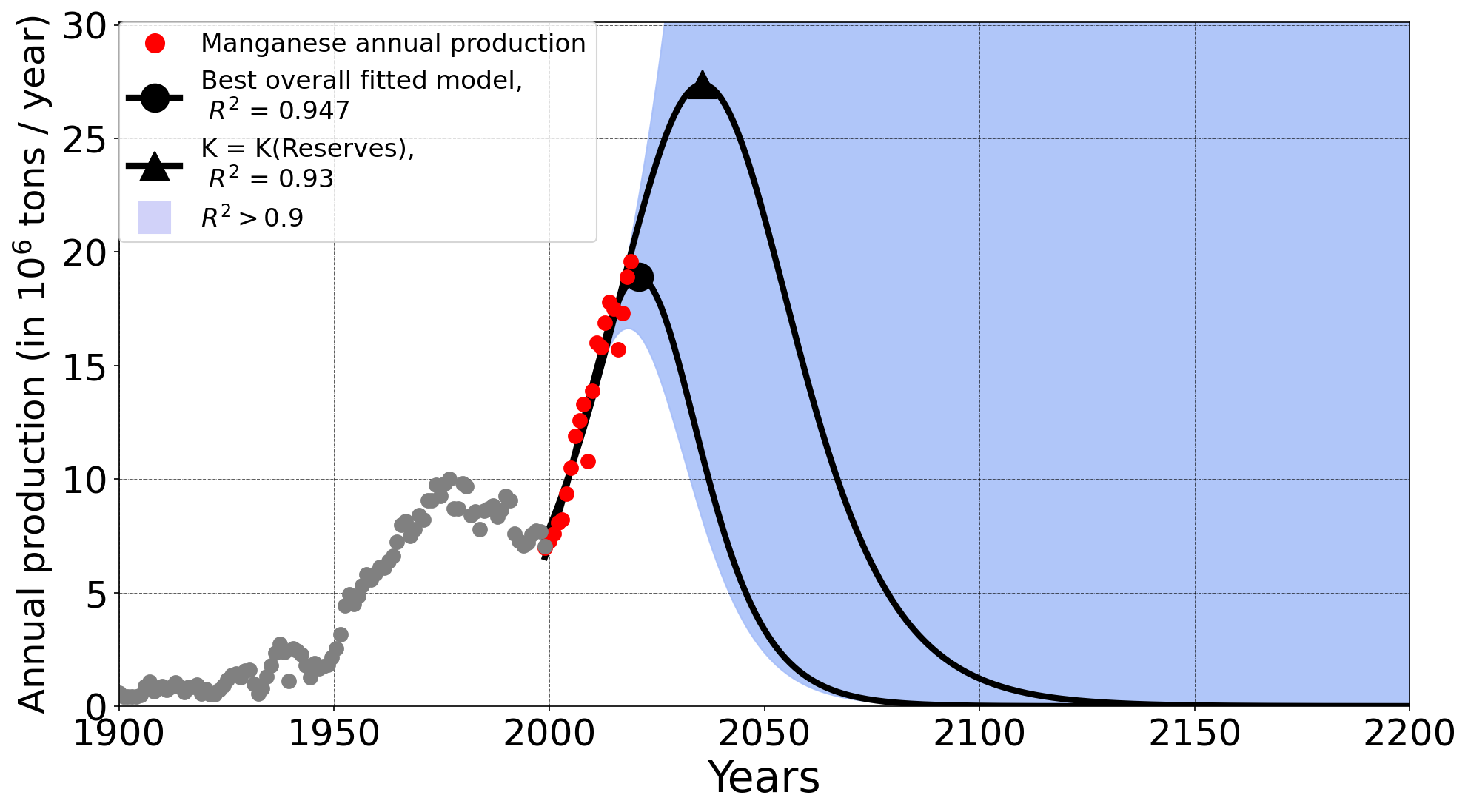}

    \includegraphics[height = 3cm]{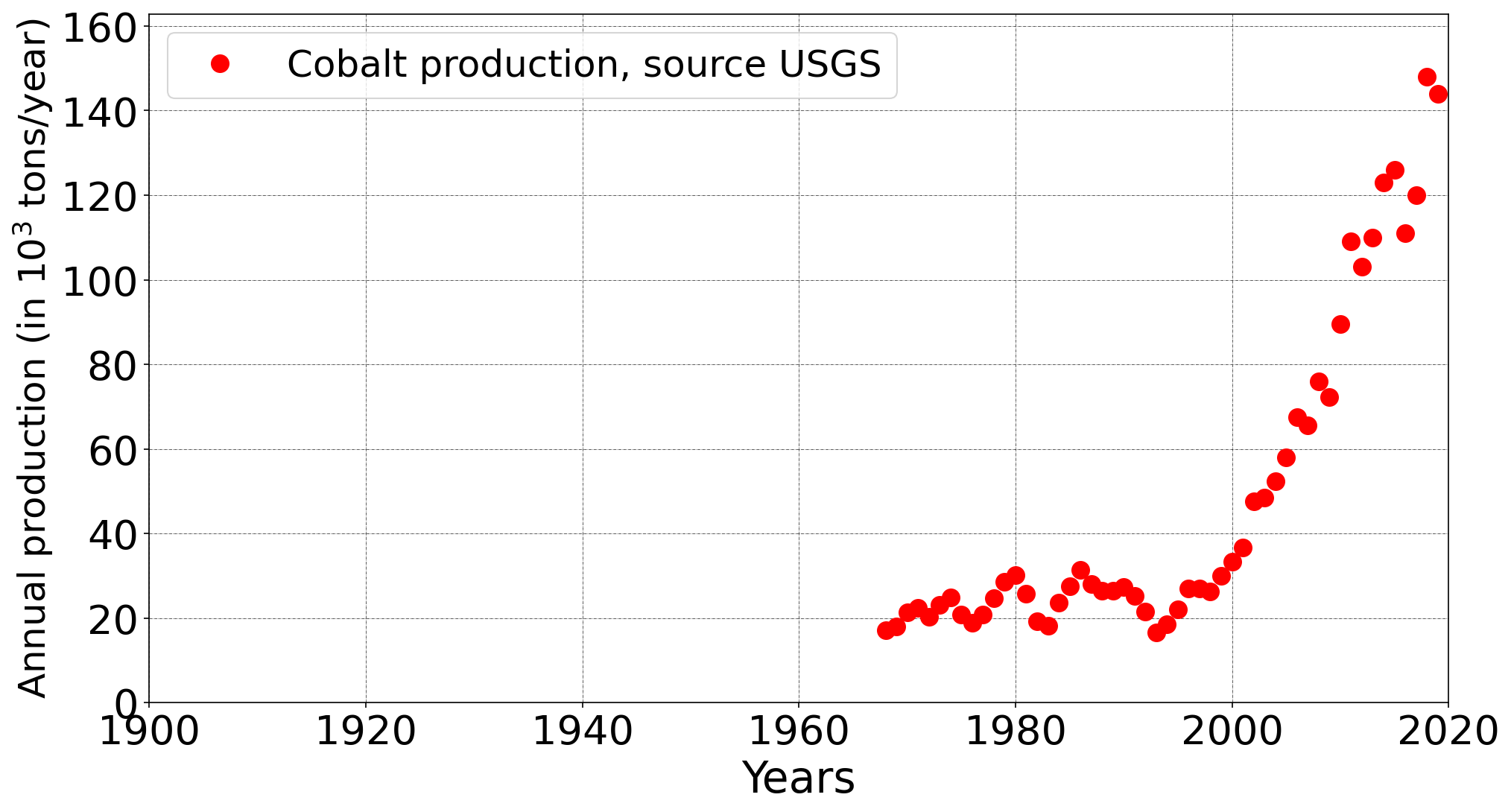}
    \includegraphics[height = 3cm]{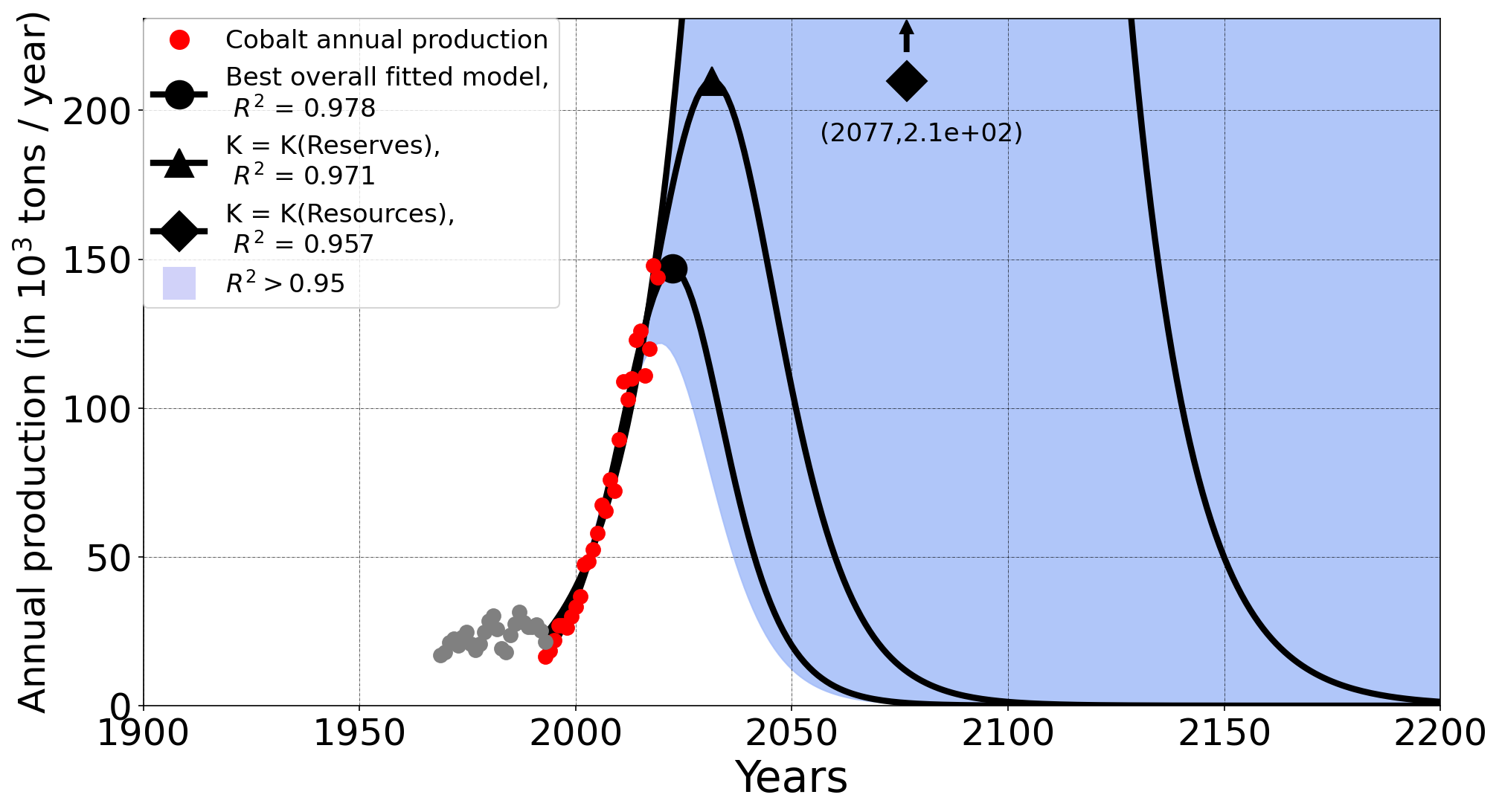}

\caption*{} \label{fig:Multiple} 
\end{figure}

\pagebreak

\section{Appendix: Silver and Rare Earths} \label{appendix:others}
\begin{figure}[H]
    \centering
    \includegraphics[height = 3cm]{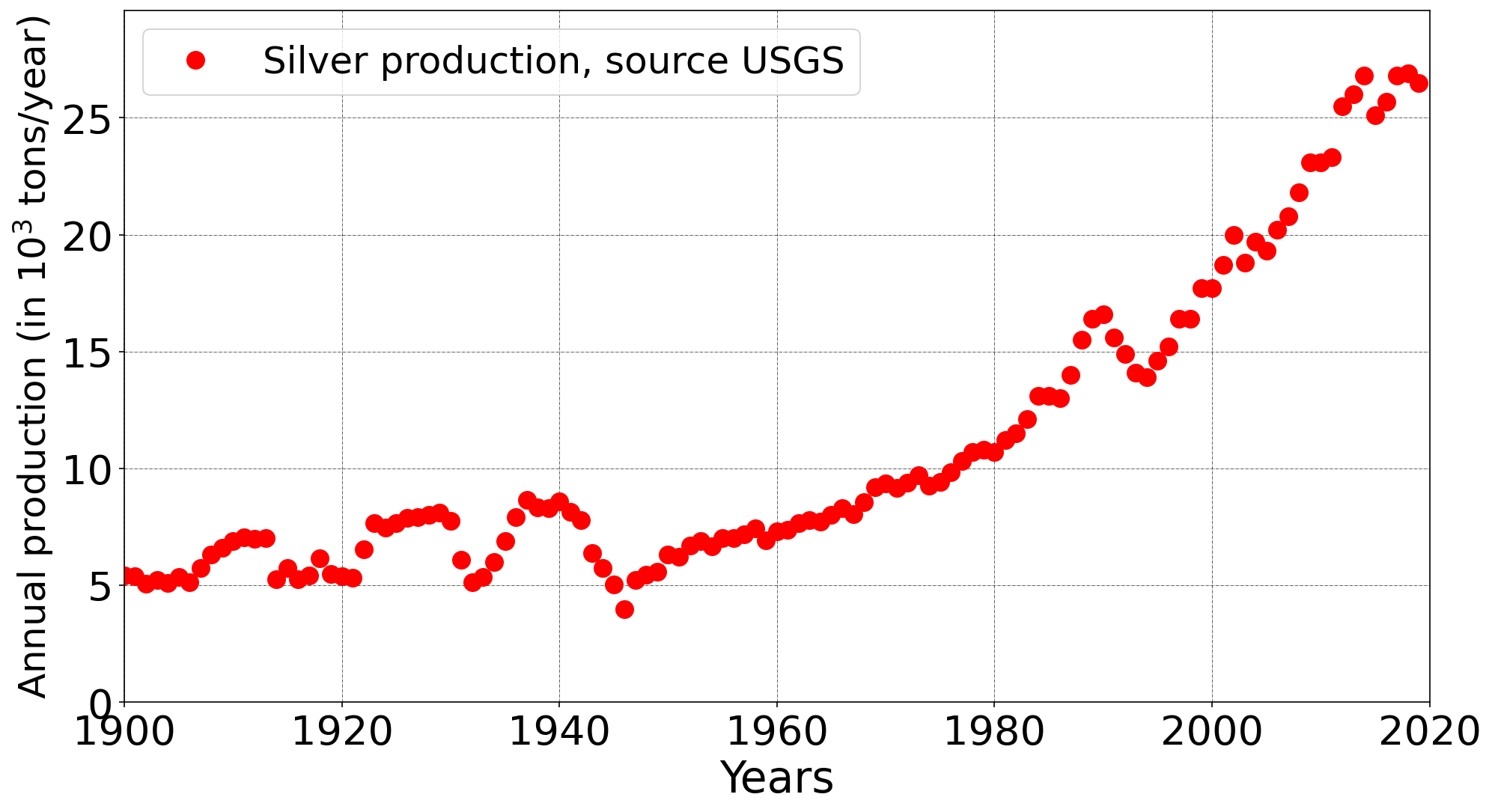}

    \includegraphics[height = 3cm]{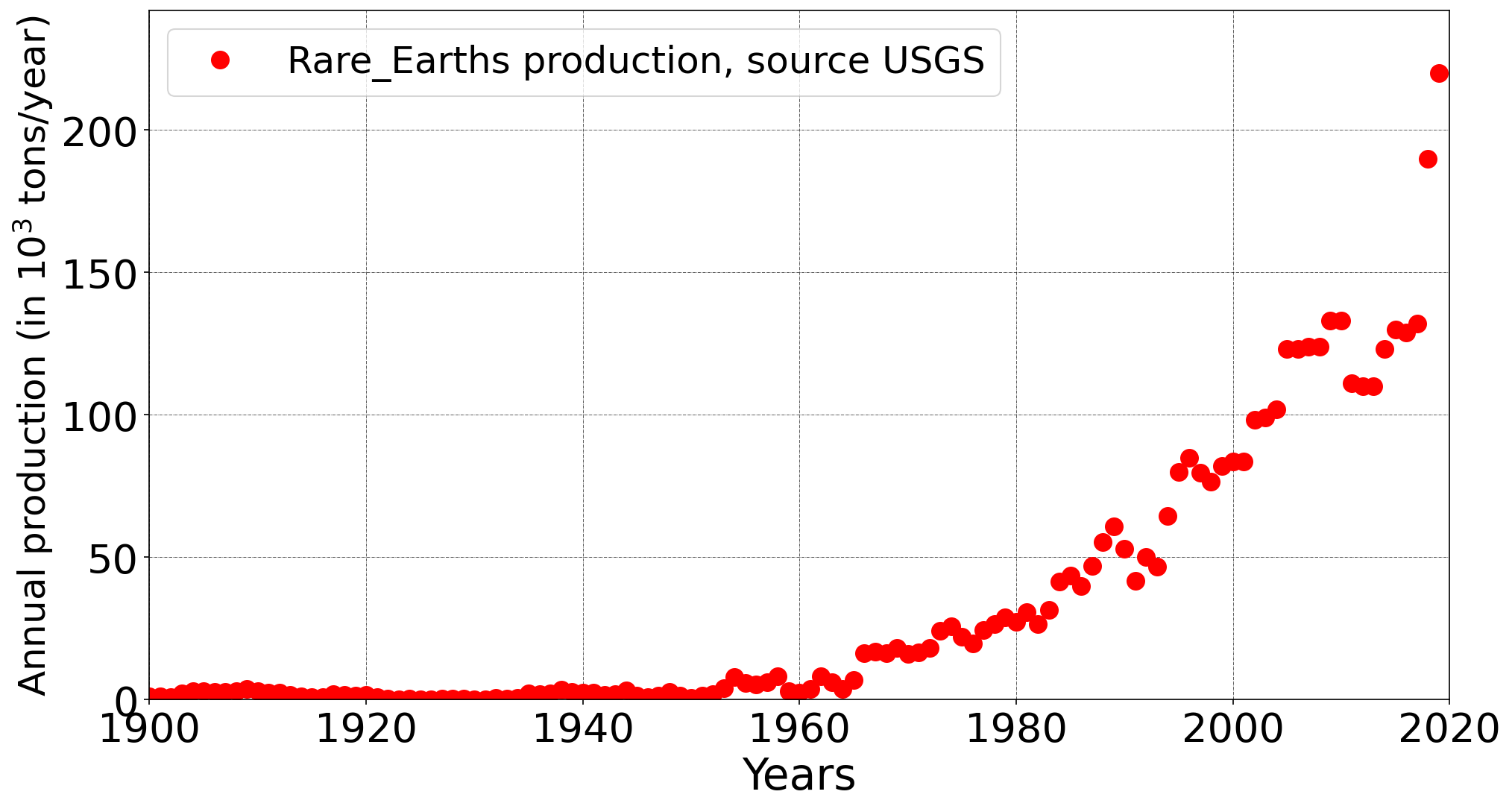}
    \includegraphics[height = 3cm]{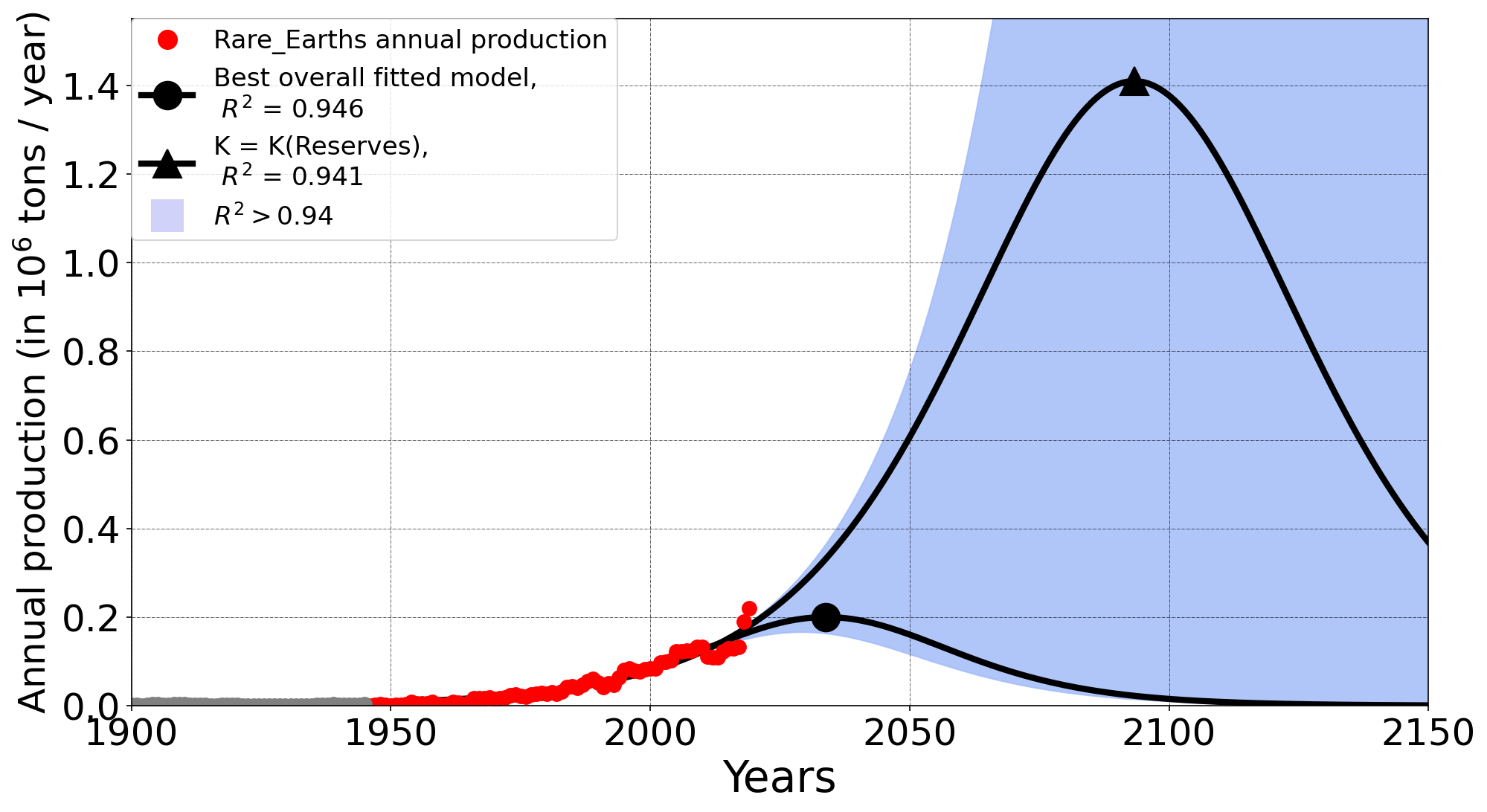}
    \caption*{} \label{fig:others}
\end{figure}

Group, so called \textit{Rare Earths}: ``17 elements composed of scandium, yttrium and the lanthanides". \cite{USGS2021} (Lanthanum, cerium, praseodymium, neodymium, promethium, samarium, europium, gadolinium, terbium, dysprosium, holmium, erbium, thulium, ytterbium, lutetium.) 


\end{document}